\documentclass[preprint,showpacs,preprintnumbers,amsmath,amssymb,superscriptaddress,nofootinbib]{revtex4}
\usepackage{graphicx,color}
\usepackage{amsmath,amssymb}
\usepackage{url}
\usepackage{epstopdf}
\newcommand{\hs}{\hspace*{0.5cm}}

\newcommand{\be}{\begin{equation}}
\newcommand{\ee}{\end{equation}}
\newcommand{\bea}{\begin{eqnarray}}
\newcommand{\eea}{\end{eqnarray}}
\newcommand{\nn}{\nonumber}
\newcommand{\crn}{\nonumber \\}

\newcommand{\bc}{\begin{center}}
\newcommand{\ec}{\end{center}}

\newcommand {\ba}{\begin{array}}
\newcommand {\ea}{\end{array}}
\newcommand{\ben}{\begin{enumerate}}
\newcommand{\een}{\end{enumerate}}

\usepackage{bm}
\usepackage{dcolumn}

\begin{document}

\title{One-loop contributions to neutral Higgs decay $h\rightarrow\mu\tau$}

\author{K.~H.~Phan}\email{phanhongkhiem@tdt.edu.vn}
\affiliation{Theoretical Physics Research Group, Ton Duc Thang University, Ho Chi
Minh City, Vietnam}
\affiliation{Faculty of Applied Sciences, Ton Duc Thang University, Ho Chi Minh City,
Vietnam}

\author{H.T. Hung}\email{hathanhhung@hpu2.edu.vn}
\affiliation{Department of Physics, Hanoi Pedagogical University 2, Phuc Yen, Vinh Phuc, Vietnam}


\author{L.T. Hue}\email{ lthue@iop.vast.ac.vn}
\affiliation{Institute for Research and Development, Duy Tan University, Da Nang City, Vietnam}
\affiliation{Institute of Physics, Vietnam Academy of Science and Technology, 10 Dao Tan, Ba
Dinh, Hanoi, Vietnam }

\begin{abstract}
 In calculating one-loop contributions to amplitudes of  the lepton-flavor violating decays of the neutral
 Higgses (LFVHD) to different flavor charged leptons, the analytic expressions  can be written in term of
 Passarino-Veltman functions. Then, they can be  computed numerically by LoopTools \cite{looptool}.
 Another approach is using suitable analytic expressions established for just this particular case.
 We compare numerical results obtained from LoopTools and those computed by different expressions that
 have been applied recently.  Then we derive  the preferable  ones that are applicable for  large ranges
 of free parameters introduced in extensions of the standard model. For illustration, the LFVHD
 in a simple model, which has been  discussed recently, will be investigated more precisely.
\end{abstract}
\maketitle
\section{Introduction}
\label{sec:intro}
In 2012, the detection of the standard model (SM) Higgs was a milestone of particle physics \cite{higgsdicovery}.
Experimental searches for lepton-flavor violating decays of neutral Higgses (LFVHD) have seen  important new improvements recently \cite{exLFVh,LFVhtauemu}.
Motivated by this, studying for signals of LFVHD at colliders in forthcoming years has been being given
attention \cite{LFVcol}. Also, the LFVHD of neutral Higgses  were  studied widely \cite{LFVgeneral,LFVgeneral2,SUSY}
where  one-loop contributions were computed in many specific frameworks such as  suppersymmetric (SUSY) \cite{SUSY},
lepton-flavored dark matter \cite{lfvhloop1}, leptoquark \cite{lfvhloop2,lfvhloop3,lfvhloop4},  seesaw  \cite{seesaw,iseesaw,apo1,apo2},
extended  mirror  fermion \cite{1loopmirror},  3-3-1 and radiative neutrino mass models \cite{hue,apo3,radiativenu1}, and
others \cite{lfvhlomore,SBaek}.   Some  recent models assume the presence of tree level lepton-flavor violating (LFV) couplings \cite{lfvhtree} in order to  explain successfully the large excess of  LFVHD of the SM-like Higgs boson noted by the LHC.

The one-loop contributions to LFVHD in SUSY models  are usually formulated by the mass insertion method \cite{SUSY}, which
will lose accuracy when the new (SUSY) scale is not much larger than the electroweak scale. There is another way that is
applied to both SUSY and non-SUSY models, in which  one-loop contributions are written in terms of the PV-functions \cite{SUSY,seesaw,iseesaw} before they are   numerically investigated using well-known computation packages \cite{PVmath}.

 Recently, there have been efforts to find  convenient analytic expressions used for calculating one-loop contributions
 to LFVHD in non-SUSY models, without using numerical packages.  The reason is that it is  advantageous for studying models with simple loops contributing to LFVHD.  In addition, it is extremely useful for qualitative estimation of particular loop contributions before making concrete investigations. The mass insertion method may not work well because these models predict new relevant scales rather close to the electroweak scale.  Refs. \cite{lfvhloop3,radiativenu1,1loopmirror} did not exhaustively solve  integrals in analytic formulas, although Ref. \cite{lfvhloop3} did cross-check   them numerically with  those built from PV-functions.  According to our experience, these expressions will not work well if the loops contain two small masses of virtual particles like active neutrinos or MeV masses of the exotic neutrinos in (inverse) seesaw models. Ref.  \cite{lfvhloop1} tried to find final analytic forms solving all  integrations, but they are only valid in very special cases, e.g.,  when masses of new particles are much heavier than the SM-like
 Higgs mass. Refs. \cite{lfvhloop4,lfvhlomore} used directly an expression containing $C_0$ functions-the simplest scalar integral in the set of one-loop-three point PV-functions.  But there were no analytic formulas for $C_0$ introduced. It was  calculated using one set of fixed values of internal masses and external momenta.  Ref. \cite{apo1} used some particular assumptions for evaluating approximate analytic forms of one-loop contributions to LFVHD in a radiative neutrino mass model.  In an effort to investigate LFVHD in a 3-3-1 model \cite{hue},
 an analytic expression for $C_0$ was introduced, needing only reasonable conditions of very small masses
 of normal charged leptons $e,\mu$ and $\tau$, corresponding to the  approximate zero on-shell momenta. This result was derived as a particular case of the general expression given in \cite{Hooft}. It is then very easy to deduce all other one-loop-three point PV-functions, which are well-known as  $C$-functions.
 One of our main purposes is proving numerically the very consistency of  LoopTools \cite{looptool} and  these analytic expressions. We then  compare  them with other formulas  that have been used recently. In particular, we will study  formulas of $C$-functions introduced in \cite{lfvhloop1}  and \cite{apo1} with the aim of  finding regions of parameter space that these two expressions are still valid.

We would like to stress that,  in calculating LFVHD at the one--loop level, the key problem in constructing  simple analytical forms of PV-functions is that the external momentum of the SM-like Higgs bosons cannot be taken to zero.  It seems  more dangerous when  decays of heavy neutral Higgs bosons in models beyond the SM are considered. Many analytic forms of the one-loop-two-point PV-functions, denoted  $B$-functions, are available and very consistent with LoopTools  \cite{Hooft,Denner}, such as expressions given in  (\ref{b0i}).  Therefore, we will not consider them in this work. But for a $C$-function, even two external momenta relating with charged leptons can be taken to zero, the momentum of the external Higgs cannot be ignored when the loop contains at least one heavy virtual particle. Hence, the analytic expressions  for three zero external momenta, e.g., those given in \cite{porod}, cannot be applied  to calculating LFVHD in general.

Another new result of this work is that  we will use the analytic formula of $C_0$ to  reinvestigate  the LFVHD in a lepton-flavored  dark matter model introduced in \cite{lfvhloop1}. In particular,  we  will focus on the ranges of  small masses of sleptons and neutral Majorana leptons that cannot apply to  expressions used in \cite{lfvhloop1}.

Our paper is arranged as follows. Section \ref{PVcheck} will check the consistency between numerical results given
by Looptools and analytical formulas introduced in \cite{hue}, concentrating on  the
$C_{0,1,2}$-functions. The other scalar factors of tensor integrals are easily derived by reduction procedures. We then use the analytic formulas to compare with those used in some recent studies. Section \ref{illustration} will restudy the LFVHD in the model given in \cite{lfvhloop1} and discuss on the possibilities of detection new particles predicted by this models at LHC and future colliders. The final section is our conclusion. The three appendices list the PV-functions discussed in this work.

\section{\label{PVcheck}PV-functions for calculating LFVHD}
\subsection{LoopTools versus  analytic expressions in \cite{hue} }
\begin{figure}[h]
  \centering
  \begin{tabular}{cc}
    \includegraphics[width=7cm]{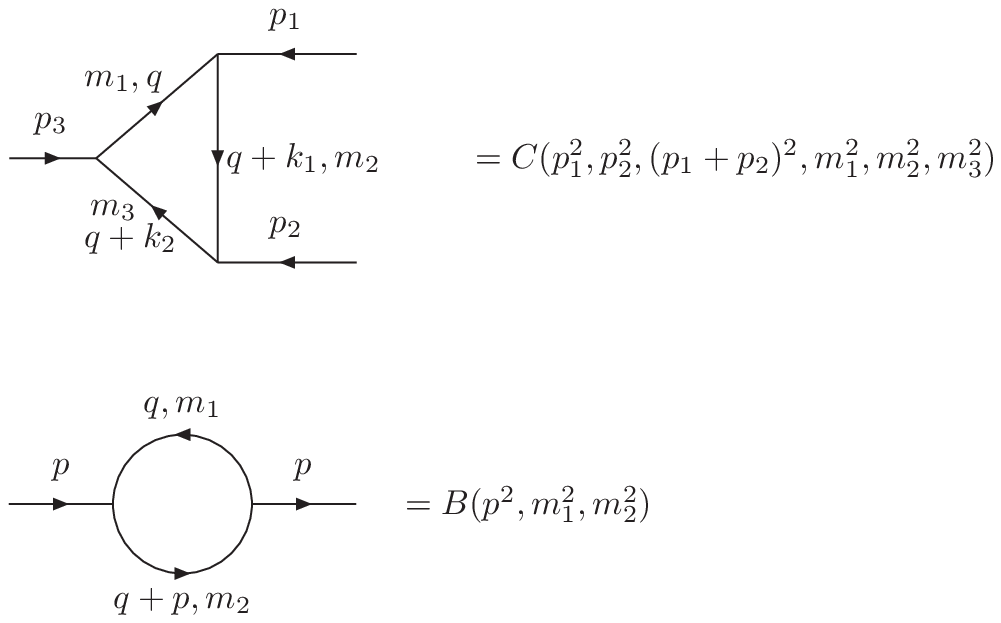} &\includegraphics[width=7cm]{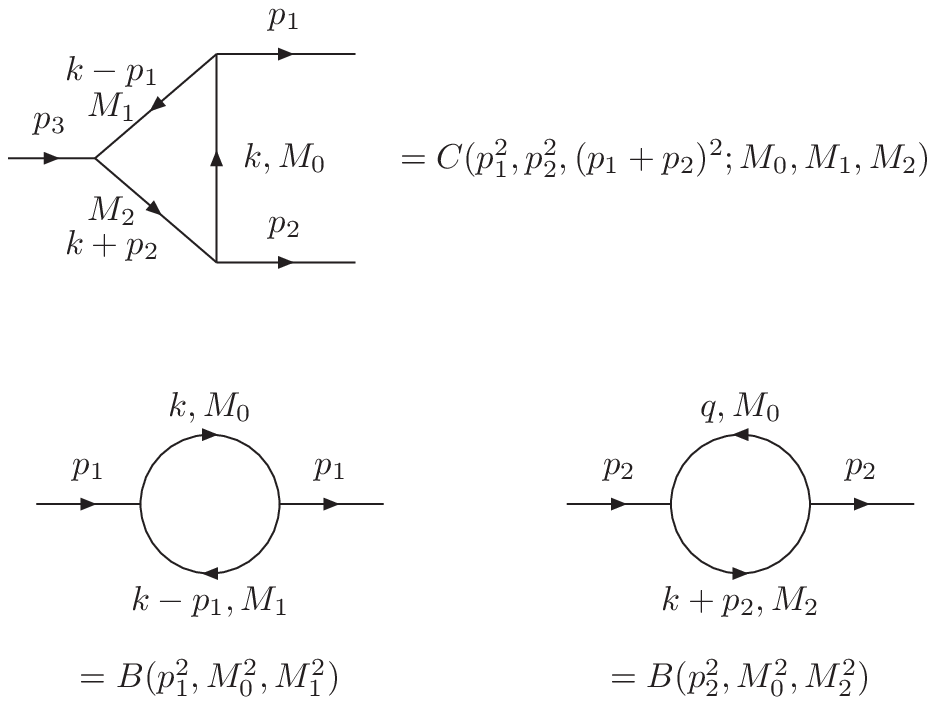}  \\
     \end{tabular}
  \caption{Notation and directions of external momenta. The left panel is from LoopTools,
  where $k_1=p_1$ and $k_2=p_1+p_2$, while the right panel is from \cite{hue}. }\label{Np12}
\end{figure}

The PV-functions relating with one-loop contributions to LFVHD are two- and three-point functions.
Conventions for external momenta are shown in Fig. \ref{Np12}.

 We use a prime to  distinguish notation between LoopTools and analytic forms, i.e.,
 PV functions computed by LoopTool are $C'$-functions with external momenta  $p_1',p_2'$, and $p_3'$.
The notation for analytic forms is unchanged, namely $C$-functions defined in Eqs. (\ref{C0fomula1}) and  (\ref{PV_func}). Relations between the two expressions for  the same one-loop-three-point functions are as follows.   The last external momentum  satisfies the on-shell condition $p_3^2=p'^2_3=m^2_h$, where $m_h$ is the mass of some neutral Higgs boson, including the SM-like Higgs boson.

  For the  scalar function $C_0$ we have
 \be C_0\equiv C_0(M_0,M_1,M_2)= C'_0(p_1^2,p_2^2, m_h^2; M_1^2,M_0^2,M_2^2)\equiv C'_0(M_1,M_0,M_2),  \label{C0func1}\ee
 where the corresponding notation is
 $p'_1\rightarrow -p_1,p'_2\rightarrow -p_2, p'_3\rightarrow p_3, m_1\rightarrow M_1,m_2\rightarrow M_0$ and $m_3\rightarrow M_2$.
 This result is proved by changing the variable $q=-k+p_1$ between the two notations. Another proof is given in Appendix \ref{C0C0p}.

 We  need only one tensor integral $C^{\mu}$ for calculating LFVHD in the 't Hooft Feynman gauge.
  The standard definition for it according to LoopTools   is
 \be  C'^{\mu}(m_1,m_2,m_3)\equiv C'^{\mu}(p'^2_1,p'^2_2,(p'_1+p'_2)^2; m_1^2, m_2^2, m_3^2)
 = C'_1 k_1^{\mu}+ C'_2 k_2^{\mu},\label{cmult}\ee
 where  the inverses of Feynman propagators in the loops are denoted as  $D'_1=q^2-m_1^2$, $D'_2=(q+p'_1)^2-m_2^2$, $D'_3=(q+p'_1+p'_2)^2-m_3^2$, $k_1=p'_1$ and $k_2=p'_1+p'_2$.
 The standard definitions for  the analytic expressions are listed in Appendix \ref{aPV}, namely
 \bea C^{\mu}&\equiv& C^{\mu}(p^2_1,p^2_2,(p_1+p_2)^2; M_0^2, M_1^2,M_2^2) = \frac{i}{\pi^2}\int\frac{d^4k \times k^{\mu}}{D_0D_1D_2}
 = C_1 p_1^{\mu}+ C_2 p_2^{\mu},\label{cmuana}\eea
 where $D_0=k^2-M_0^2$, $D_1=(k-p_1)^2-M_1^2$, $D_2=(k+p_2)^2-M_2^2$.

 Now we  try to find  the relation between $C'^{\mu}$ and $C^{\mu}$ that have relations among momenta, $p'_1=-p_1$, $p'_2=-p_2$, $p'_3=p_3$, and masses $\{m_1,m_2,m_3\}=\{M_0,M_1,M_2\}$. Because $C'^{\mu}$ is finite,
  changing the integral variable $q \rightarrow -k+p_1$ in Eq.  (\ref{cmult}) does not affect the integral value. Therefore,  we get a new expression for $C'^{\mu}$:
\bea C'^{\mu}(m_1,m_2,m_3)&=&\frac{i}{\pi^2}\int\frac{d^4k \times (-k+p_1)^{\mu}}{D'_0D'_1D'_2}
= C'_1 k_1^{\mu}+ C'_2 k_2^{\mu},\label{cmult1}\eea
where  $D'_0=k^2-m_2^2$,  $D'_1=(k-p_1)^2-m_1^2$, and $D'_2=(k+p_2)^2-m_3^2$. Fixing $m_2=M_0$, $m_1=M_1$ and $m_3=M_2$
and comparing Eq. (\ref{cmult1}) with Eq.  (\ref{cmuana}), we obtain an  important equality
\bea  C'^{\mu}(M_1,M_0,M_2)&=&-C^{\mu}+ C_0 p_1^{\mu}\hs \mathrm{or}\crn
-p_1^{\mu} C_1'(M_1,M_0,M_2)&-&(p_1+p_2)^{\mu}C_2'(M_1,M_0,M_2)=-(C_1p_1^{\mu}+ C_2p_2^{\mu})+ C_0p_1^{\mu}. \label{relaCCp}\eea
As a result, the relations between scalar functions are
\be C'_1(M_1,M_0,M_2)=C_1-C_2-C_0, \hs C'_2(M_1,M_0,M_2)=C_2. \label{sfunction}\ee
Now we will  check numerically the consistency between LoopTools and analytic
expressions based on the three equalities shown in (\ref{C0func1}) and (\ref{sfunction}), where  $C'_{0,1,2}(M_1,M_0,M_2)$ is computed by LoopTools;  and  $C_0$, $C'_1=C_1-C_2-C_0$, and $C_2$ are computed using $C_{0,1,2}$ given in Appendix \ref{aPV}.

In many models,  the case of $M_1 = M_2$ often occurs in LFVHD calculations, where  $M_1\;(M_2)$ may be masses of  charged Higgs bosons; fermions including active neutrinos, exotic leptons or  quarks; charged gauge bosons and their Goldstone
bosons. Therefore, we  will check with $M_1=M_2$ in the two following cases: (i) all $M_{0,1,2}$  are new particles from beyond the SM,
(ii) at least one of  these masses is an active neutrinos mass. On the other hand some models, such as the one introduced in Ref. \cite{lfvhloop1},  and contain three different internal masses. Therefore we will consider two more cases: (iii) $M_1=M_0$,
and (iv) $M_1=2 M_0$.  In the first case,  $M_0$ (or $M_1=M_2$) will be fixed with two small and large values of $100$ and $1000$ GeV, respectively. The remaining $M_1=M_2$ (or $M_0$)  will vary from $150$ GeV to $2000$ GeV. In the second case, the lightest active neutrino mass will be fixed as $M_0=10^{-10}$ (or $M_1 = M_2 = 10^{-10}$) GeV with varying $M_1$ (or $M_0$) from $80$ GeV to $2000$ GeV. The virtual mass ranges  chosen here cover all cases of new particles or SM particles like $W^\pm$ gauge bosons in  seesaw models or top quarks in leptoquark models. Regarding the last two  cases, we will fix $M_0=100;\; 1000$ GeV
and change $M_2$ in the range of $(150\; \mathrm{GeV}, 2000\; \mathrm{GeV})$. As an illustration, we  consider only the PV functions relating to LFVHD of the SM-like Higgs boson with  $m_h=125.1$ GeV.

In order to estimate the discrepancy between the analytical results and
LoopTools, we define the relative error as follows:
\begin{eqnarray}
\label{error}
\delta [\%] = \frac{|\mathrm{LoopTools}|
               - |\mathrm{This work}|}{| \mathrm{This work}|}
                \times 100.
\label{delta}\end{eqnarray}
The following results are presented only for the real parts of
the functions $C_0,\; C_1,\; C_2$ in comparison with the corresponding
ones in LoopTools. Although the imaginary parts of these functions are
not shown in this paper,  it is noticeable that they are in perfect
agreement with LoopTools as well (with relative errors all smaller
than $\mathcal{O}(10^{-6}\; \%)$).  The  figures in this section will be  plotted on a logarithmic scale.
\begin{figure}[htpb]
\begin{center}$
\begin{array}{cc}
\vspace*{-1.7cm}
\includegraphics[width=5cm,height=8cm, angle=-90]{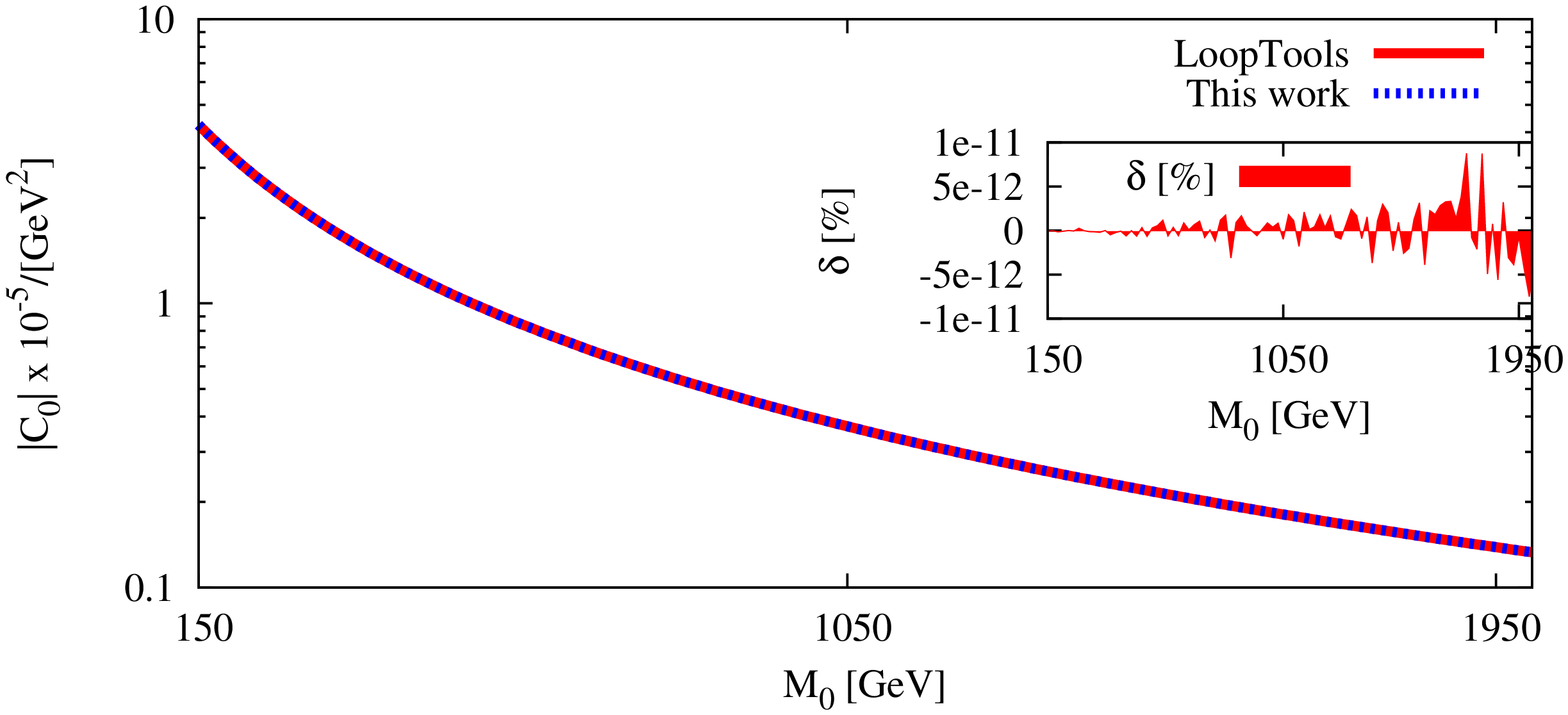} &
\includegraphics[width=5cm,height=8cm, angle=-90]{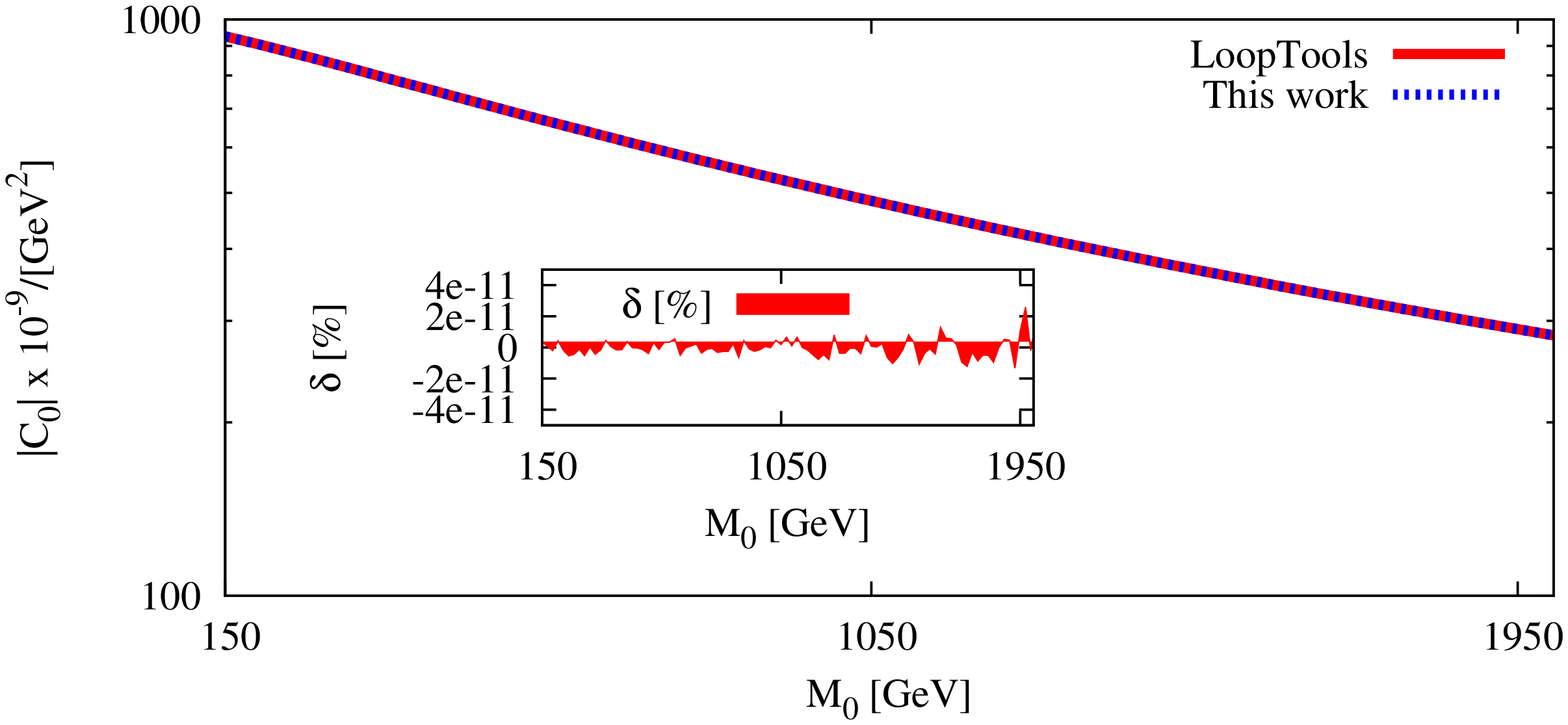}\\
\vspace*{-1.7cm}
\includegraphics[width=5cm,height=8cm, angle=-90]{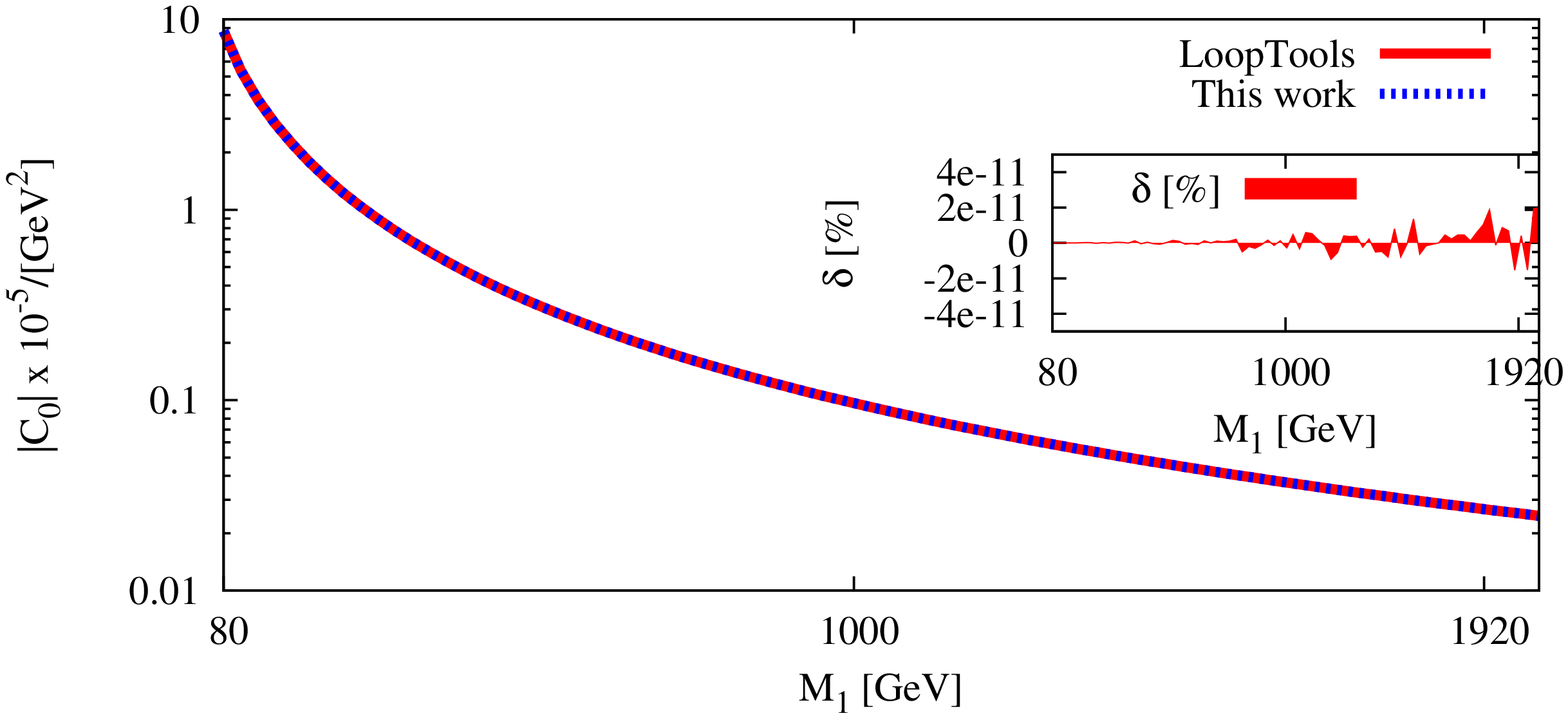} &
\includegraphics[width=5cm,height=8cm, angle=-90]{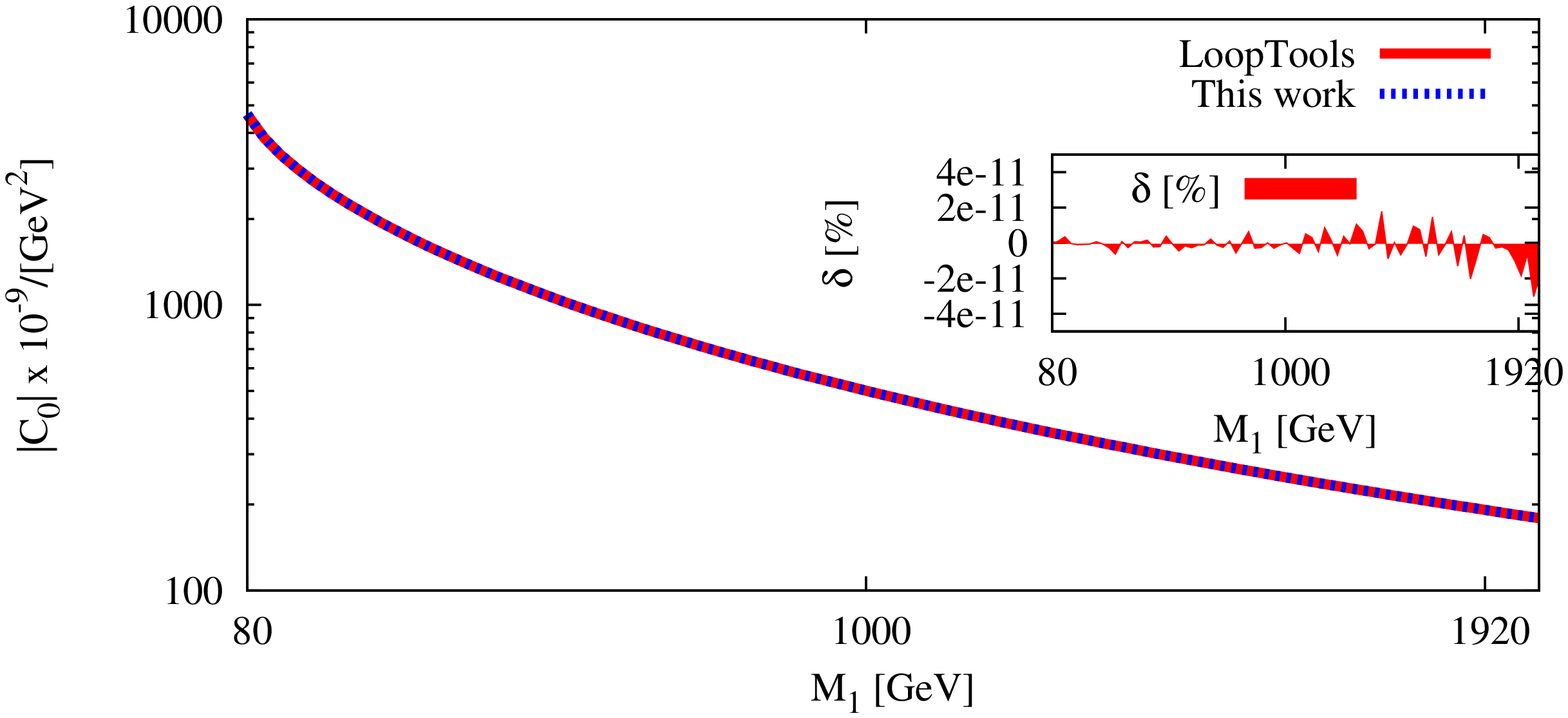}\\
\vspace*{-1.7cm}
\includegraphics[width=5cm,height=8cm, angle=-90]{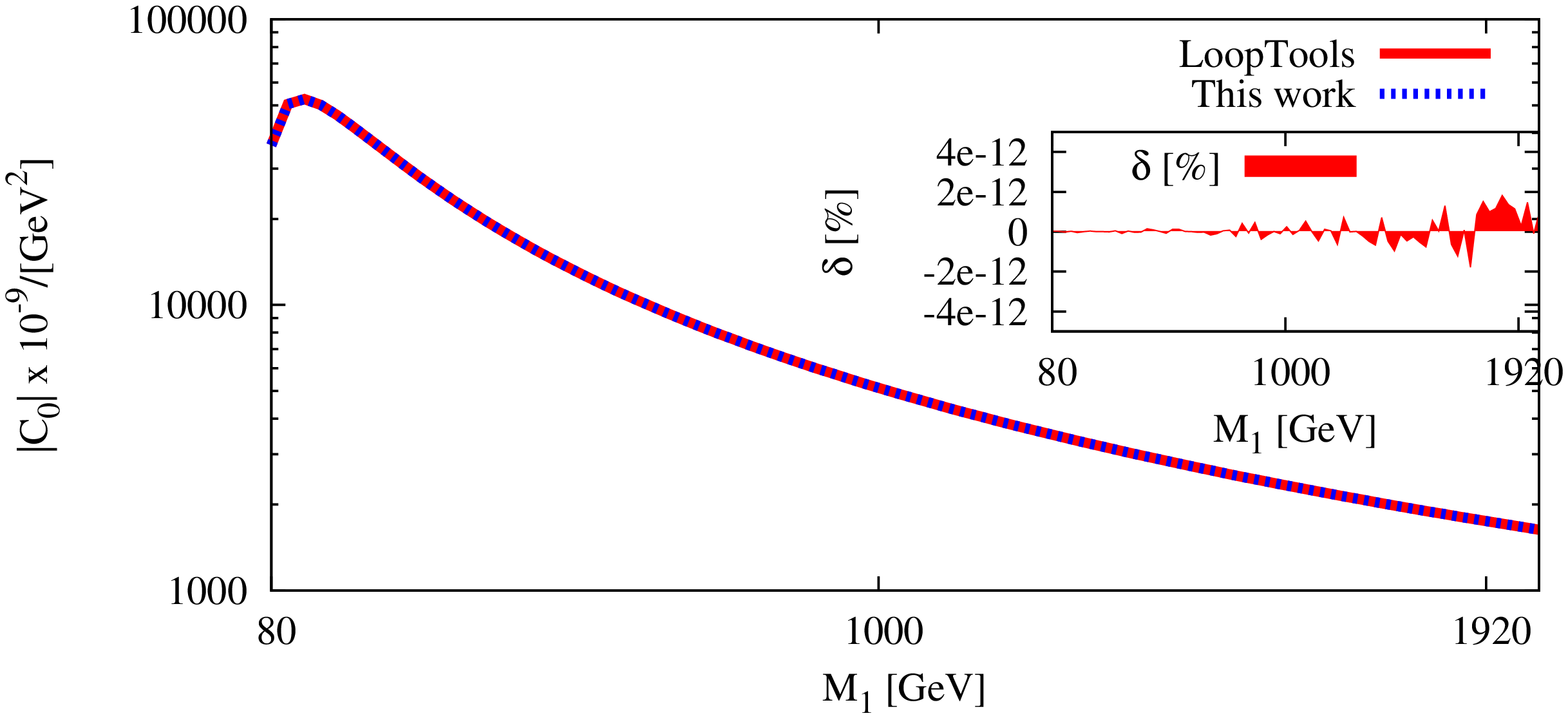} &
\includegraphics[width=5cm,height=8cm, angle=-90]{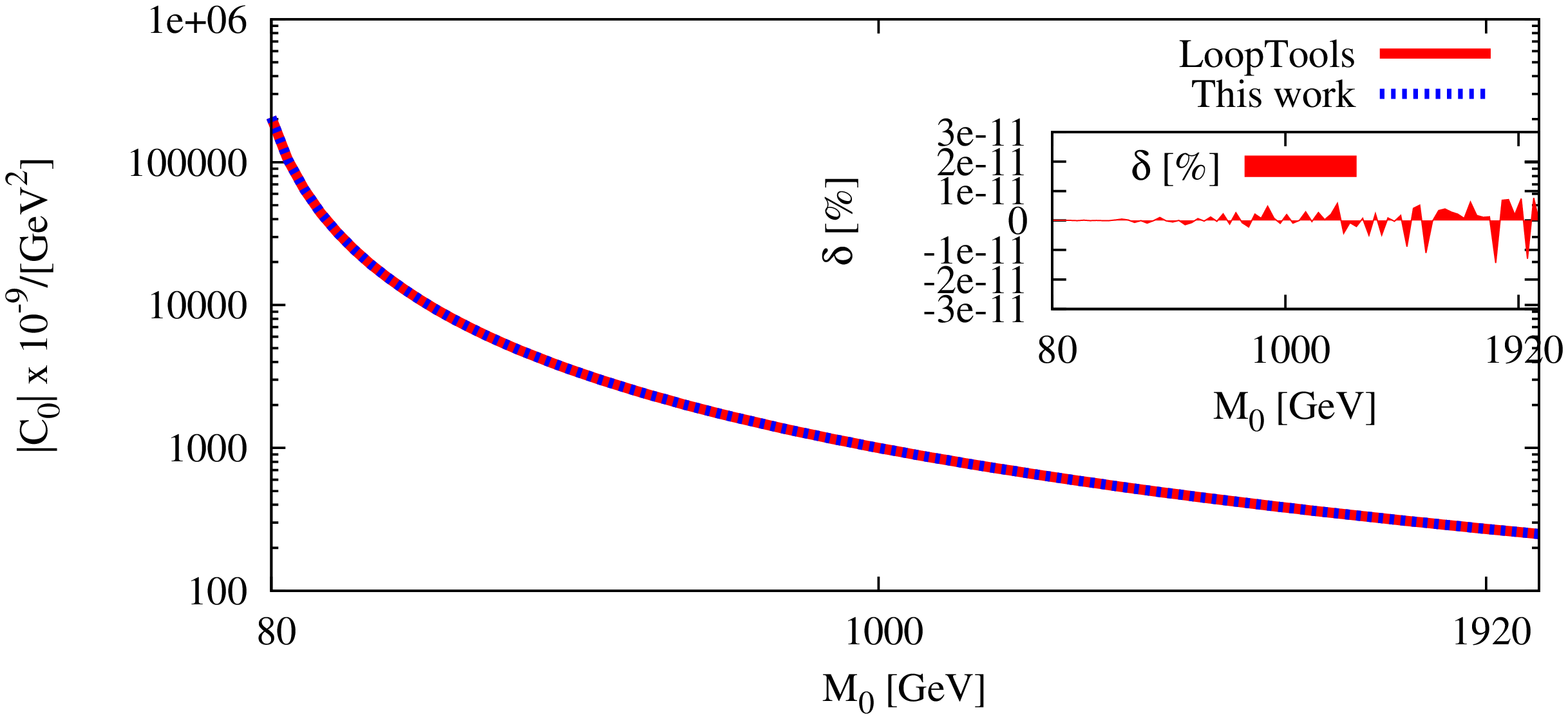}\\
\vspace*{-1.7cm}
\includegraphics[width=5cm,height=8cm, angle=-90]{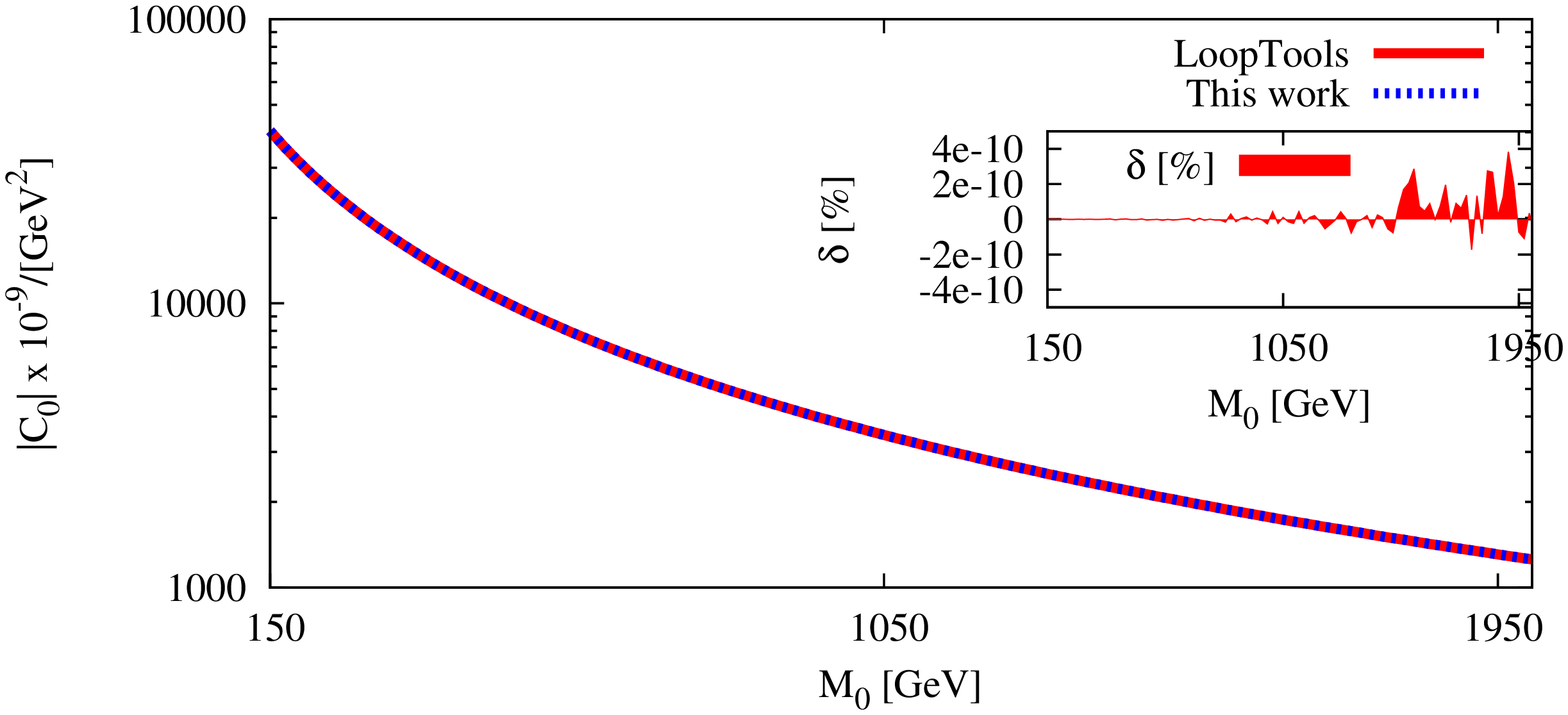} &
\includegraphics[width=5cm,height=8cm, angle=-90]{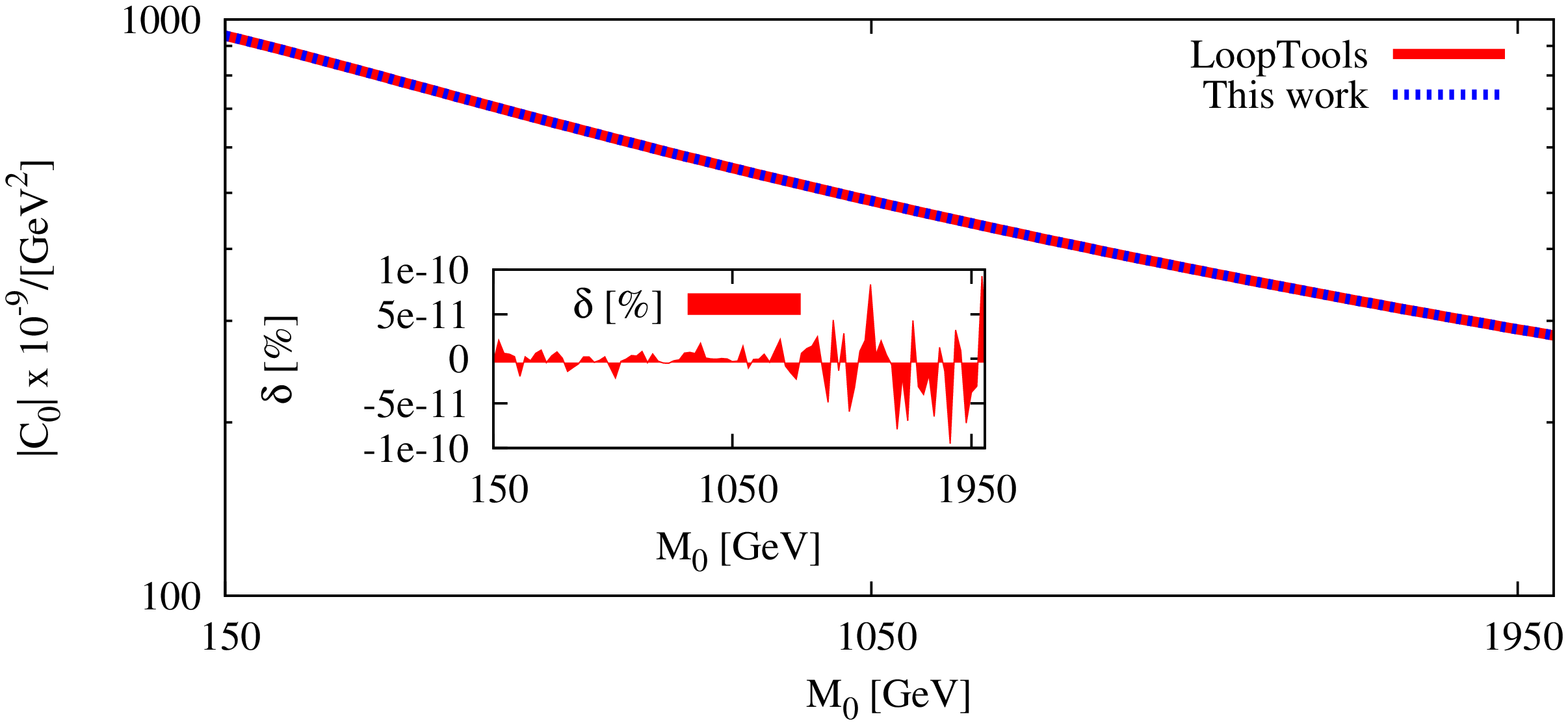}\\
\vspace*{-1.7cm}
\includegraphics[width=5cm,height=8cm, angle=-90]{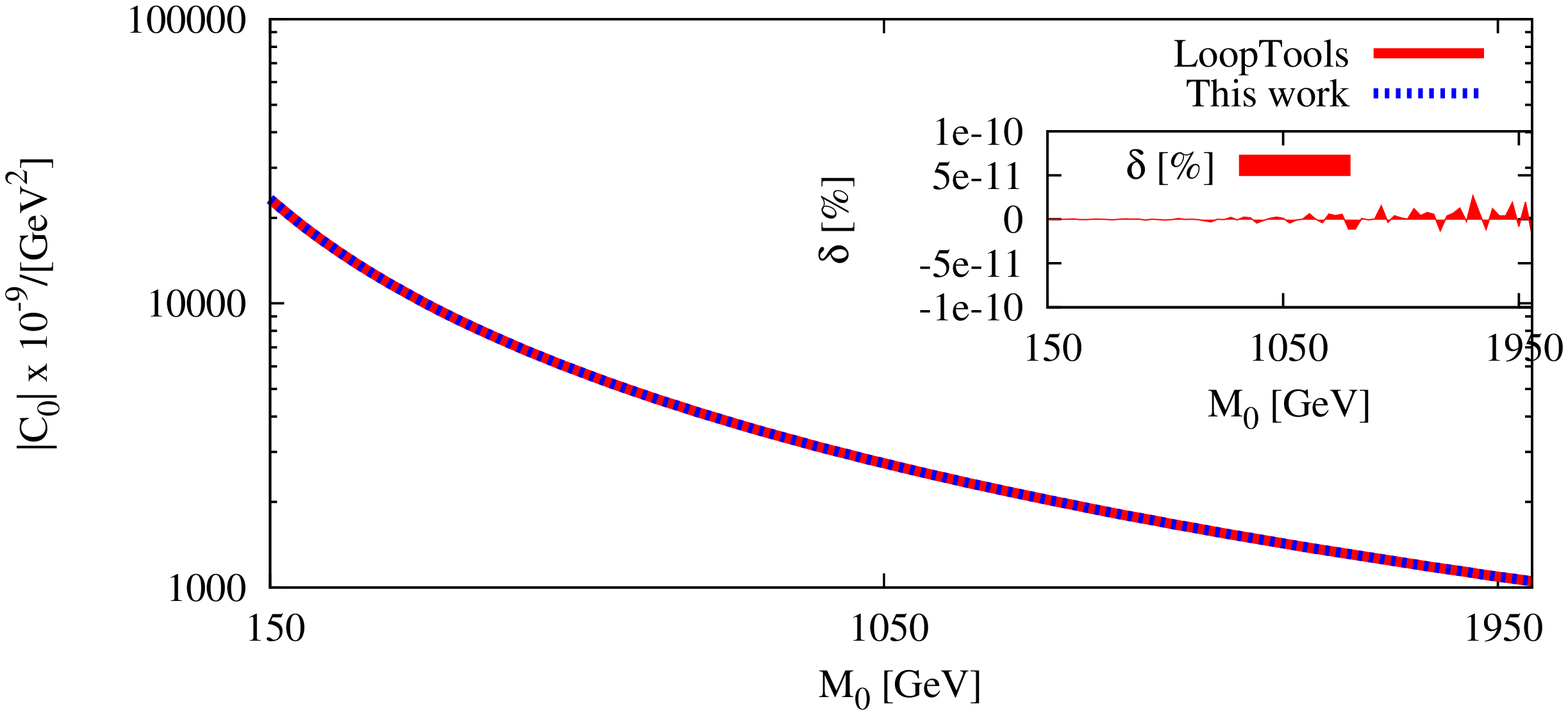} &
\includegraphics[width=5cm,height=8cm, angle=-90]{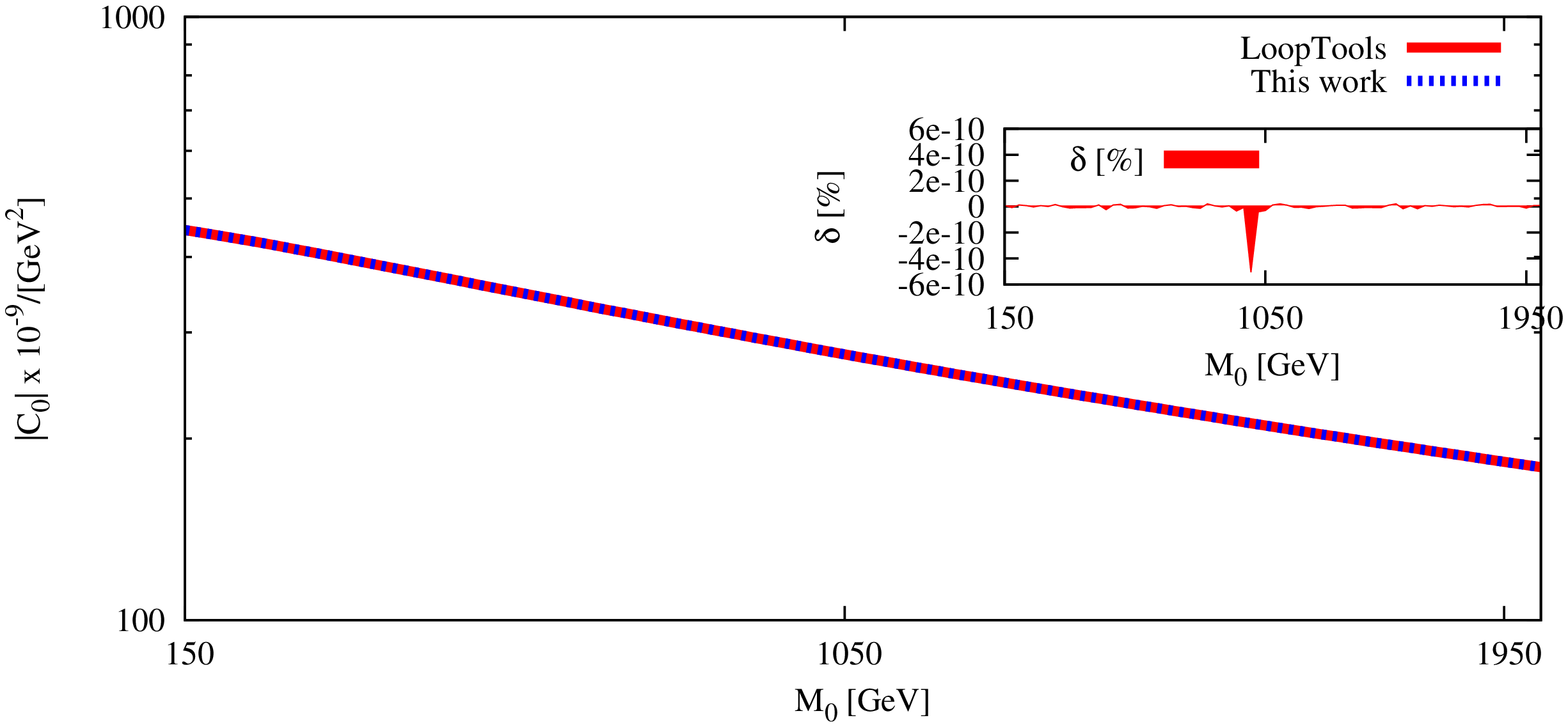}\\
\end{array}$
\end{center}
\caption{\label{C0} Checking numerically  the consistency of  the $C_0$ expression with LoopTools.  The four plots in the  first two rows refer to  the first case. The three remaining rows show  cases of (ii), (iii),  and (iv), respectively. }
\end{figure}

\begin{figure}[htpb]
\begin{center}$
\begin{array}{cc}
\vspace*{-1.7cm}
\includegraphics[width=5cm,height=8cm, angle=-90]{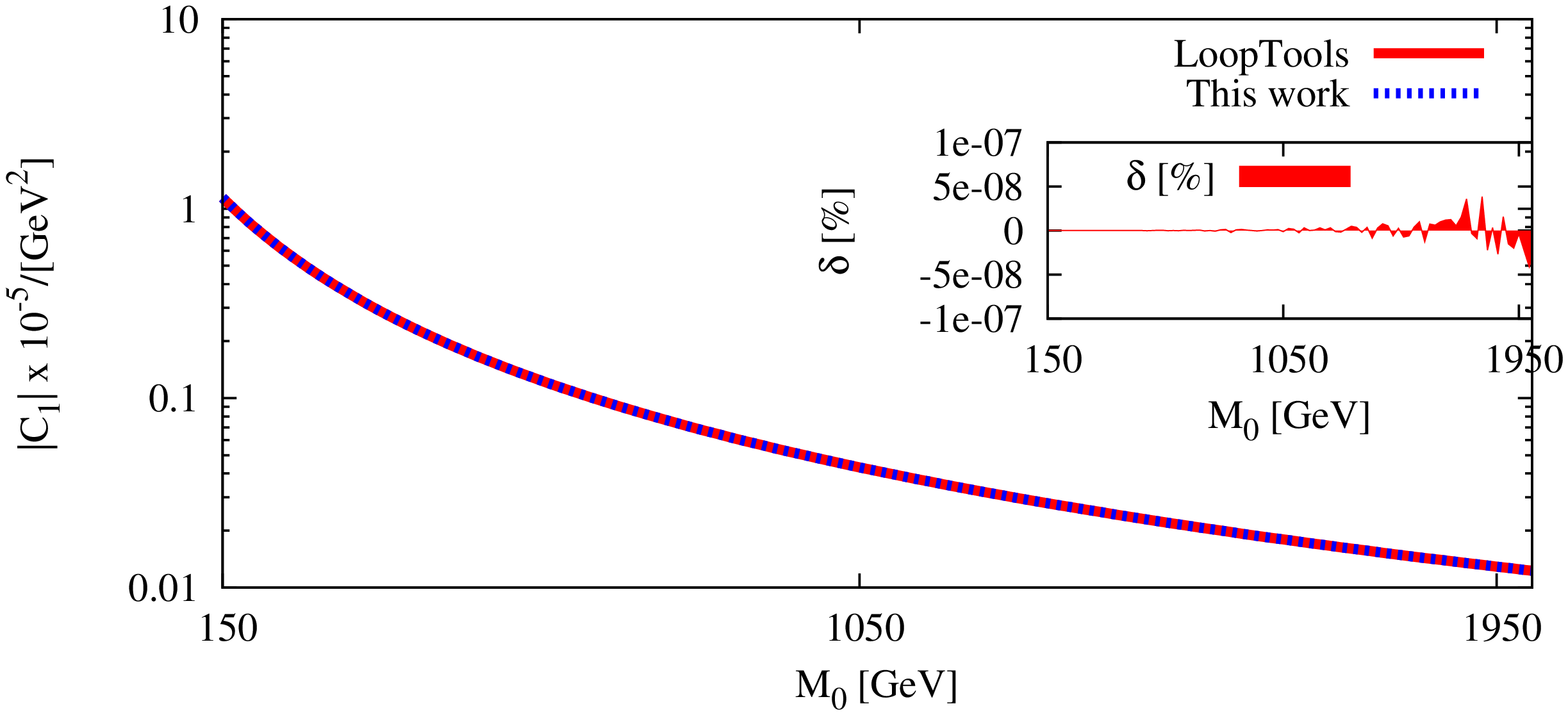} &
\includegraphics[width=5cm,height=8cm, angle=-90]{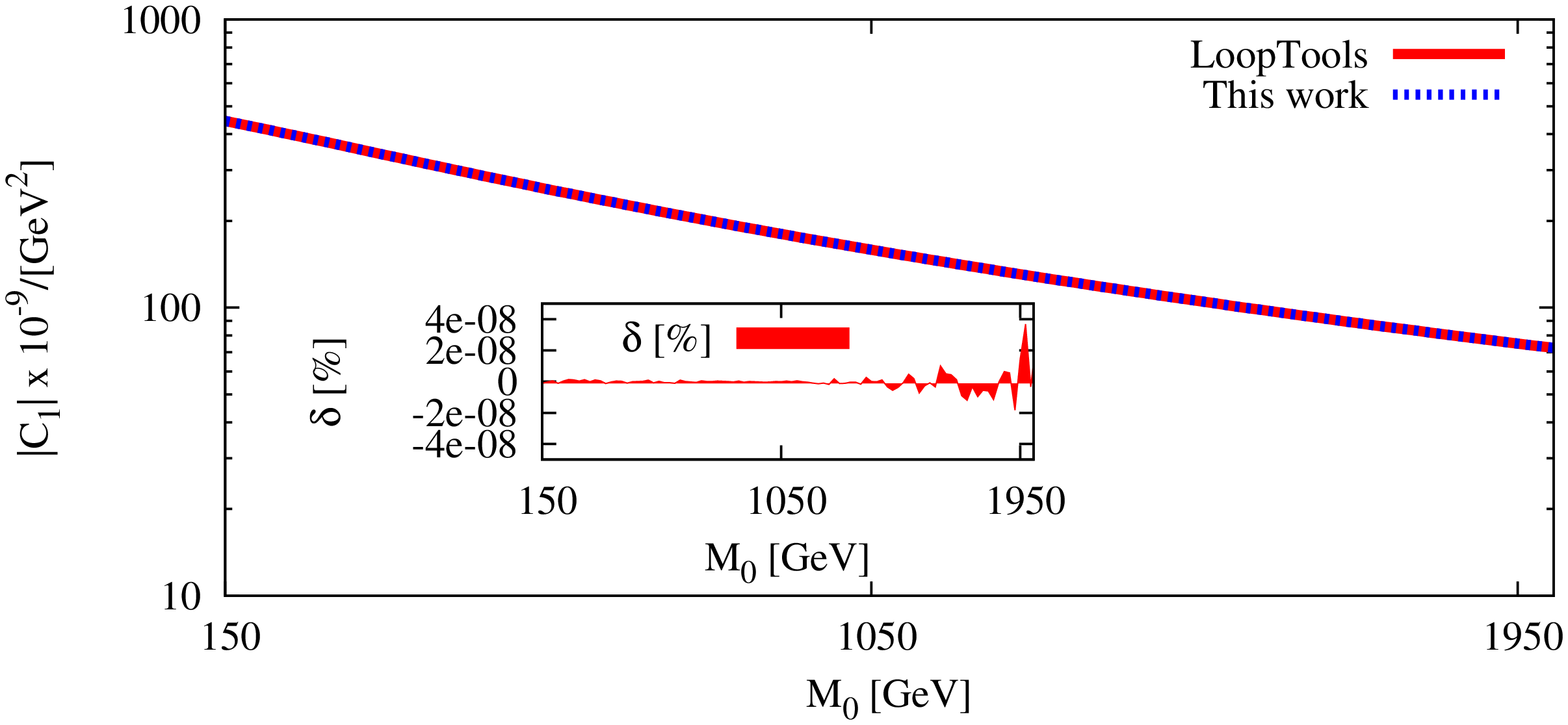}\\
\vspace*{-1.7cm}
\includegraphics[width=5cm,height=8cm, angle=-90]{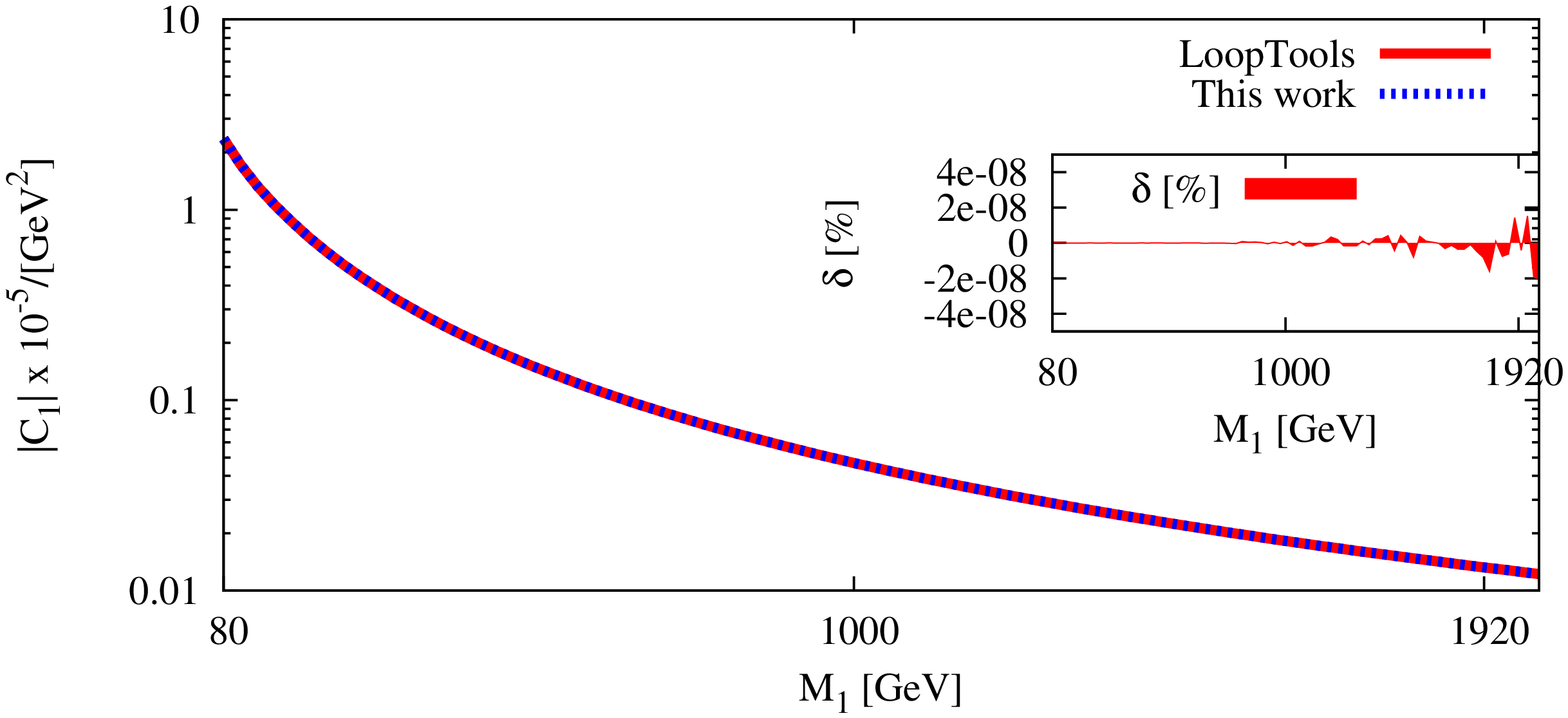} &
\includegraphics[width=5cm,height=8cm, angle=-90]{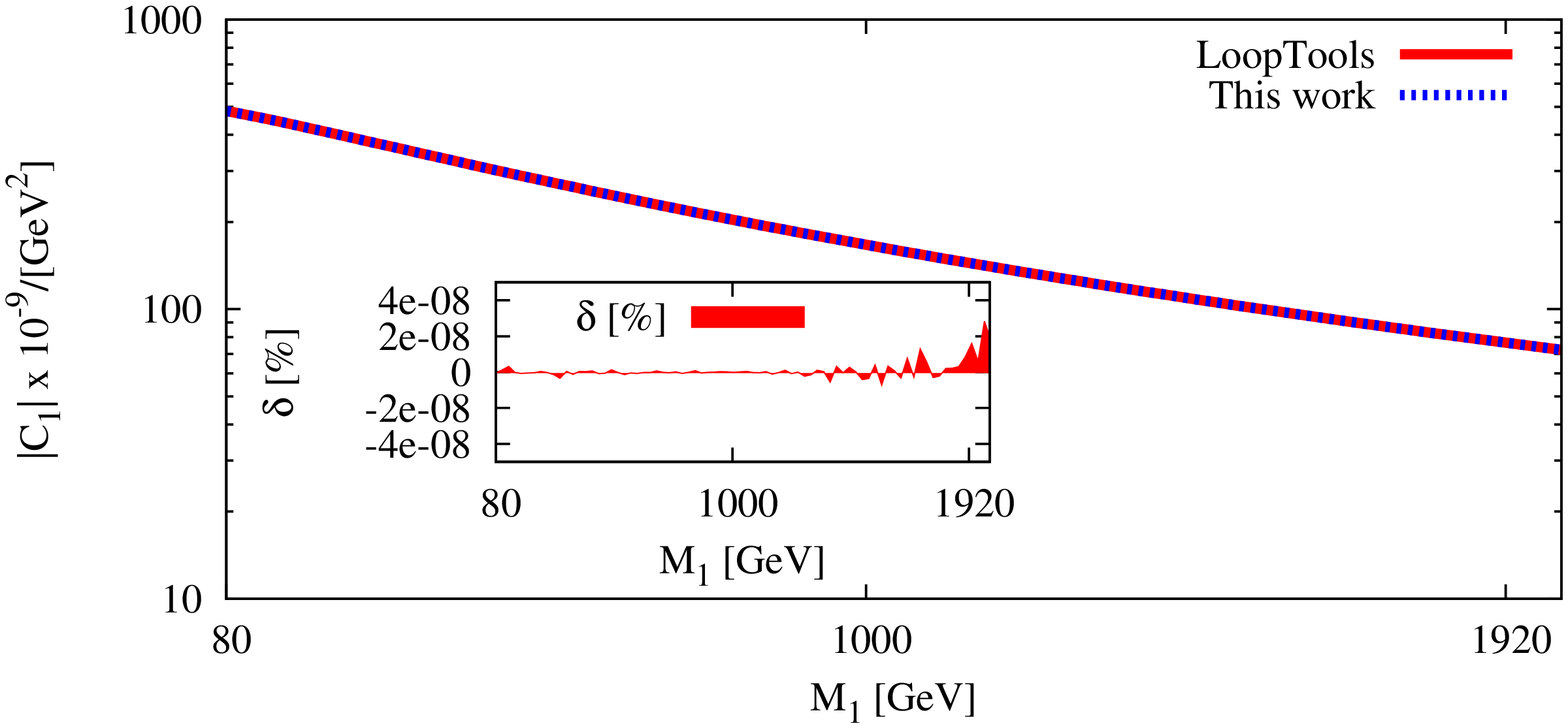}\\
\vspace*{-1.7cm}
\includegraphics[width=5cm,height=8cm, angle=-90]{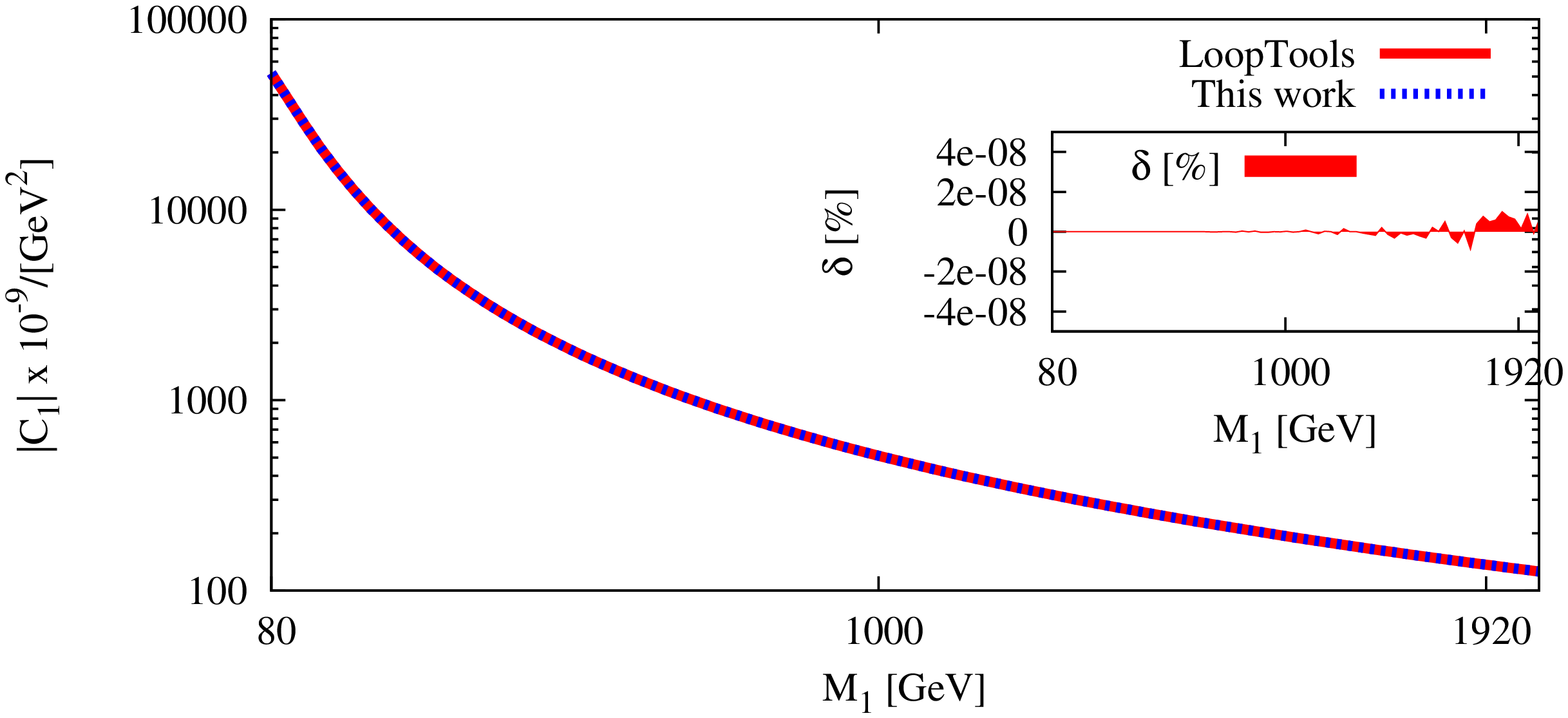} &
\includegraphics[width=5cm,height=8cm, angle=-90]{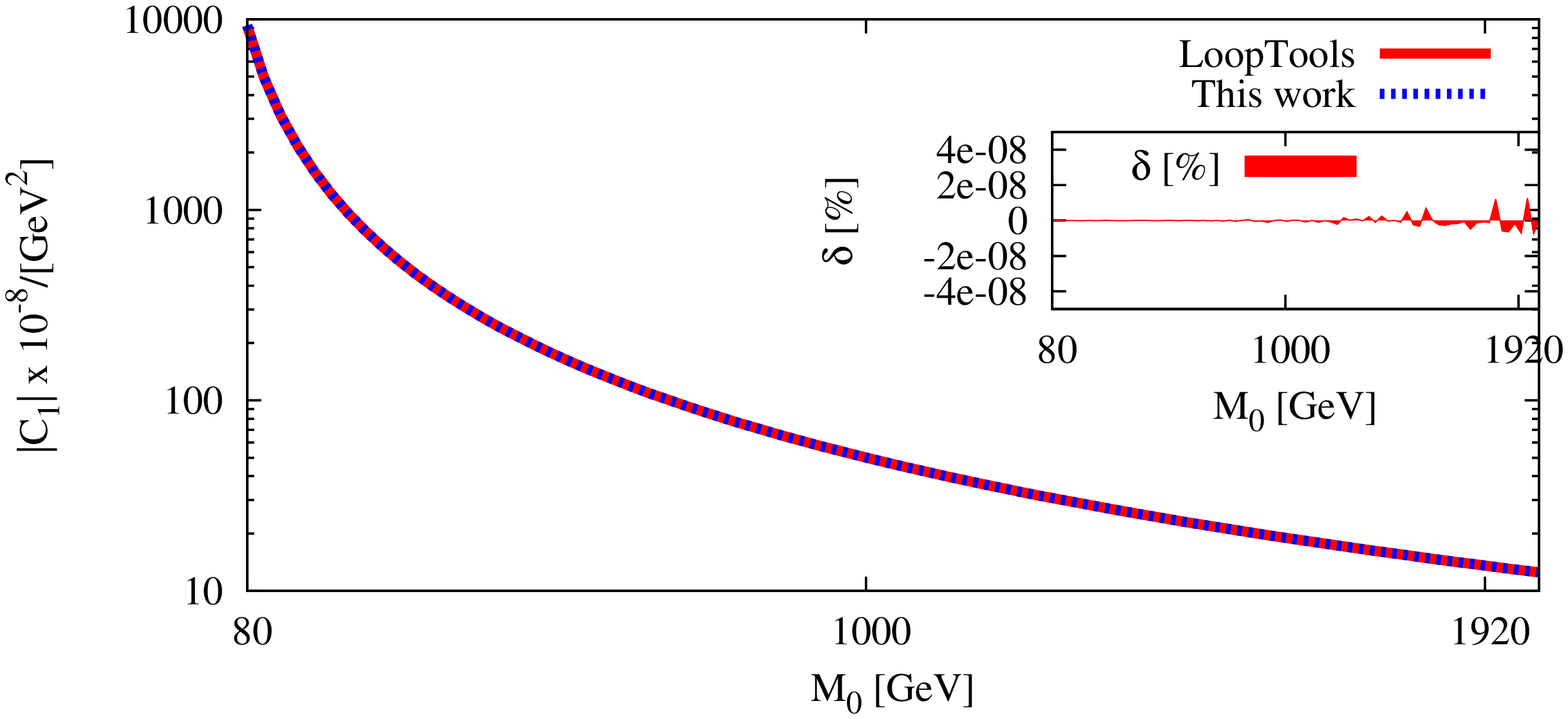}\\
\vspace*{-1.7cm}
\includegraphics[width=5cm,height=8cm, angle=-90]{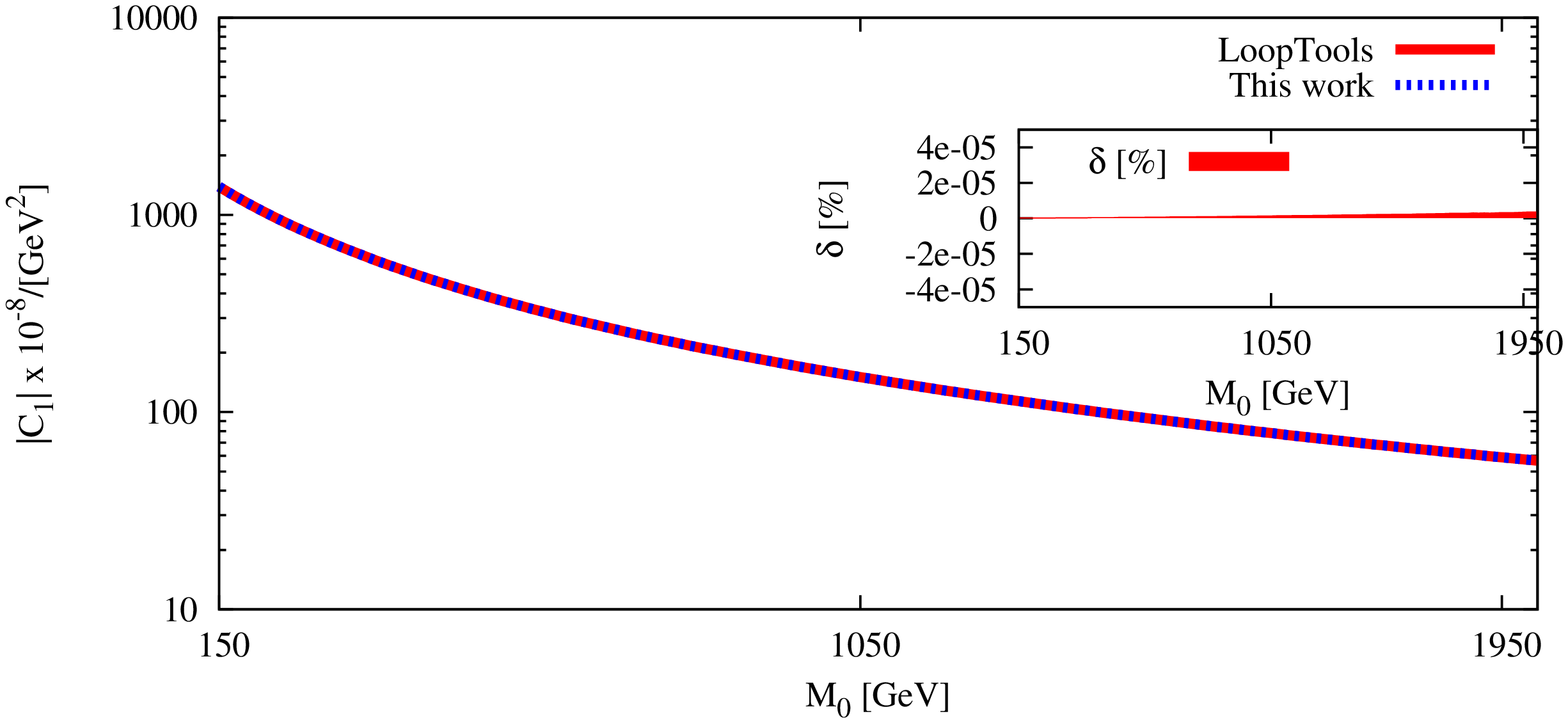} &
\includegraphics[width=5cm,height=8cm, angle=-90]{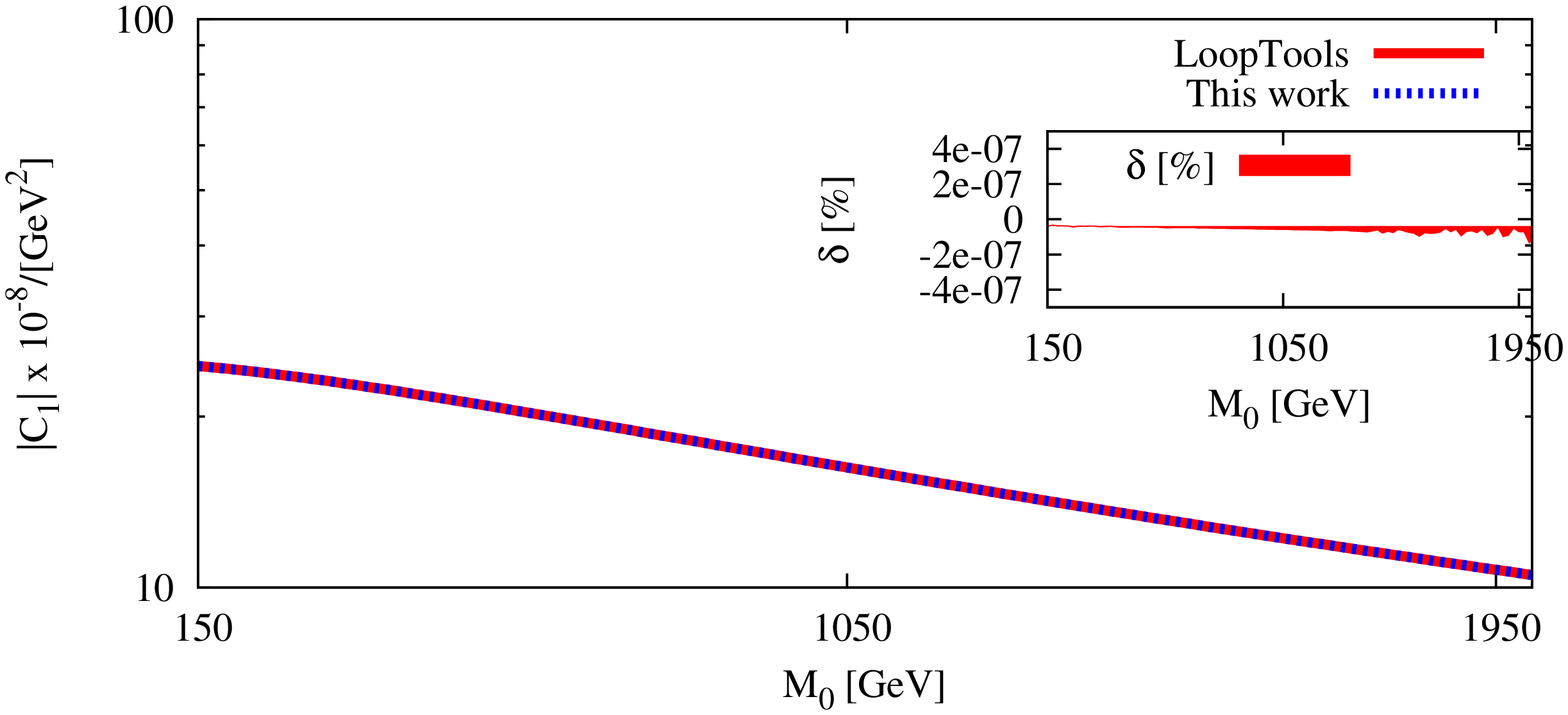}\\
\vspace*{-1.7cm}
\includegraphics[width=5cm,height=8cm, angle=-90]{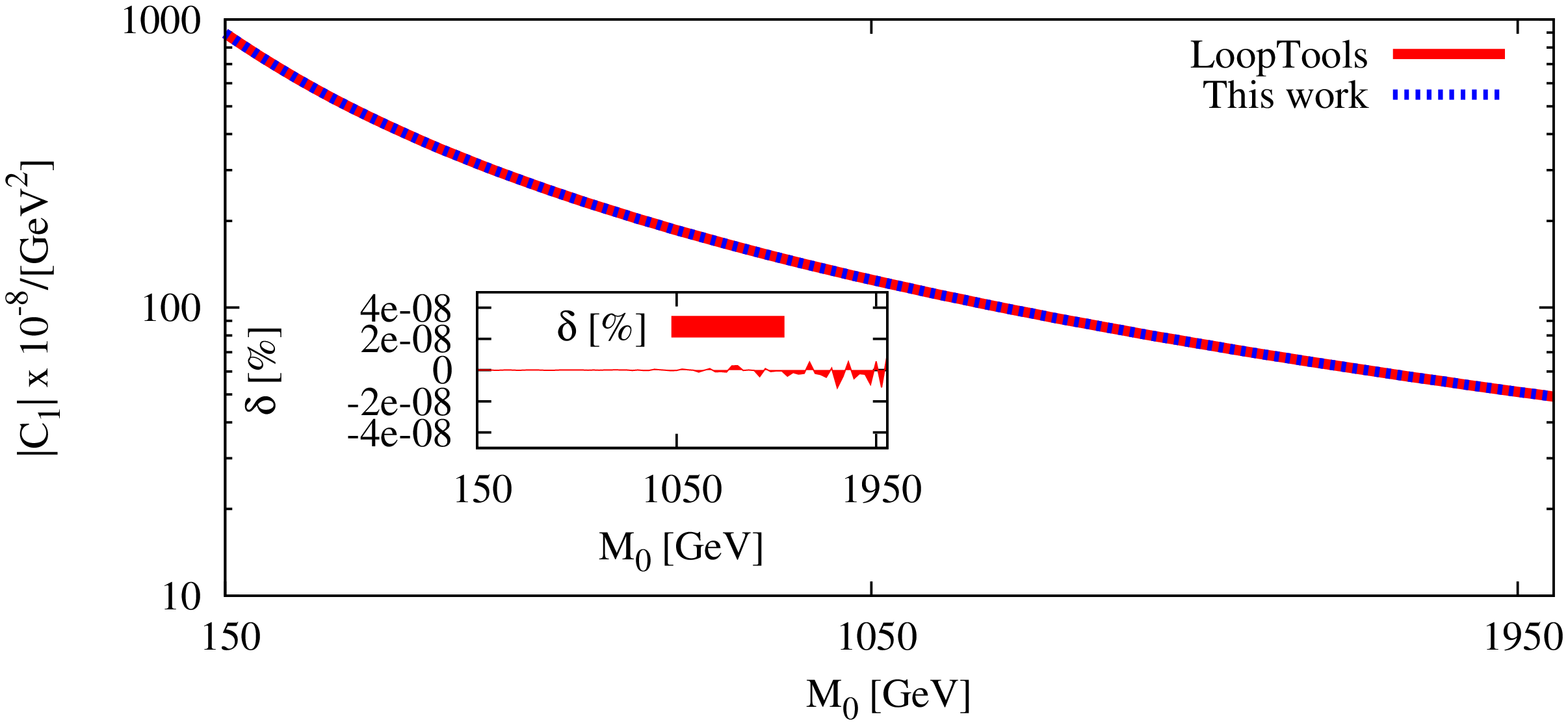} &
 \includegraphics[width=5cm,height=8cm, angle=-90]{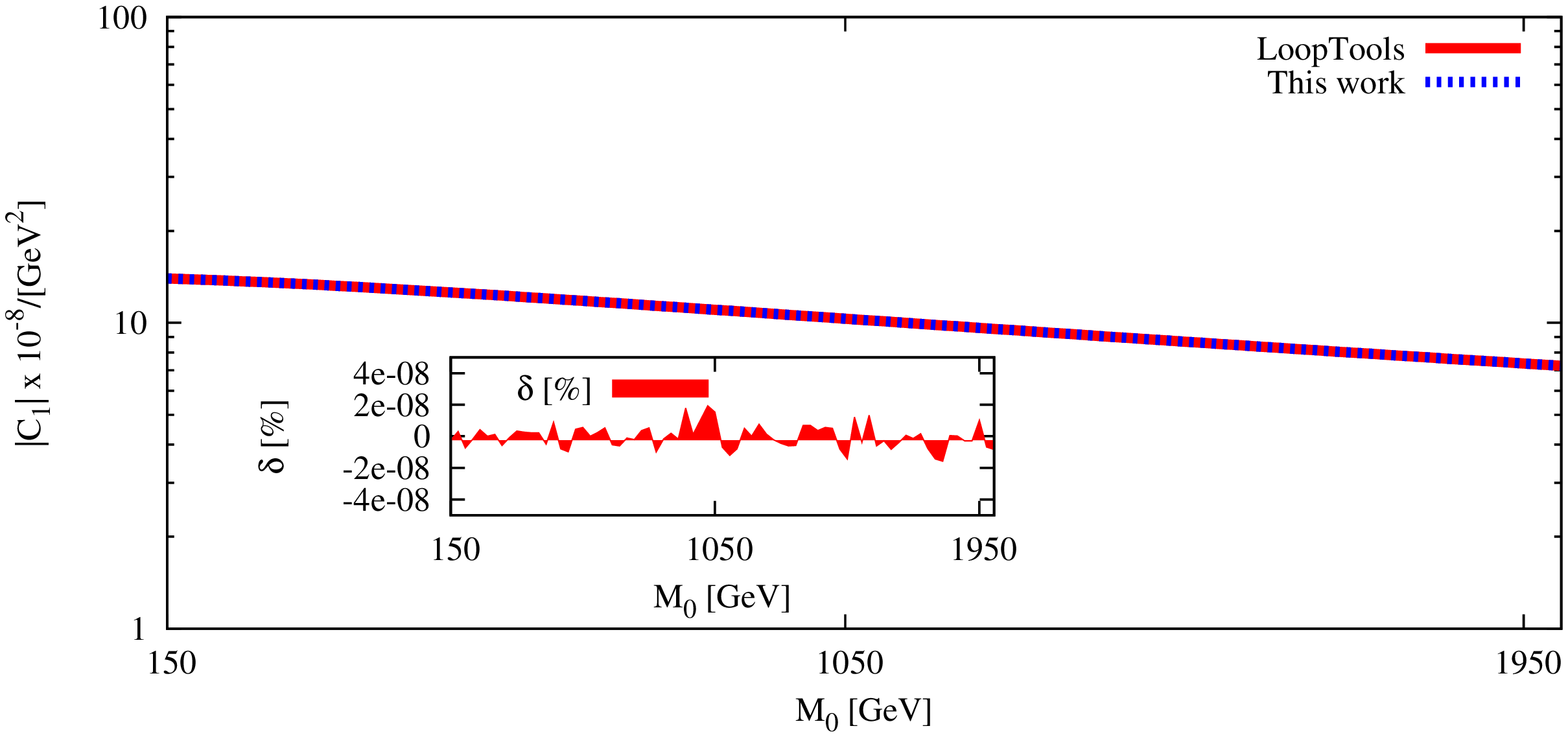}\\
\end{array}$
\end{center}
\caption{\label{C12A} The function $C_1$ in this work is numerically
cross-checked  with LoopTools for all cases, with the same orders mentioned in Fig.  \ref{C0}. }
\end{figure}

\begin{figure}[htpb]
\begin{center}$
\begin{array}{cc}
\vspace*{-1.7cm}
\includegraphics[width=5cm,height=8cm, angle=-90]{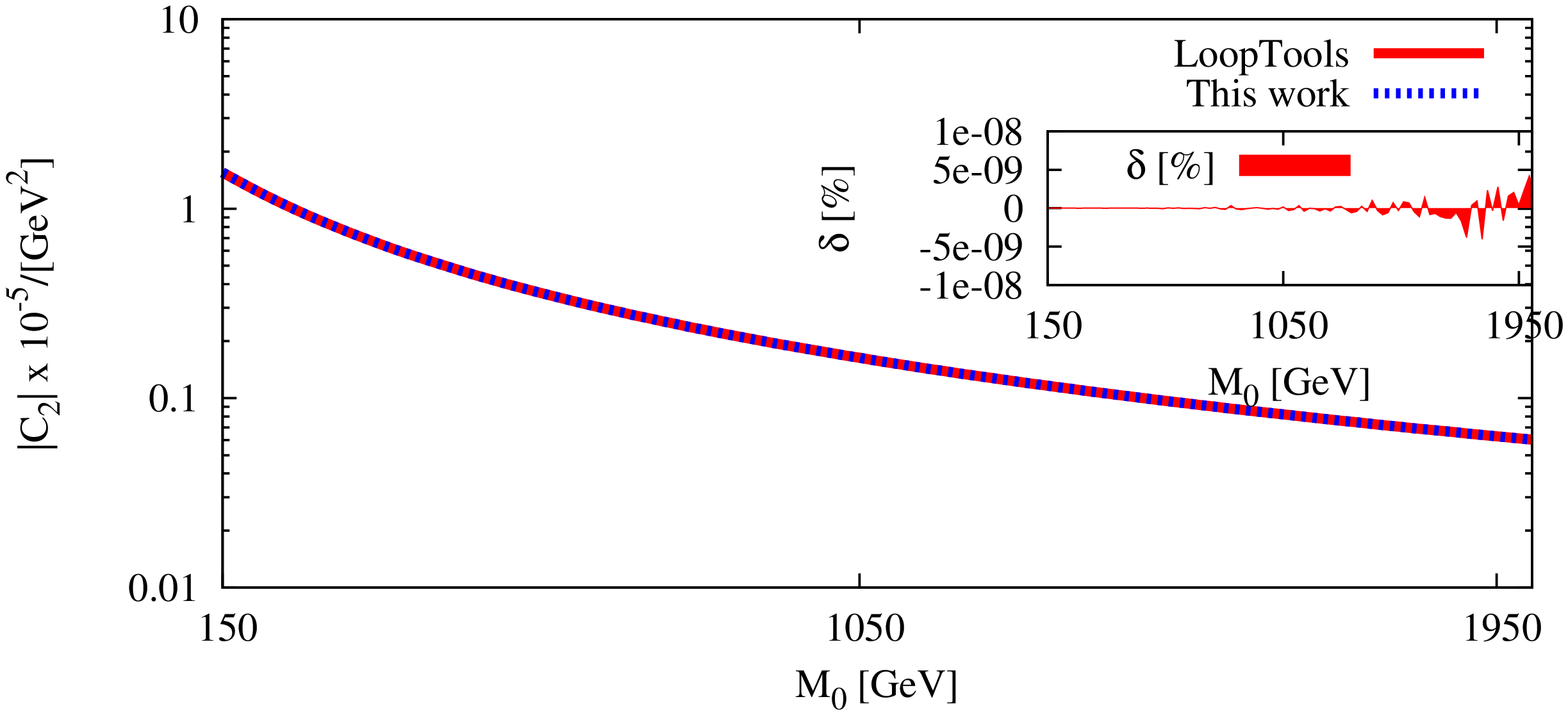} &
\includegraphics[width=5cm,height=8cm, angle=-90]{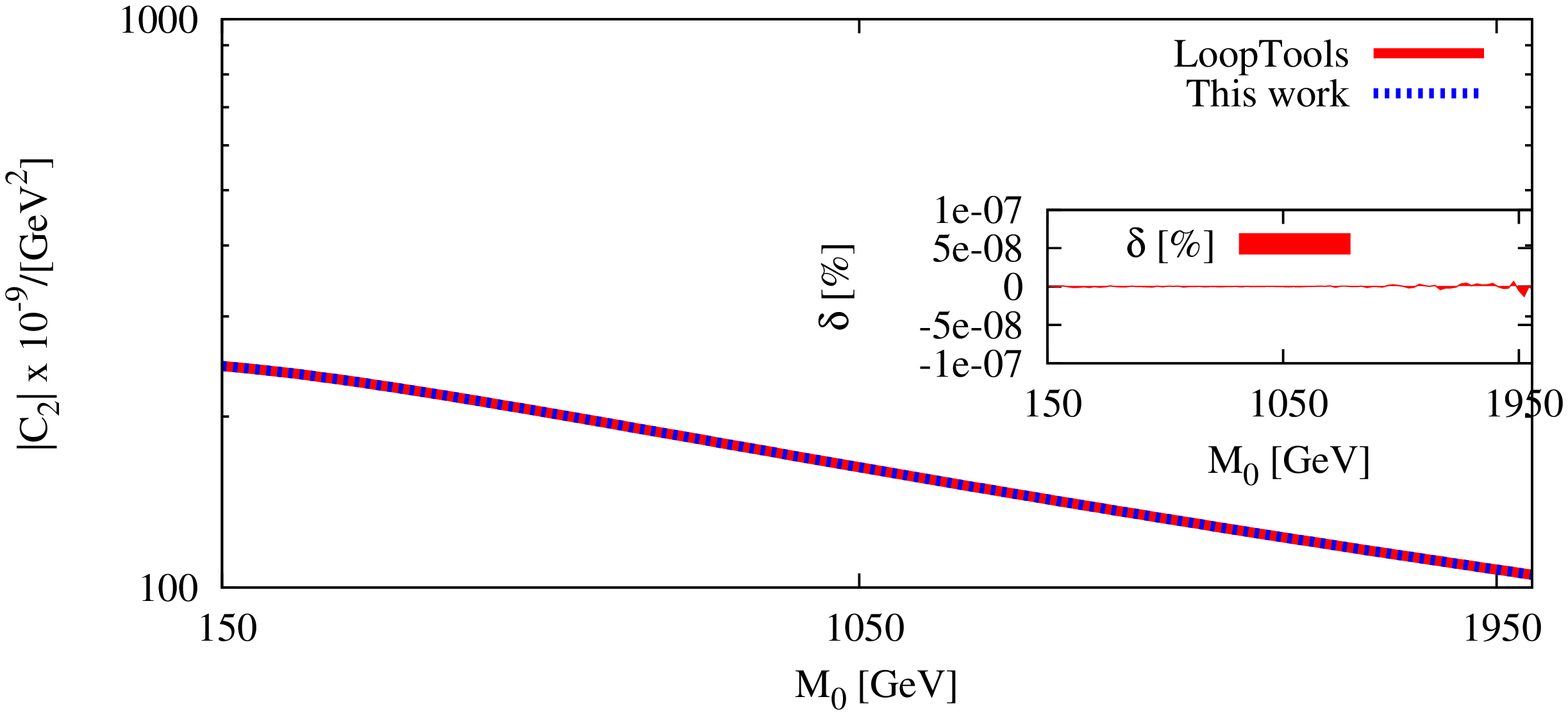}\\
 \vspace*{-1.7cm}
\includegraphics[width=5cm,height=8cm, angle=-90]{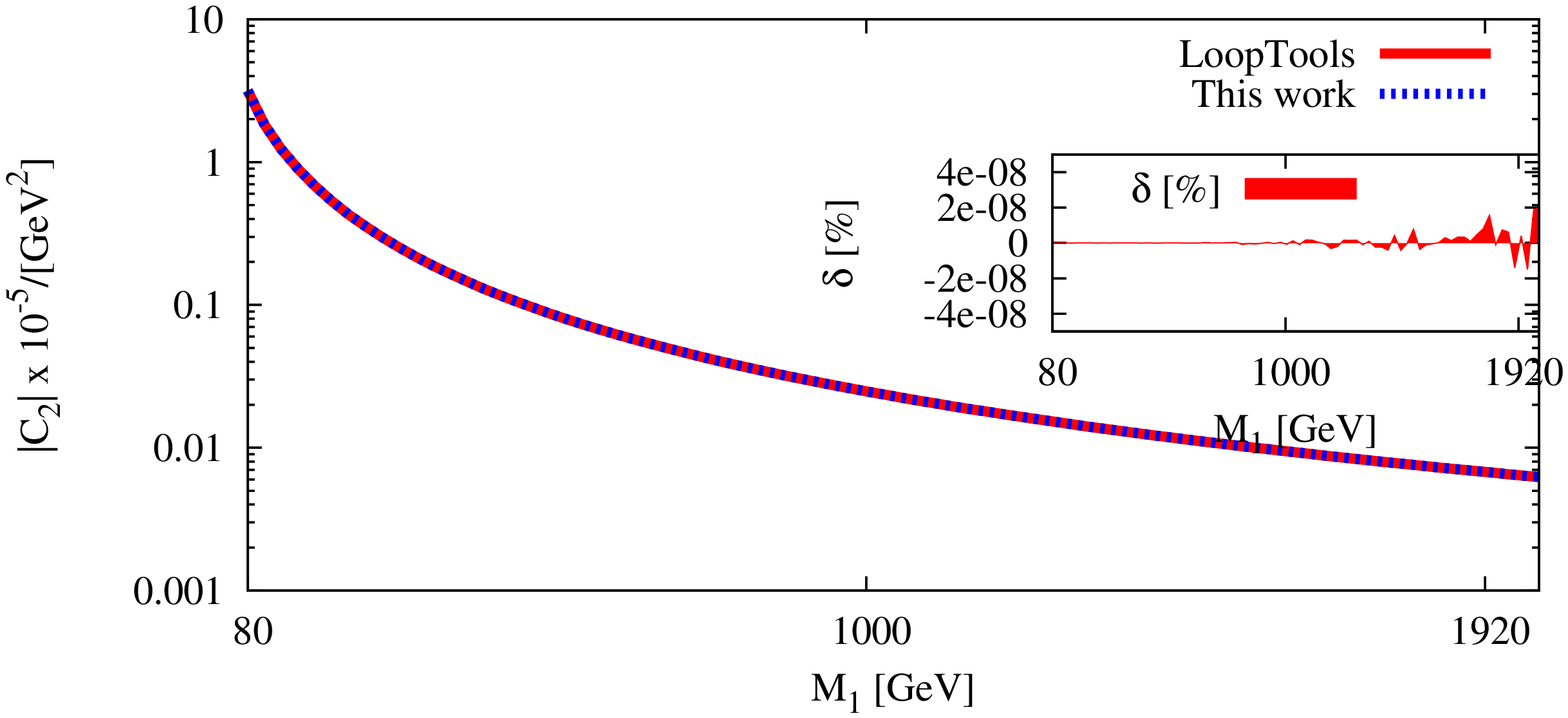} &
\includegraphics[width=5cm,height=8cm, angle=-90]{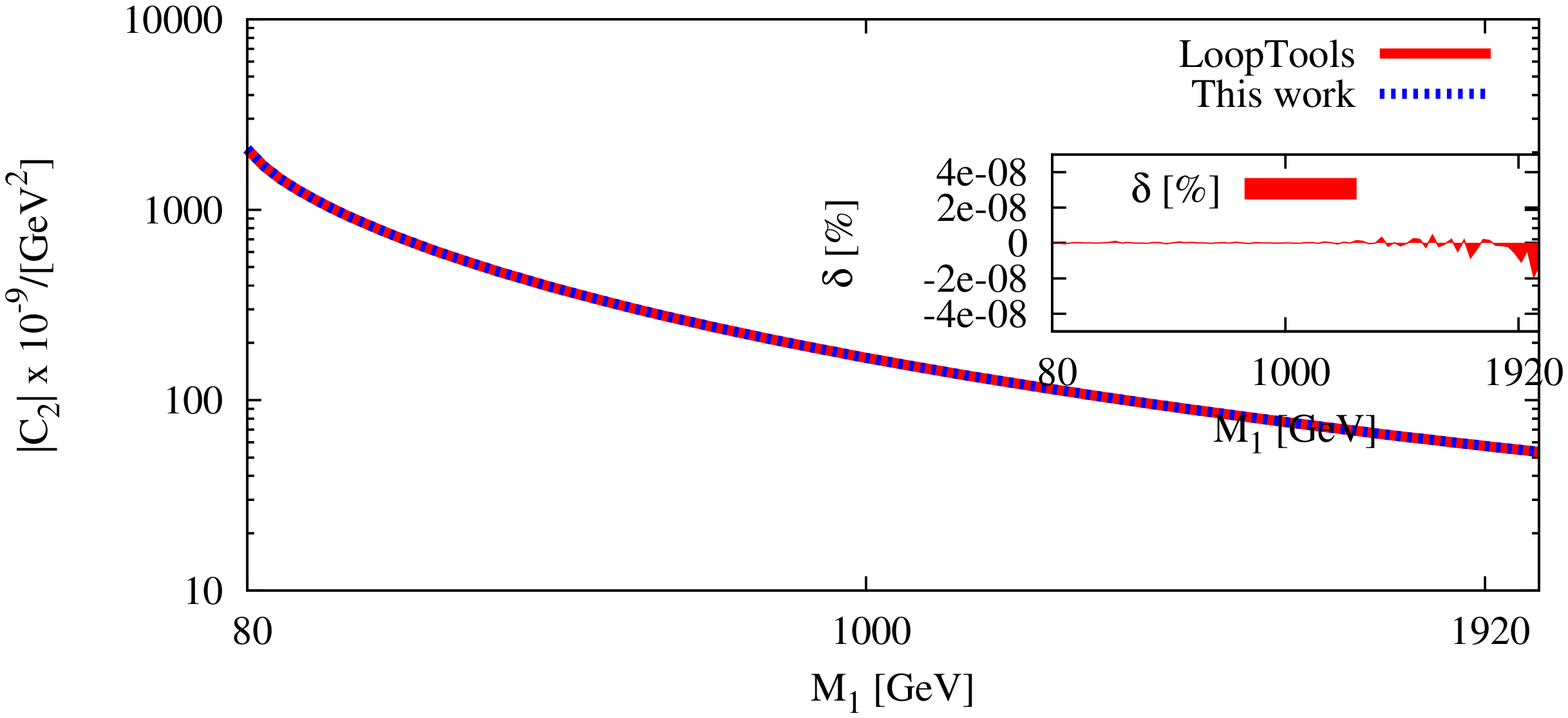}\\
\vspace*{-1.7cm}
\includegraphics[width=5cm,height=8cm, angle=-90]{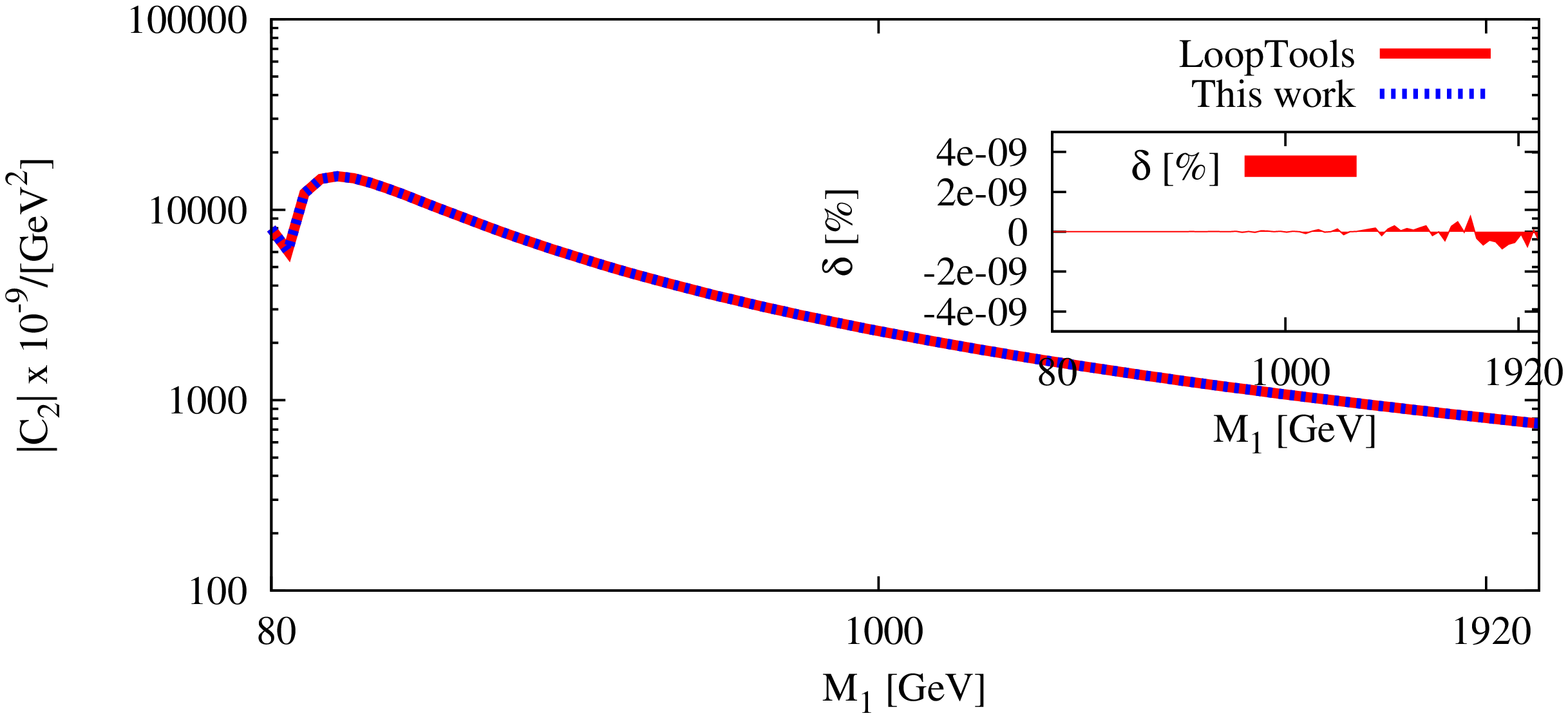} &
\includegraphics[width=5cm,height=8cm, angle=-90]{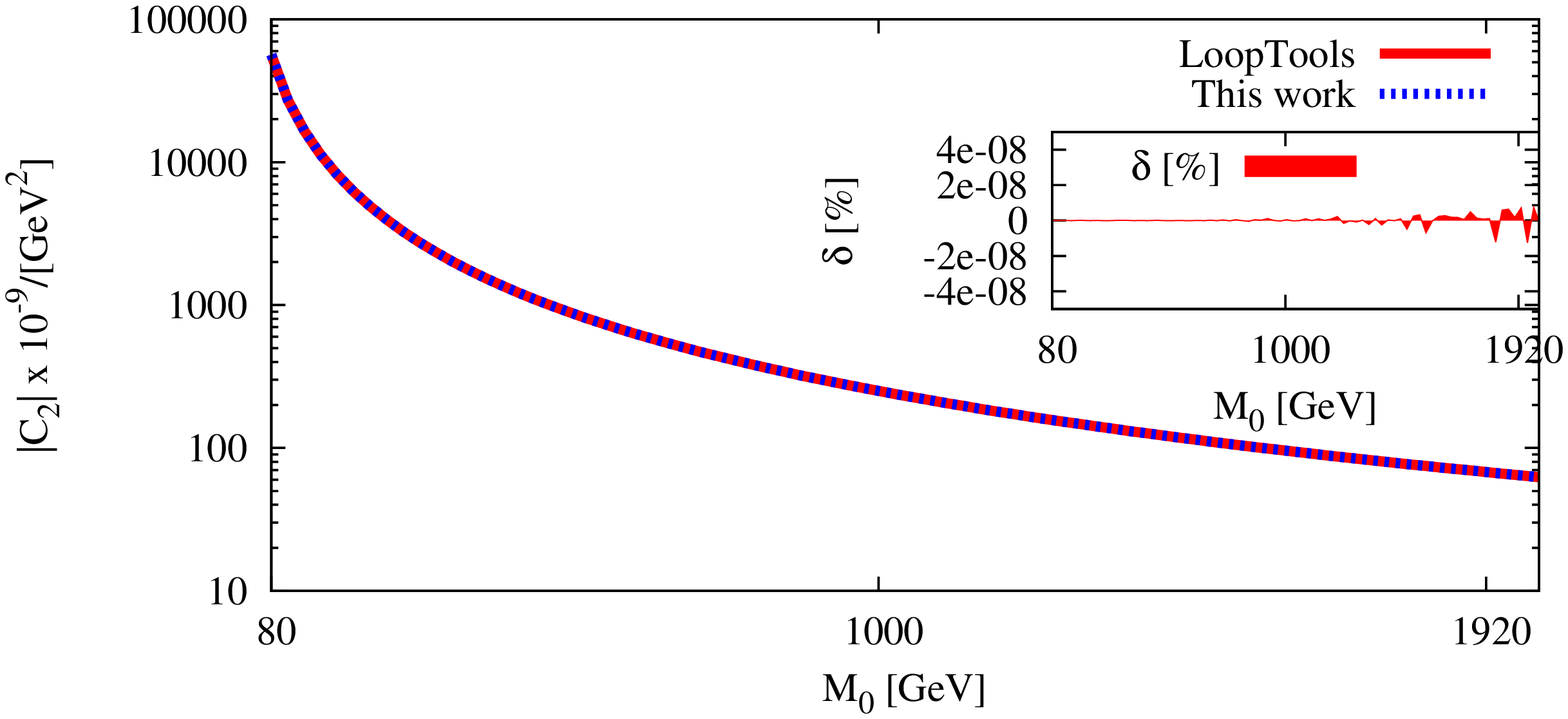}\\
\vspace*{-1.7cm}
\includegraphics[width=5cm,height=8cm, angle=-90]{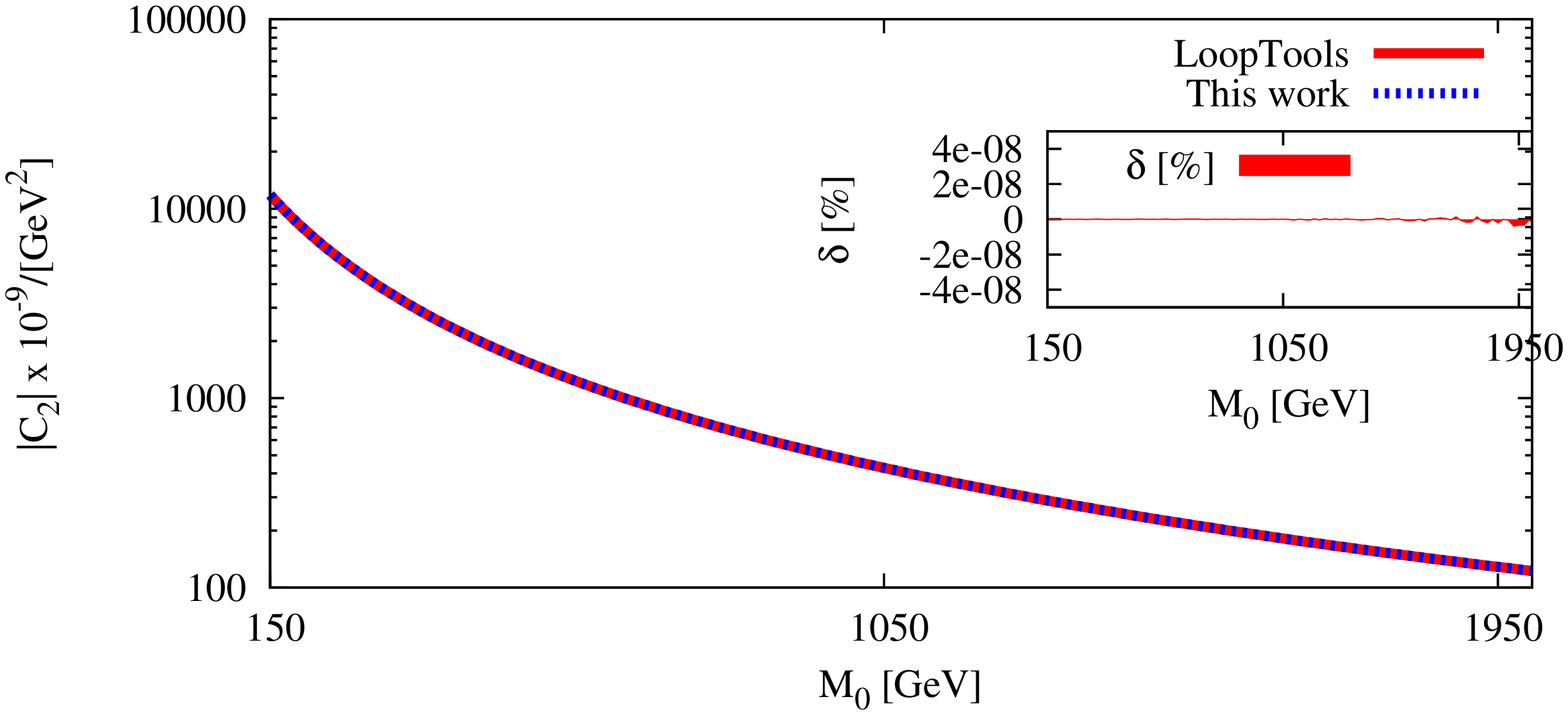} &
 \includegraphics[width=5cm,height=8cm, angle=-90]{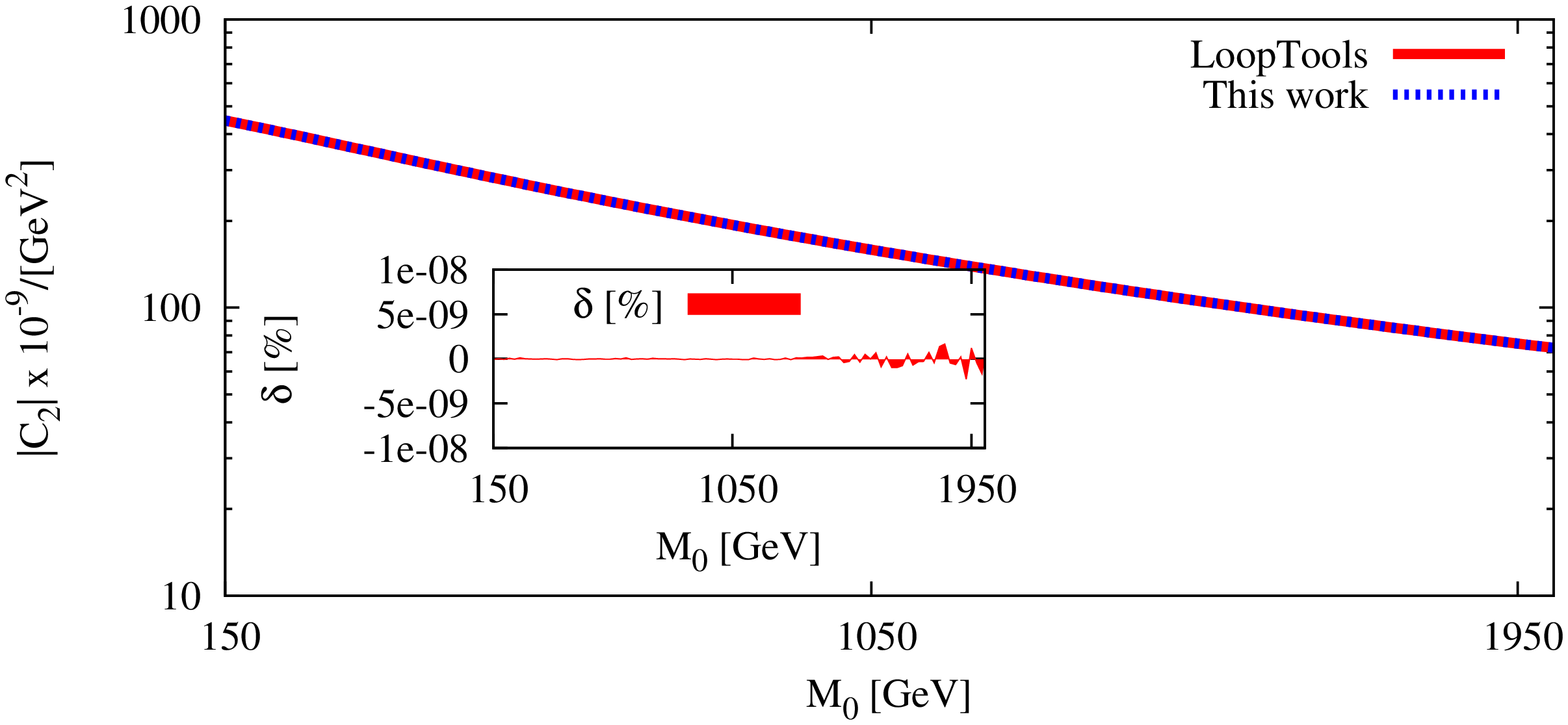}\\
 \vspace*{-1.7cm}
\includegraphics[width=5cm,height=8cm, angle=-90]{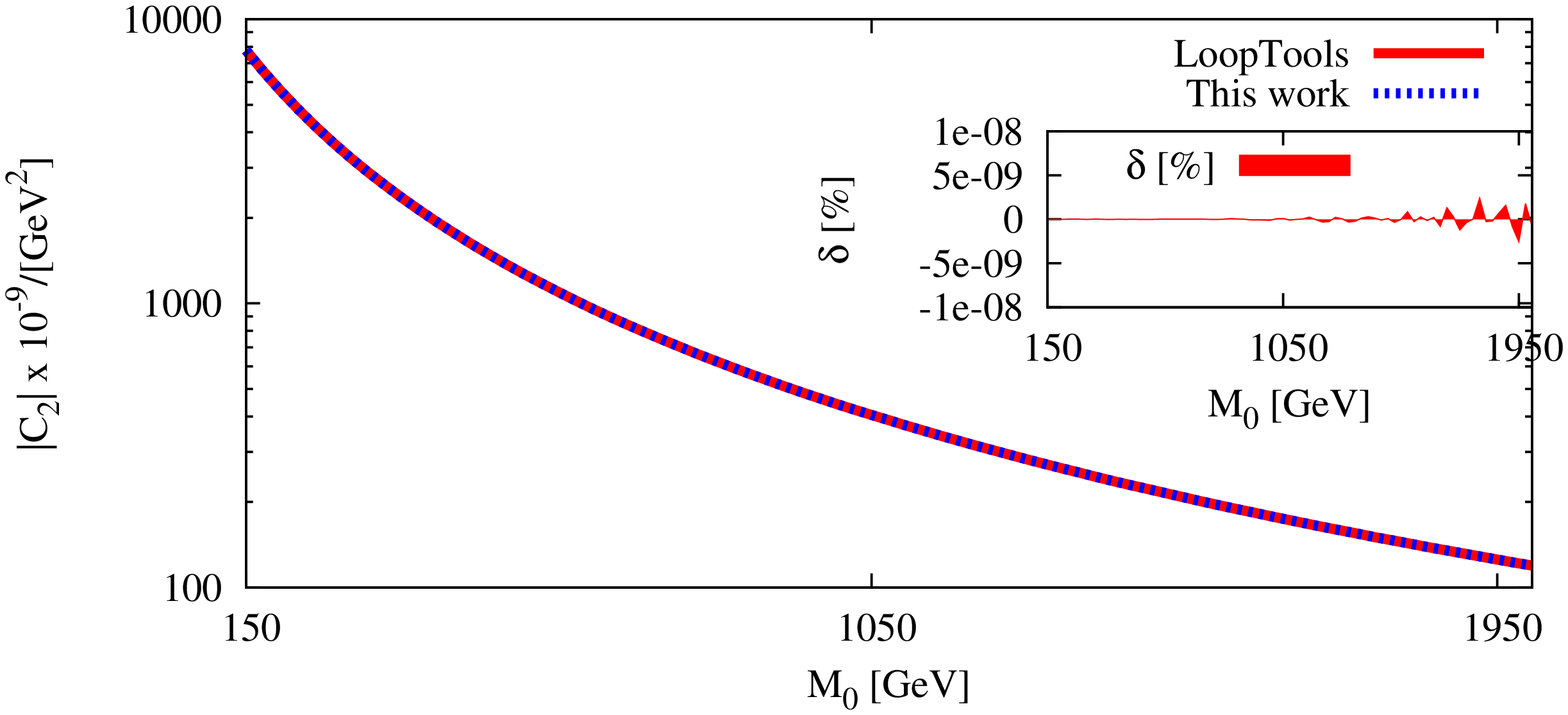} &
 \includegraphics[width=5cm,height=8cm, angle=-90]{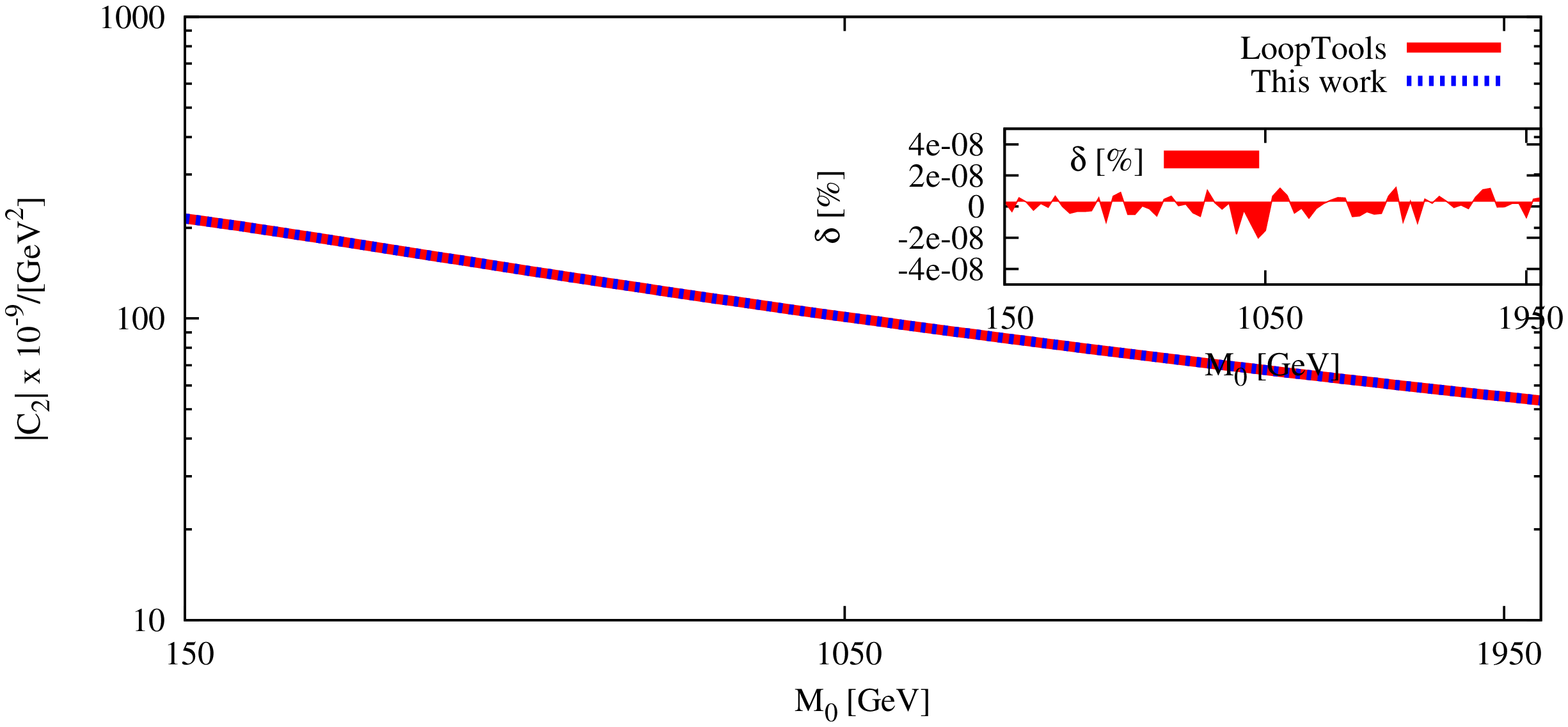}\\
\end{array}$
\end{center}
\caption{\label{C12B} The function $C_2$ in this work is numerically
cross-checked  with LoopTools for all cases, with the same orders mentioned in Fig. \ref{C0}.}
\end{figure}

 Figure \ref{C0}  shows a numerical comparison of the function $C_0$,  where the dotted blue and red curves represent  analytic results and  LoopTools, respectively. One finds that they are consistent with all relative errors being  smaller than $\mathcal{O}(10^{-6}\; \%)$. Similarly,  Figs.  \ref{C12A} and \ref{C12B} illustrate the cases of the $C_1$ and $C_2$ functions.  Again, we find the same conclusion as the case of the $C_0$-function.

To finish the comparison with LoopTools, we emphasize that the analytic results mentioned here can be successfully  applied to  calculating one-loop contributions to LFVHD of  heavy neutral Higgs bosons beyond the SM such as the CP-odd Higgs boson in the minimal supersymmetric model (MSSM), heavy neutral Higgs bosons in the 3-3-1 models, or even the 750 GeV Higgs boson that has been widely discussed  recently. Other one-loop contributions to decays of  heavy particles  to  pairs of very light particles such as leptons, light quarks can also expressed as functions of the above $C$-functions, without any inconsistencies with the results obtained from using numerical packages. The complete set of analytic expressions of $C$-functions needed for calculating one-loop contributions in the unitary and 't Hooft Feynman gauges introduced in \cite{hue}.

In the next subsection, we discuss other analytic forms used for calculating LFVHD at the one-loop level.
\subsection{ Discussion of other expressions for C-functions }
\subsubsection{ Comparison with results in Refs. \cite{lfvhloop4,lfvhlomore}}
In this subsection we would like to compare the numerical  results of analytic forms of
$C$-functions in \cite{hue} with other recent  expressions. Refs.\cite{lfvhloop4,lfvhlomore} used directly a formula containing $C_0$-functions but with only one set of fixed  values of internal masses and external momenta. The relevant loops  are defined in the formula
\bea g_1(\lambda,m^2_{\Delta})&=&(m^2_{\Delta}+m^2_t)C_0(0,0,m_h^2,m_t^2,m^2_{\Delta},m^2_t)+ B_0(m_h^2,m_t^2,m_t^2)- B_0(0,m_t^2,m_{\Delta}^2) \crn
&+& \lambda v^2 C_0(0,0,m_h^2,m_t^2,m^2_{\Delta},m^2_t),    \label{flfvhloop4}\eea
where $m_{\Delta}$ is the mass of the leptoquark in the loop, $m_h=125.1$ GeV, $m_t=173$ GeV, $v=246$ GeV and $\lambda$
is the trilinear Higgs-self-coupling.  The important property of LFVHD in this model is that the top quarks play role of LFV mediators in the loop, hence analytic results in \cite{lfvhloop1,apo1} cannot be applied. Using the expressions in \cite{hue}, the corresponding notation translations are $M_0=m_{\Delta}, M_1=M_2=m_t$ ($ M_0=m_t, M_1=M_2=m_{\Delta}$)
in the first (second) line of Eq. (\ref{flfvhloop4}), $B_0(m_h^2,m_t^2,m_t^2)=B^{(12)}_0$ and $B_0(0,m_t^2,m_{\Delta}^2)=B^{(1)}_0=B^{(2)}_0$. For $m_{\Delta}=650$ GeV
we get $ C_0(0,0,m_h^2,m_t^2,m^2_{\Delta},m^2_t)=-4.866\times 10^{-6}$, $ C_0(0,0,m_h^2,m_{\Delta}^2,m^2_t,m^2_{\Delta})=-2.04\times 10^{-6}$
and $B^{(12)}_0- B^{(1)}_0=1.941$. As a result, $g_1(\lambda,650\mathrm{GeV})=-(0.26+0.12\lambda)$, being consistent with the value given in \cite{lfvhloop4,lfvhlomore}.

\subsubsection{ Approximation of $C_0$-function in Refs. \cite{lfvhloop1} and \cite{apo1} }
\begin{figure}[h]
 \centering
\begin{tabular}{cc}
   \includegraphics[width=7cm]{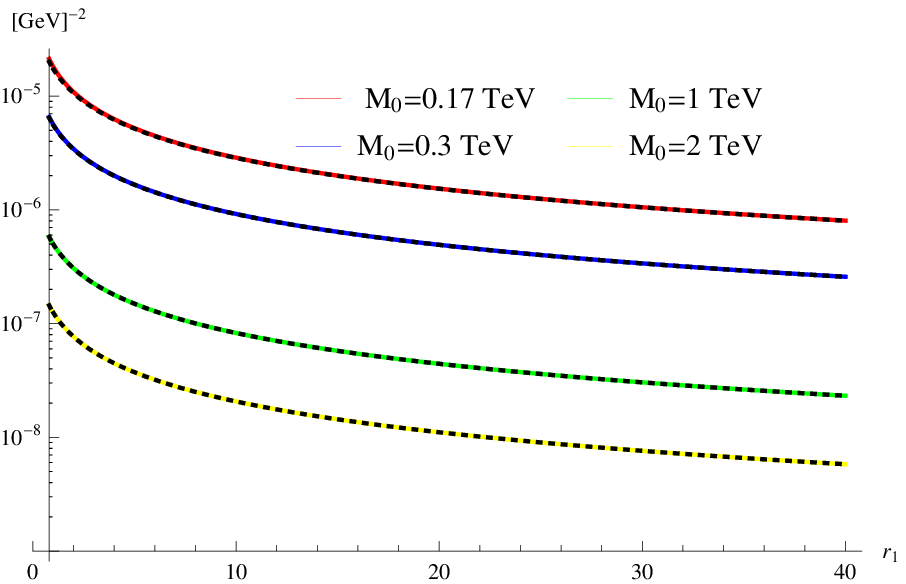} & \includegraphics[width=7cm]{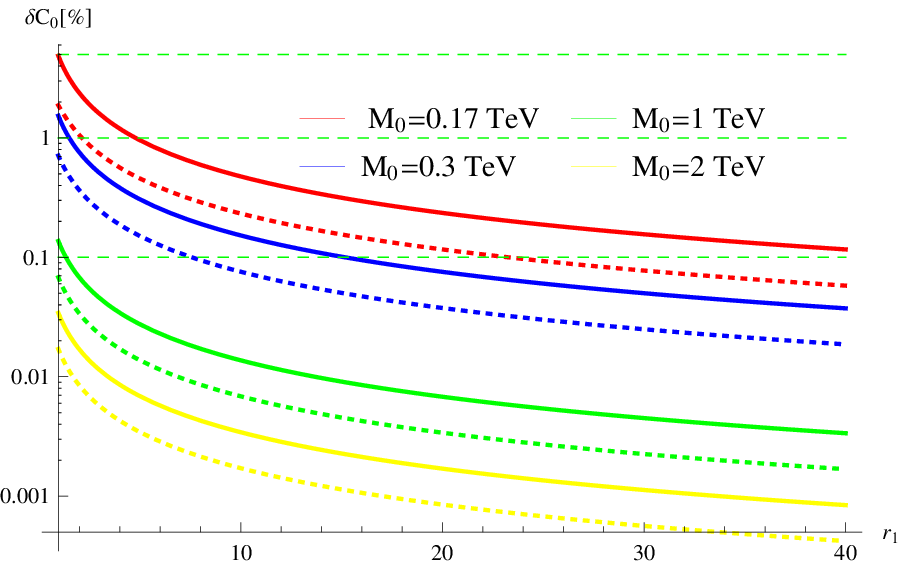} \\
   \includegraphics[width=7cm]{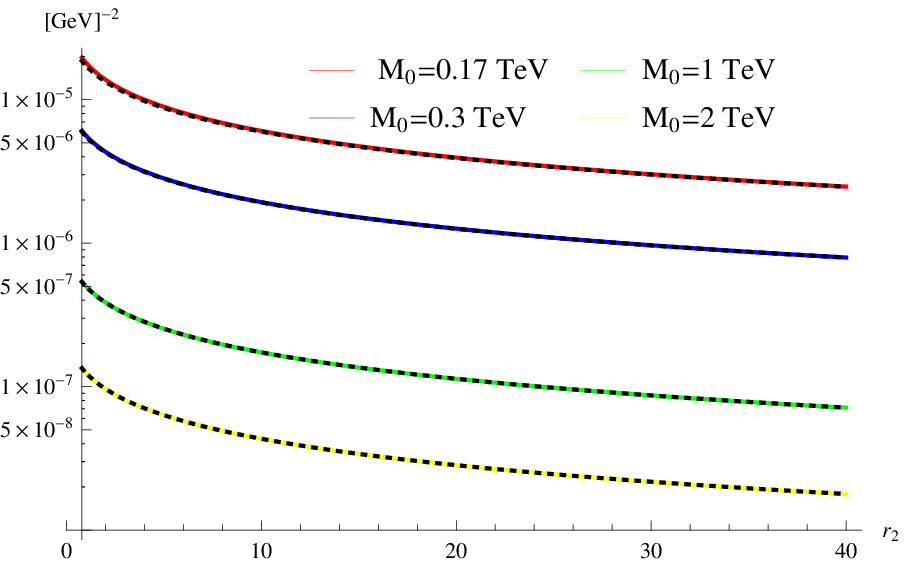} & \includegraphics[width=7cm]{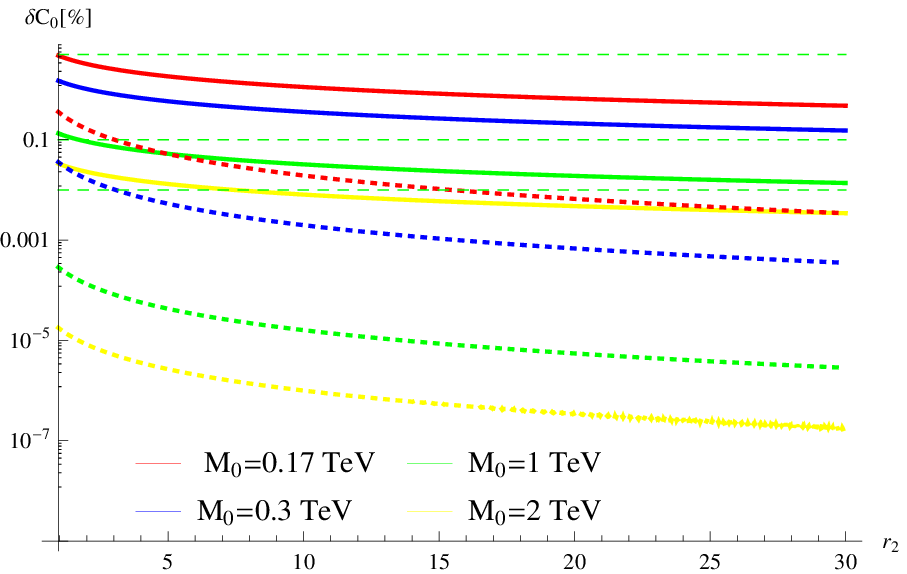} \\
 \end{tabular}
   \caption{Comparison among expressions for $C_0$  introduced in Refs. \cite{hue}, \cite{lfvhloop1} and \cite{apo1}
   in the case of $M_1=M_2$ ($M_1=M_0$) corresponding to the upper (lower) panels. In the left panels, the solid (dotted)
   curves show the numerical results from \cite{hue} (\cite{lfvhloop1}) as functions of $r_1=M_i^2/M_0^2$.
   In the right panels, the solid (dotted) curves show the relative difference between  Ref. \cite{lfvhloop1} (\cite{apo1})
   and Ref. \cite{hue}.  The highest dashed green lines imply values of  5(\%). }\label{fc0r1}
\end{figure}

  Now we consider special cases used in Ref. \cite{lfvhloop1}, where the notation of the  $C_0$-function is the same as that in \cite{hue}, in particular  $C_0(0,0,m_0,m_1,m_2)\equiv C_0(M_0,M_1,M_2)$. Apart from the approximation
  $p^2_{\mu},p^2_{\tau}\simeq 0$, calculation in \cite{lfvhloop1} assumed a very special limit where
  $M^2_{0,1,2}\gg m_h^2=(125.1\mathrm{GeV})^2$. In our notation,  the $C_0$-function derived from \cite{lfvhloop1}  is as follows
  \bea C''_0(M_0,M_1,M_2)\equiv -\frac{1}{M_0^2}G(r_1,r_2)=\left\{
                                 \begin{array}{ll}
                                   \frac{1}{M_0^2(r_1-r_2)}\left( \frac{r_1\ln r_1}{r_1-1}- \frac{r_2\ln r_2}{r_2-1}\right)  , & r_1\neq r_2\neq1; \\
                                    -\frac{1}{2M_0^2} , & r_1= r_2=1; \\
                                  -\frac{1}{M_0^2} \frac{r_1-1-\ln r_1}{(r_1-1)^2}, & r_1= r_2\neq 1;\\
                                 -\frac{1}{M_0^2} \frac{1-r_2+r_2\ln r_2}{(r_2-1)^2} , & r_2\neq r_1\rightarrow 1;\\
                                -\frac{1}{M_0^2} \frac{1-r_1+r_1\ln r_1}{(r_1-1)^2} , & r_1\neq r_2\rightarrow 1,\\
                                 \end{array}
                               \right.
   \label{Gx12}\eea
where $r_i\equiv M_i^2/M^2_0$, $i=1,2$. The relative difference between $|C_0|$ and $|C''_0|$ is defined by the quantity
$|\delta C_0|$  given in Eq. (\ref{delta}). The two cases of $r_1=r_2$ and $r_2\gg r_1=1$ are shown in Fig.
\ref{fc0r1}. Here we also include an approximate function for $C_0(m_N,m_V,m_V)$ used in Ref. \cite{apo1}.
The precise formula  is collected in Appendix \ref{apo1}. We just show the relative difference with the
main analytic formula   in the right panels of  Fig. \ref{fc0r1} with the dotted curves.

  Fig. \ref{fc0r1} shows that two analytic forms in  \cite{apo1} and \cite{hue} are more consistent than the expression in \cite{lfvhloop1}, if  internal masses are few hundred GeV. If all internal masses are as large as TeV scale or more,  all three results are well consistent with the relative differences being smaller than $0.1\%$.   %
\subsubsection{Approximation of $C_2$-function in Ref. \cite{apo1}}
\begin{figure}[h]
 \centering
\begin{tabular}{cc}
    \includegraphics[width=8cm]{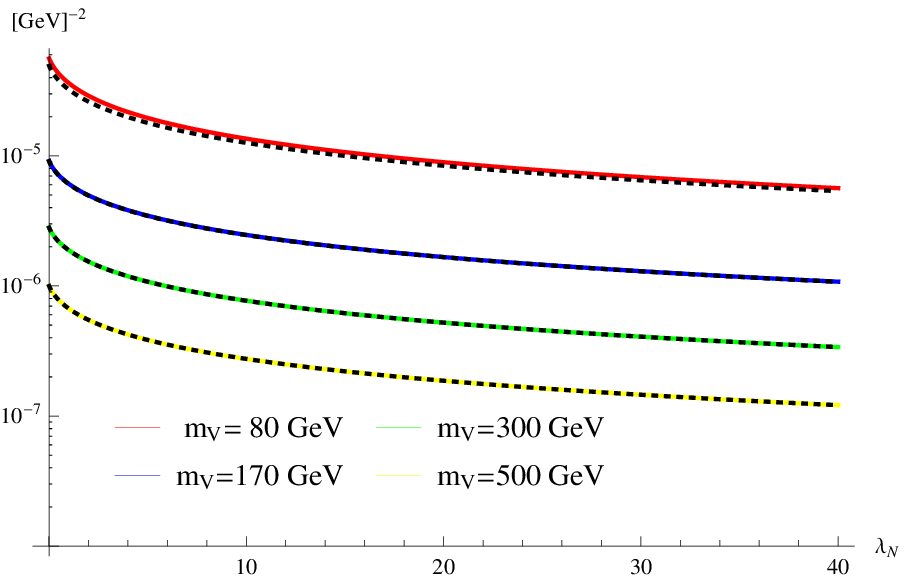} & \includegraphics[width=8cm]{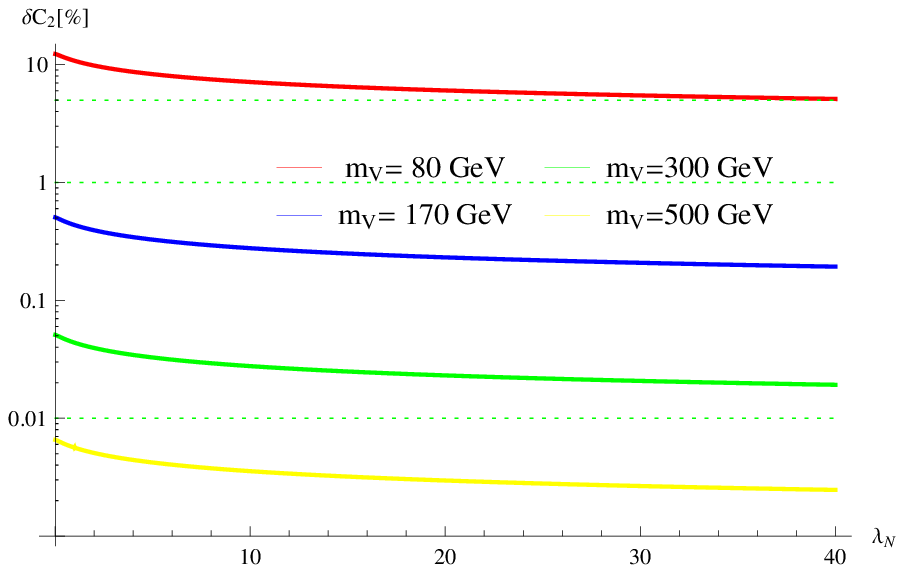} \\
 \end{tabular}
   \caption{Comparison between $|C_2(m_N,m_V,m_V)|$ and $|C'''_2(m_N,m_V,m_V)|$ in case of $m_h=125.1$ GeV.
   In the left panel, the solid (dotted) curves represent $|C_2|$ ($|C''_2|$) as functions of $\lambda_N=m_N^2/m_V^2$. }\label{capo}
\end{figure}

The LFVHD was also investigated in Ref. \cite{apo1} with some special conditions.
In the light of today's experimental data, though the analytic expressions of one-loop contributions from diagrams with $W^\pm$ mediations may give large errors compared with LoopTools,
they are still applicable to diagrams with new particle mediations such as new heavy charged scalars, gauge bosons
and fermions  in models beyond the SM \cite{lfvhloop1,hue,1loopmirror}. Comparing two analyses for particular diagrams in the 't Hooft Feynman gauge, e.g.,  Ref. \cite{apo1}, diagram 1 a), we can derive an approximate formula for the $C_2$ function denoted $C'''_2$.  It  is listed  in Appendix
\ref{apo1}. Here new notation is  $\lambda_N\equiv m_N^2/m_V^2$ ,  $M_W\rightarrow m_V$, and $M_H\rightarrow m_h$.
The $m_V$ now can be considered as the mass of some new particle playing the role of $W^\pm$ bosons in the loops.  All of the assumptions given in \cite{apo1} are still valid in this case, specifically $ \frac{m_h^2}{4 m_V^2}, \frac{m_h^2}{m_N^2}\ll 1$. Similarly, the analytic expression for $C_2(m_N,m_V,m_V)=C_2(m_h,\lambda_N,m_V)$  is  derived  from  (\ref{PV_func}).
The comparison is shown in  Fig.  \ref{capo}.
We can see that for seesaw models with loops containing $W^{\pm}$ bosons and neutral exotic leptons, the expressions in \cite{apo1} are not good, with the relative discrepancy around $5\%$. In the models with top quarks in the loop, the relative discrepancy is better, with values smaller than $1\%$.

 We  obtain the same conclusion for other $C$-functions where the analytic formulas are shown in Appendix \ref{apo1}.  In general, these formulas are very well consistent with all internal masses larger than 300 GeV.

\section{\label{illustration} LFVHD in  a model with lepton-flavored dark matter}
\subsection{The model and results of LFVHD from the previous work}
To illustrate an  effort to find how large the branching ratio (BR) Br$(h\rightarrow\mu\tau)$ can reach from one-loop contributions,
in this section we will consider a  model constructed in \cite{lfvhloop1} with a simple kind of  LFVHD loop.
This model extends the SM by adding a Majorana DM candidate $N$ and scalar partners of both left- and right-handed leptons,
called sleptons. The simplest model contains only one Majorana lepton with mass $M$, one slepton doublet $\phi_\ell=(\phi^+_\ell,\; \phi^0_\ell)^T$, and a slepton singlet $\phi_e$. These scalars couple directly with leptons through the Yukawa interactions, and therefore give new LFV couplings. Unlike the models where Yukawa couplings relate with active neutrinos, these new couplings  may be large and result in  large BRs of  LFVHD.

The details of the model are given in \cite{lfvhloop1}, we collect here only the ingredients relating  to the LFVHD.

The Lagrangian containing all LFVHD couplings  is
\bea -\mathcal{L}&=&-\mathcal{L}_{\mathrm{SM}}+m^2_{\phi_\ell}|\phi_\ell|^2 + m^2_{\phi_e}|\phi_e|^2+ \frac{1}{2}M\overline{N}N
+ \left( -y_{L_a}\bar{l}_a P_R N \tilde{\phi}_\ell + y_{R_a}\bar{e}_a P_L N \phi^-_e +\mathrm{h.c.} \right)\crn
&+& \left( -\mu H^\dagger\tilde{\phi}_\ell \phi^*_e \mathrm{+h.c.} \right)+ \lambda_{-1}|\phi_e|^2|\phi_\ell|^2+\lambda_0|H|^2|\phi_\ell|^2+V_{\mathrm{2HDM}},\label{Lagsca}
\eea
where $\tilde{\phi}_\ell=i\sigma_2\phi^*_\ell$, $H$ is the SM Higgs doublet, and  $V_{2\mathrm{HDM}}$, which is the same as the Higgs potential of two Higgs doublet models (2HDM),  can be found in \cite{lfvhloop1}. The  slepton doublet  $\phi_\ell$ and the SM Higgs doublet have the same $U(1)_Y$ charge; therefore $\phi_\ell$ can be regarded as the second Higgs doublet in the 2HDM, except that the neutral component has zero vacuum expectation value  (VEV). This is similar to the case of extension the SM Higgs sector, which is one of the necessary conditions to get large Br$(h\rightarrow\mu\tau)$ without any inconsistencies with the experimental constraint of LFV of charged lepton decays \cite{lfvhlomore,LFVgeneral2}.

 Neutrino masses in this model are originally from radiative  corrections \cite{lfvhloop1}. We will ignore contributions from  active neutrinos  to LFVHD because they are  much smaller than those from new LFV couplings \cite{hue}. Only charged sleptons and $N$ involve as LFV mediators. After symmetry breaking, these new sleptons get mass from the two mass terms of $\phi_\ell$ and $\phi_e$, as well as the part coming from the trilinear coupling $\mu v\phi_e\phi^+_\ell/\sqrt{2}+\mathrm{h.c.}$ They are all new physics beyond the SM, leading to new free parameters of the model. The original and mass eigenstates
$(\phi^\pm_\ell, \phi_e^\pm)$ and $(\tilde{e}_1^\pm,\; \tilde{e}_2^\pm)$ relate to each other through the following relations
\be \tilde{e}^\pm_1=\cos\theta \phi^\pm_\ell -\sin\theta \phi^\pm_e, \hs  \tilde{e}^\pm_2=\sin\theta \phi^\pm_\ell+\cos\theta \phi^\pm_e,\label{sleptonstate}\ee
where
\be \tan\theta=\frac{1}{\sqrt{2}v\mu}\left[ \Delta m^2_{\phi} + \sqrt{(\Delta m^2_\phi)^2+2v^2\mu^2}\;\right], \label{deftheta} \ee
and $\Delta m^2_{\phi}\equiv  m^2_{\phi_\ell}-m^2_{\phi_e}$.

The  masses of $\tilde{e}_{1,2}$ are
\be  m^2_{\tilde{e}_{1,2}}=\frac{1}{2} \left[ m^2_{\phi_\ell}+m^2_{\phi_e} \mp\sqrt{(\Delta m^2_\phi)^2+2v^2\mu^2}\; \right]. \label{slepmass}\ee
The mixing angle $\theta$ can be also read  as
\be \sin\theta\cos\theta=\frac{\mu v}{\sqrt{2}\left( m^2_{\tilde{e}_2}- m^2_{\tilde{e}_1}\right)}\rightarrow  m^2_{\tilde{e}_2}
= m^2_{\tilde{e}_1}+\frac{ \sqrt{2}v\mu }{ \sin2\theta}. \label{cstheta}\ee
  Eq. (\ref{cstheta}) implies that $m^2_{\tilde{e}_2}$,  $\theta$, $\mu$, and $m^2_{\tilde{e}_{1}}$
  are not independent of each other, i.e., one of them must be treated as a function of the remaining ones. The strict decoupling condition
  is $\mu=0$ and $\theta= 0,\pm \pi/2$.  For convenience, it is enough to assume that $\mu>0$, $0\leq \sin\theta\leq \frac{1}{\sqrt{2}}$,
  leading to the consequence that $m_{\tilde{e}_2}\geq m_{\tilde{e}_1}$. The signs of $\mu$ and $ \sin2\theta$ will be commented on if needed. The LFVHD amplitude contains only the functions $C_0(M,m_{\tilde{e}_i},m_{\tilde{e}_j})$ with $i,j=1,2$ for two sleptons $\tilde{e}_1$ and $\tilde{e}_2$.  Interestingly, the partial decay width of this decay is proportional to the following part \cite{lfvhloop1},
  \be \Gamma(h\rightarrow \mu\tau) \sim \frac{m_h}{16\pi}\times \left|\frac{M}{16\pi^2} \times \frac{\mu}{\sqrt{2}}\times C_0(-p_2,p_1-p_2,M,m_{\tilde{e}_i},m_{\tilde{e}_j})\right|^2, \label{HDCW}\ee
 where $p_2$ and $(p_1-p_2)$ are external momenta of the $\mu$ and $\tau$ leptons. Note that the factor $M$ comes from the propagator of  the neutral lepton $N$ in the loop. Because of the appearance of trilinear coupling $\mu$ in (\ref{Lagsca}), let us discuss a very interesting property of the LFVHD  suggested by Eq. (\ref{HDCW}), where  the most interesting case  is  $m^2_{\tilde{e}_2}\gg m^2_{\tilde{e}_1},M^2$. Normally, we have  $C_0\sim 1/m^2_{\tilde{e}_2}$ and $\mu\sim m^2_{\tilde{e}_1}\sin2\theta/(v \sqrt{2})$, implying that the product $|\mu C_0(-p_2,p_1-p_2,M,m_{\tilde{e}_i},m_{\tilde{e}_j})|^2$ might be finite even with a very large new scale.  As a result, an increasing value of $M$ will enhance the Br$(h\rightarrow \mu\tau)$. This property of the LFVHD was shown even in the inverse seesaw model \cite{iseesaw}, where only new mass terms of new heavy neutral leptons are added in the SM. But the LFVHD predicted by this model was still small because the exotic neutrino masses are constrained from the condition of Yukawa couplings  that must satisfy the pertubative limit.  In the model under consideration, the LFVHD is not affected from this constraint. The LFV decay of $\tau\rightarrow\mu\gamma$ does not have this property, hence the corresponding BR will decrease with  increasing values $M$ and $m_{e_{1,2}}$.

For the convenience of readers, we will review the main results shown in \cite{lfvhloop1} before going on to our main investigation. Apart from the LFVHD, new scalars and neutral leptons give new contributions to LFV decays of charged leptons, loop-induced decay $h\rightarrow\gamma\gamma$, and DM problems.   Hence the experimental data relating to these was investigated for prediction of large LFVHD.  The constraint from the decay $h\rightarrow\gamma\gamma$ allows two regions of parameters $m_{\tilde{e}_1}$ and $m_{\tilde{e}_1}$: (i) $m_{\tilde{e}_1}$ should be sufficiently heavy or nearly degenerate with $m_{\tilde{e}_2}$; and (ii) $m_{\tilde{e}_1}$ should be small for consistent values of $\mu$, which should not be too large.

    One-loop contributions to LFV decays of the SM-like Higgs boson and charged leptons were constructed from new  functions  $G(x_1,x_2)$, $G(x_1)$, $G(x_2)$,  and $F(x_{1,2})$, which are derived from  the  $C_0$-function based on different conditions of external momenta and internal masses.  New variables  are defined as  $x_{1,2}\equiv m^2_{\tilde{e}_{1,2}}/M^2$. The Br$(h\rightarrow\mu\tau)$ was estimated from the rate $R_{\tau}\equiv \mathrm{Br}(h\rightarrow\mu\tau)/\mathrm{Br}(\tau\rightarrow\mu\gamma)$ which is proportional to $G(x_1,x_2)/(F(x_1)-F(x_2))$ or $(G(x_1)+G(x_2))/(F(x_1)-F(x_2))$ in the decoupling or maximal mixing limit. They are denoted by a common function $r(x^{-1}_1,x^{-2}_2)$. According to \cite{lfvhloop1}, under the constraint of Br$(\tau\rightarrow\mu\gamma)< 4.4\times 10^{-8}$, the value of  $r(x^{-1}_1,x^{-2}_2)$ should be  large enough to explain the current experimental value of Br$(h\rightarrow\mu\tau)$. In particular, the preferable regions of parameter space are as follows. In the maximal limit, masses of the two sleptons should be degenerate. In the decoupling limit, there are three regions: (i) $m^2_{\tilde{e}_2}\simeq m^2_{\tilde{e}_1}\gg M^2$  and large  $\mu\sim\mathcal{O}(10)$ TeV; (ii) $m^2_{\tilde{e}_2}\geq \mathcal{O}(10)\times m^2_{\tilde{e}_1}\simeq M^2$  and $M^2\sim\mathcal{O}([1\mathrm{TeV}]^2)$, and iii)  $m^2_{\tilde{e}_2}\geq m^2_{\tilde{e}_1}$ and  $m^2_{\tilde{e}_1}$ should not be too much larger than $M^2$. The relic density of DM in this model can be explained with a few hundred GeV of $M$.

In the next section we will use many analytic expressions constructed in \cite{lfvhloop1} to discuss the more interesting aspects of LFVHD; in particularly we will pay attention to the regions of parameter space with few hundred GeV masses of new particles.

\subsection{New results for  LFVHD}

First, we consider the function  $G(x_1,x_2)$ defined in (\ref{Gx12}), where
$G(x_1)\equiv G(x_1,x_1)$ and $G(x_2)\equiv G(x_2,x_2)$, and
\be x_{1,2}=\frac{m^2_{\tilde{e}_{1,2}}}{M^2}. \label{x12}\ee
Our   numerical investigation shows that the difference between the  results produced from the two analytic
 expressions in \cite{lfvhloop1} and \cite{hue} does significantly increase with
 small  masses of  $M$ and $m^2_{\tilde{e}_{1,2}} $. Especially  if all of them are around 300 GeV,  not far away from $m_h=125.1$ GeV.

In the following investigation, we will use the formulas for
Br$(h\rightarrow\mu\tau)$, Br$(\tau\rightarrow\mu\gamma)$ and the deviation $c_{\gamma}$ of the
$h\gamma\gamma$ coupling that are established in \cite{lfvhloop1}, except that the $G(x_1,x_2)$-function
is replaced with the  accurate $C_0$-function mentioned above.

Regrading the estimation of  Br$(h\rightarrow\mu\tau)$ with the ratio $R_\tau$, which is
defined as $\mathrm{Br}(h\rightarrow\mu\tau)\equiv R_{\tau}\times \mathrm{Br}(\tau\rightarrow\mu\gamma)$
\cite{lfvhloop1}, seems not very good, for the  following reasons. First,  even with very
 large $R_{\tau}$, a tiny value of Br$(h\rightarrow\mu\tau)$ may correspond to  a very small
Br$(\tau\rightarrow\mu\gamma)$, and vice versa.  Second, it does not show the allowed regions satisfying the bound
Br$(\tau\rightarrow\mu\gamma)<4.8\times 10^{-8}$, because this constraint may rule out the regions with large
Br$(h\rightarrow\mu\tau)$.  We will  give a numerical discussion of these points  after reviewing the formulas required from  \cite{lfvhloop1}.

 The BR of LFVHD can be written as \cite{lfvhloop1}
\bea  \mathrm{ Br}(h\rightarrow\mu\tau)&=&1.2\times 10^{-2}\times \left(\frac{\mu}{5\mathrm{TeV}}\right)^2  \left(\frac{1\mathrm{TeV}}{M}\right)^2 \times \left(\frac{|y_{R_\tau}y^*_{L_\mu}|}{1}\right)^2\crn
 &\times&\left|\frac{\left[(G(x_1)+G(x_2)\right]\sin^22\theta}{0.4}+\frac{ G(x_1,x_2)\cos^22\theta}{0.2}\right|^2 ,\label{GBrhmuta}\eea
where the two  particular forms for this BR are  $\theta\rightarrow 0$ in the decoupling limit and $\theta\rightarrow \pi/4$ in the maximal mixing limit. The parameters $\mu$,  $y_{R_\tau}$, $y_{L_\mu}$ and  $M$ are introduced in the Lagrangian  (\ref{Lagsca}).   The Yukawa couplings can  be fixed as $|y_{R_\tau}y^*_{L_\mu}|=1$ because they are independent of charged slepton masses. In contrast, the parameter $\mu$ affects the masses of sleptons through Eq.  (\ref{slepmass}), implying that it affects $G(x_1,x_2)$. Hence we believe that $\mu$ and $G(x_1,x_2)$  do not  independently affect Br$(h\rightarrow\mu\tau)$, as discussed in \cite{lfvhloop1}. In addition, the increasing $\mu$,  corresponding to decreasing $x_{1,2}^{-1}$,  changes all values of $G(x_1),G(x_2)$, and $G(x_1,x_2)$.  So  the BR in  (\ref{GBrhmuta}) depends complicatedly on $\mu$.

\begin{figure}[h]
 \centering
\begin{tabular}{cc}
 \includegraphics[width=8cm]{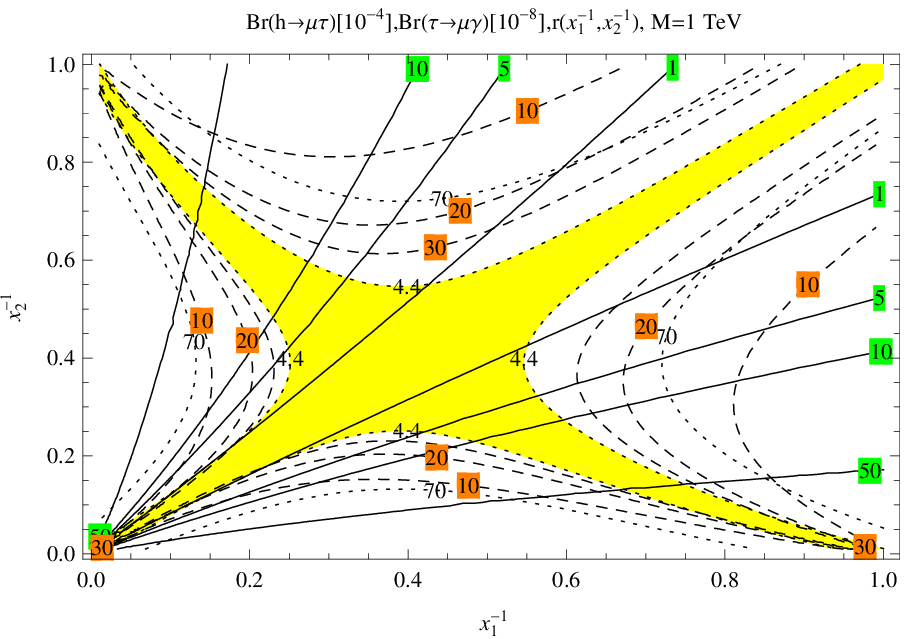} & \includegraphics[width=8cm]{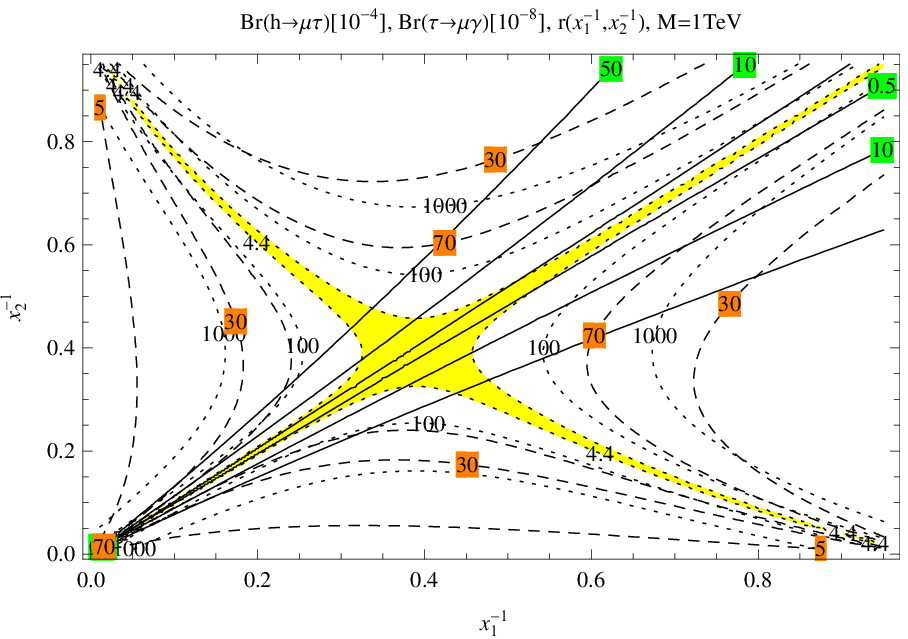} \\
  \includegraphics[width=8cm]{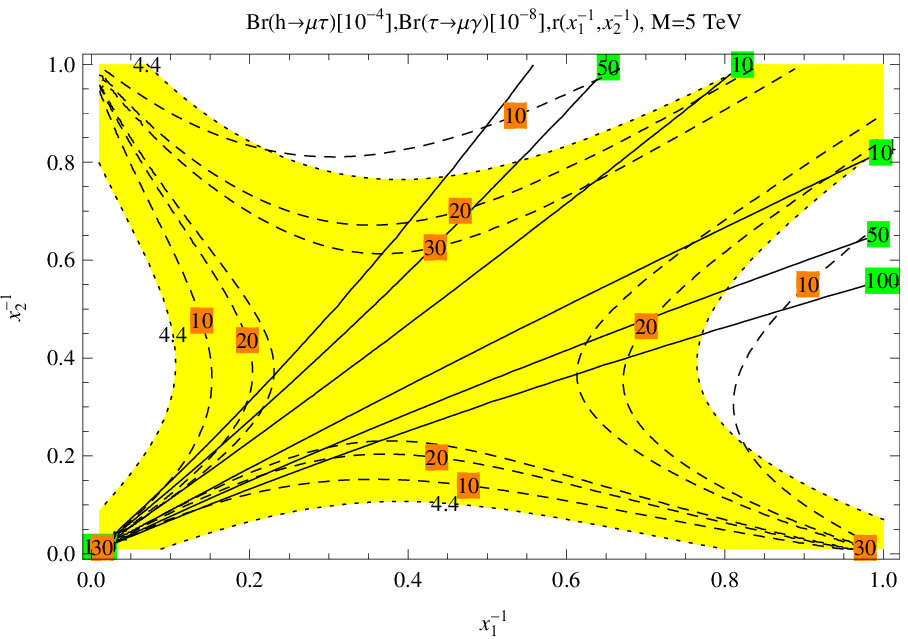} & \includegraphics[width=8cm]{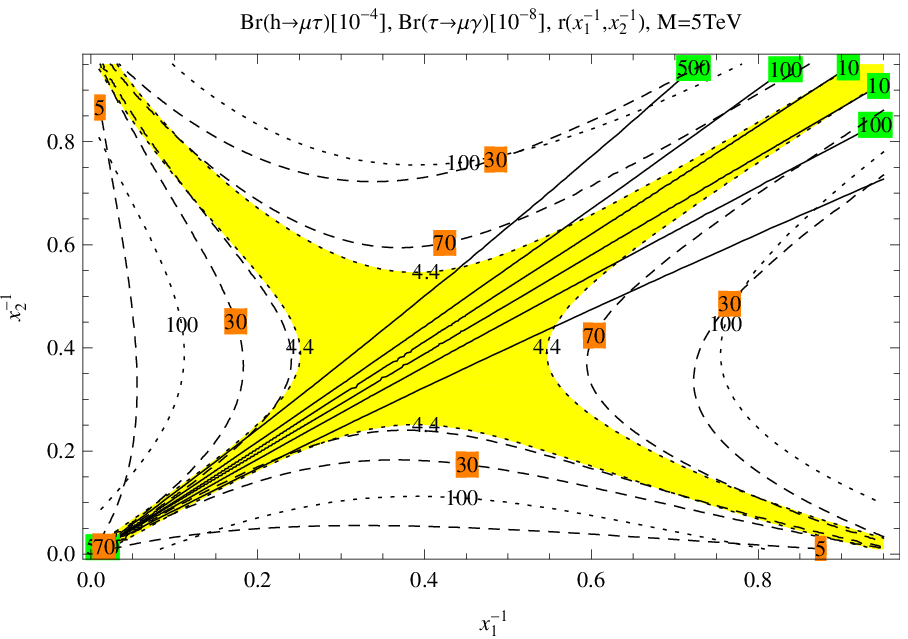} \\

  \end{tabular}
   \caption{Contour plots as functions of $x_1^{-1}$ and $x_2^{-1}$ . The yellow regions satisfy the upper bound of Br$(\tau\rightarrow\mu\gamma)<4.4\times 10^{-8}$. The left (right) panel corresponds to the decoupling (maximal mixing) limit. The solid, dotted and dashed curves represent the constant values of Br$(h\rightarrow\mu\tau)$, Br$(\tau\rightarrow\mu\gamma)$, and $r(x^{-1}_1,x^{-1}_2)$, respectively. }\label{clfvvsH}
\end{figure}

The relation between Br$(\tau\rightarrow\mu\gamma)$ and Br$(h\rightarrow\mu\tau)$  is given by
{\footnotesize \bea  \mathrm{Br}(\tau\rightarrow\mu\gamma)&=& \frac{10^{-5}}{2.8}\left( \frac{5\mathrm{TeV}\sin2\theta}{2\mu}\right)^2\crn
&\times&  \left|\frac{ 400(F_2(x_2)-F_2(x_1))}{\left[G(x_1)+G(x_2)\right]\sin^22\theta+ 2G(x_1,x_2)\cos^22\theta}\right|^2\times \mathrm{Br}(h\rightarrow\mu\tau) \label{ratiogen}\eea
}
with \cite{lfvhloop1}
\be F_2(x)\equiv \frac{-1+x^2-2x\ln x}{2x(1-x)^2}. \label{f2x}\ee
  Eq. (\ref{ratiogen}) contains two specific limits of decoupling and maximal mixing, which are separately considered in  \cite{lfvhloop1}. Here we use the ratio $1/R_{\tau}$ instead of $R_{\tau}$. Recall  that  these ratios cancel  all Yukawa couplings appearing in both expressions of the branching ratios. If we consider simultaneously both  (\ref{GBrhmuta}) and  (\ref{ratiogen}),  the discussion in  Ref.  \cite{lfvhloop1} for large LFVHD is  illustrated in another way, as shown in Fig. \ref{clfvvsH}, where $\sin\theta=0.1$ for  the decoupling limit. The three quantities Br$(h\rightarrow\mu\tau)$, Br$(\tau\rightarrow\mu\gamma)$ and $r(x^{-1}_1,x^{-1}_2)$ are represented in the same figure. We emphasize that in this investigation the $\mu$ parameter is expressed as a function of $m_{\tilde{e}_i}$ and $\theta$, given by  (\ref{cstheta}).
\footnote{We guess that Ref. \cite{lfvhloop1} did not pay attention to this point.}
We can see that the constraint of Br$(\tau\rightarrow\mu\gamma)$ seems to favour small Br$(h\rightarrow\mu\tau)$ in the region with degenerate slepton masses.

 Fig. \ref{clfvvsH} also shows two  interesting points: i) a large $r(x^{-1}_1,x^{-1}_1)$ does not always correspond to a large Br$(h\rightarrow\mu\tau)$ when the experimental constraint of Br$(\tau\rightarrow\mu\gamma)$ is considered, and ii) the allowed regions with large Br$(h\rightarrow\mu\tau)$  are sensitive to the variation of $M$ but seem not sensitive to the change of $r(x^{-1}_1,x^{-1}_2)$. Furthermore, an increasing $M$ enhances  Br$(h\rightarrow\mu\tau)$, but causes Br$(\tau\rightarrow\mu\gamma)$ to decrease. As a result,   the allowed region expands wider.  These conclusions are, in general, different from those indicated in \cite{lfvhloop1}. An illustration of the first point is that with $x_1\simeq x_2$,  increasing $r$ will cause  a decrease in the value of Br$(h\rightarrow\mu\tau)$; see the lower-left and upper-right of all panels in Fig. \ref{clfvvsH}.  For the second point, enhancement of Br$(h\rightarrow\mu\tau)$ can be explained by considering formula (\ref{GBrhmuta}). With large $M>1$ TeV and $x_2\gg x_1\simeq 1$,  the approximate expressions (\ref{Gx12}) are applicable and  very convenient  for qualitative  estimation. Fig. \ref{clfvvsH} suggests two allowed regions giving  large Br$(h\rightarrow\mu\tau)$: (i)  $x_1=x_2\gg1$, and (ii) $x_1\rightarrow1$ while $x_2\rightarrow\infty$ assuming that $x_2>x_1$. To understand, using $\mu=\sin2\theta \times M^2(x_2-x_1)/(\sqrt{2}v)$ obtained from (\ref{cstheta}), with $v\simeq0.25$ TeV,  $m^2_{\tilde{e}_i}=M^2 x_i$, we get a formula  for LFVHD derived from (\ref{HDCW}) and (\ref{GBrhmuta}):
\bea  \mathrm{Br}(h\rightarrow\mu\tau) &\simeq& 9.6\times 10^{-2} \left| \frac{y_{R_\tau}y^*_{L_\mu}}{1}\right|^2\times \left|\frac{M\sin2\theta}{1\mathrm{TeV}}\right|^2\crn
&\times&\left|(x_2-x_1)\left(\frac{1}{2}\sin^22\theta [G(x_1)+G(x_2)] +\cos^22\theta G(x_1,x_2)\right)\right|^2. \label{hmutaudegen}\eea

 Assuming that $M$ and $\theta$ are fixed, in the first region with $x_1\rightarrow x_2$, the total factor relating to $x_i$ goes to zero in the limit, then the BR of LFVHD will go  to zero  too. So the large LFVHD is not caused by this degeneration between slepton masses. Instead the enhancement of BR of  LFVHD  originates from the large product $|M\sin2\theta/(1\mathrm{TeV})|^2$.  Comparing the upper and lower panels of  Fig. \ref{clfvvsH}, the effect of  $M$  is clearly illustrated, while the effect of $\sin\theta$  can be seen in the left and right panels,  where $\sin\theta=0.1$ and $1/\sqrt{2}$, respectively.

In the second region, where $x_1\rightarrow1$, we have $\left|(x_2-x_1)G(x_1,x_2)\right|= \left|\frac{1-x_2+x_2\ln x_2}{(x_2-1)}\right|$ and $\left|(x_2-x_1)\left[G(x_2)+G(x_1)\right]\right|=\left|(1 - x_2^2 + 2 \ln x_2)/(2 - 2 x_2)\right|$. Both of them can be arbitrary large if  $x_2$ is not constrained.  Combining with the factor of $|M\sin2\theta/(1\mathrm{TeV})|$, it is easy to derive that the value of Br$(h\rightarrow\mu\tau)$ will take arbitrary values with large $M$ and nonzero $\sin2\theta$.  In contrast, in this case the Br$(\tau\rightarrow\mu\gamma)$ is very suppressed, which  can be explained as follows. From  Eq. (\ref{ratiogen}), or the precise  form of the partial decay width of the LFV process $\tau\rightarrow\mu\gamma$ shown in \cite{lfvhloop1}, we can see that
$$\mathrm{Br}(\tau\rightarrow\mu\gamma)\sim  \frac{\sin^22\theta}{M^2}\times|F_2(x_2)-F_2(x_1)|^2,$$
with $F_2(x)$ given in (\ref{f2x}). It is easy to prove that $\lim_{x_1\rightarrow 1}F_2(x_1)=0$ and $\lim_{x_2\rightarrow \infty}F_2(x_2)=0$. Hence, if $x_2$ or $M$ is large enough, the Br$(\tau\rightarrow\mu\gamma)$ always satisfies the experimental bounds.  Because $x_2=m^2_{\tilde{e}_2}/M^2$ and with relation (\ref{cstheta}),  a large $m^2_{\tilde{e}_2}$ will correspond to a large $\mu$, leading to a very narrow allowed region with large $\mu$ values, as we will show below.  In this case, the experimental data such as LFVHD and $h\gamma\gamma$ coupling will give information on the parameters of the model.

 In the next section, we focus just  on the small values of $M$ below 1 TeV, where $N$ can be detected by experiments and  addressed  with dark matter candidates \cite{lfvhloop1}. Another reason is that small values of $M$  can be  accurately investigated using the analytic expressions we mentioned above.

 In this investigation, we will combine two constraints of BR$(\tau\rightarrow\mu\gamma)$ and $h\gamma\gamma$ coupling to estimate Br$(h\rightarrow\mu\tau)$. The free parameters chosen are $M$, $\theta$, $\mu$, and $m_{\tilde{e}_1}$, apart from fixed Yukawa couplings. As an  illustration, the parameter $M$  will be fixed in two cases:  small $M=300$ GeV and large $M=1$ TeV. The value of $\theta$ will be chosen the same as the assumption from Ref. \cite{lfvhloop1}, with the two limits for decoupling $\sin\theta=0.1$  and   maximal mixing $\sin\theta=1/\sqrt{2}$.

The loop-induced coupling of decay $h\rightarrow\gamma\gamma$ is  nonnegative and constrained by \cite{lfvhloop1}
\be 0\leq \delta_{\gamma}\equiv\frac{\delta_{c_{\gamma}}}{c_{\mathrm{SM},\gamma}}=\frac{1}{48\times 0.81}\times \frac{(\mu v)^2}{m^2_{\tilde{e}_1}(m^2_{\tilde{e}_1}+\sqrt{2}\mu v\sin2\theta)} <0.20,  \label{h2ga} \ee
where we have used $ \left(m^2_{\tilde{e}_2}-m^2_{\tilde{e}_1}\right)\sin2\theta =\sqrt{2}\mu v$.  In contrast to Ref.  \cite{lfvhloop1}, where $m_{\tilde{e}_2}$ is ignored, here we include this mass in (\ref{h2ga}). The interesting consequence is that $\delta_{\gamma}$ is always positive, unlike the conclusion about the two allowed regions indicated in previous work. It is easy to see that the $h\gamma\gamma$ coupling deviation  gives an upper bound on $\mu$.

\begin{figure}[h]
 \centering
\begin{tabular}{cc}
 \includegraphics[width=8cm]{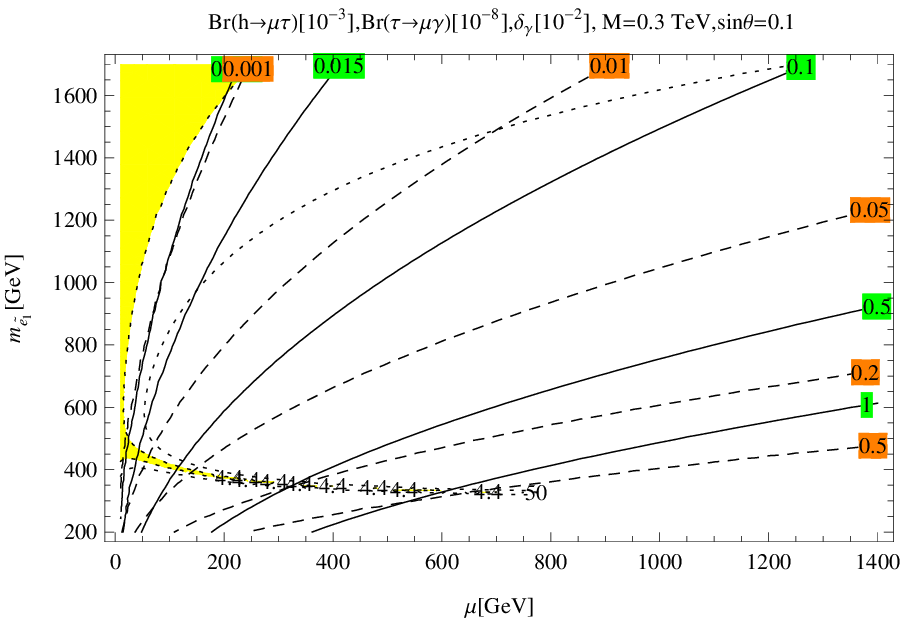} & \includegraphics[width=8cm]{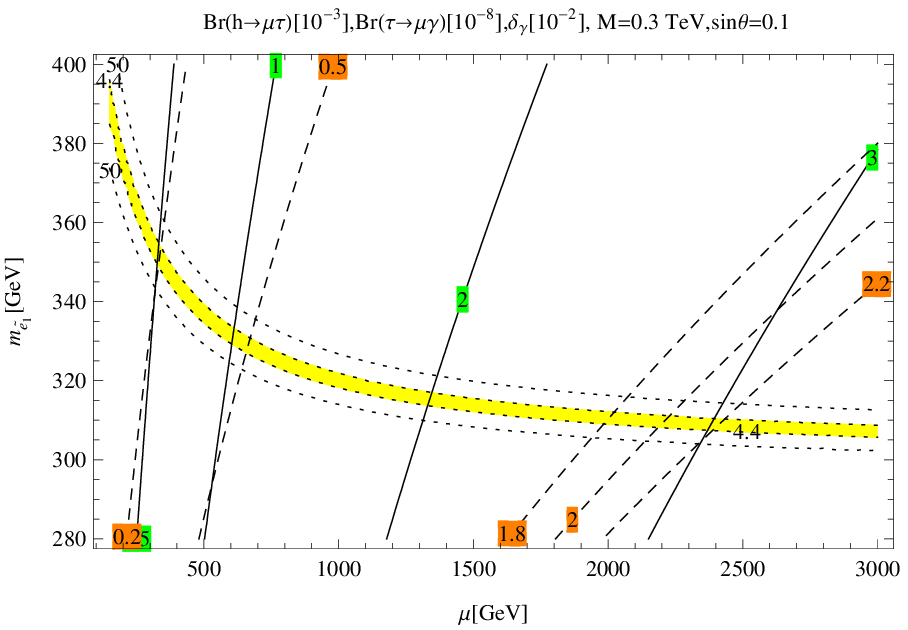} \\
 \includegraphics[width=8cm]{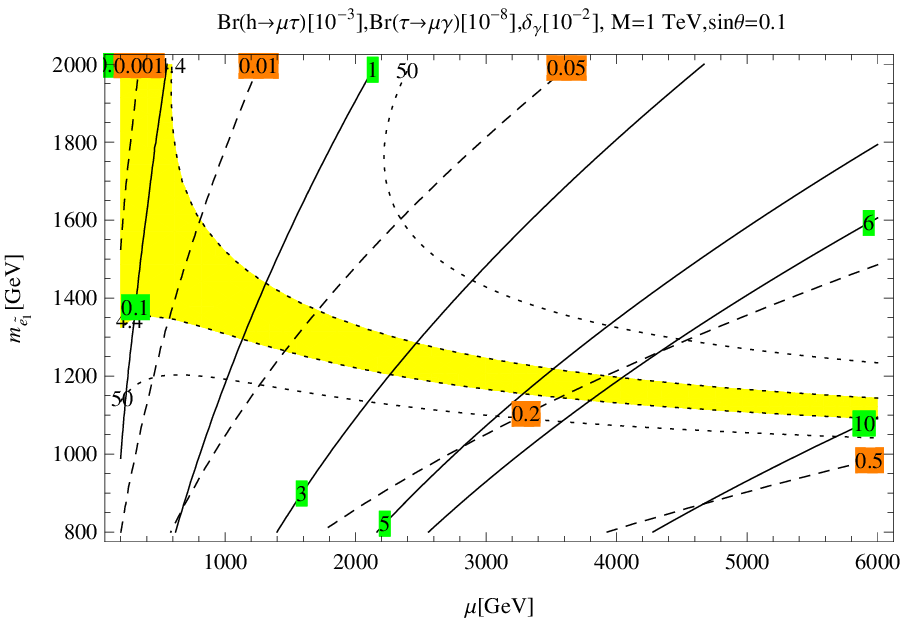} & \includegraphics[width=8cm]{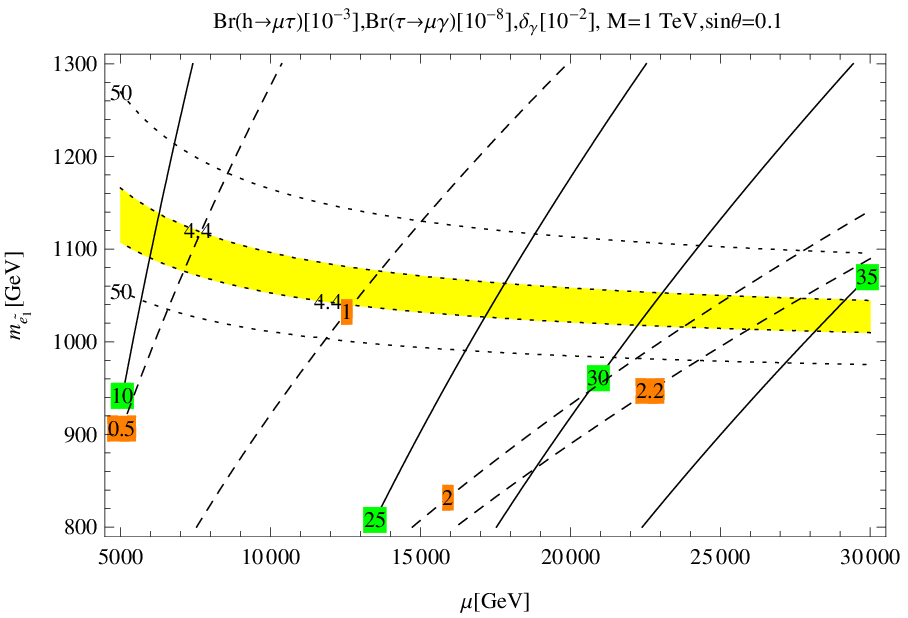} \\
  \end{tabular}
   \caption{Contour plots as functions of $\mu$ and $m_{\tilde{e}_1}$ in the decoupling limit. The yellow regions satisfy the upper bound of Br$(\tau\rightarrow\mu\gamma)<4.4\times 10^{-8}$.  The solid, dotted, and dashed curves represent the constant values of Br$(h\rightarrow\mu\tau)$, Br$(\tau\rightarrow\mu\gamma)$ and $\delta_{\gamma}$, respectively.}\label{conmum1d}
\end{figure}

\begin{figure}[h]
 \centering
\begin{tabular}{cc}
\includegraphics[width=8cm]{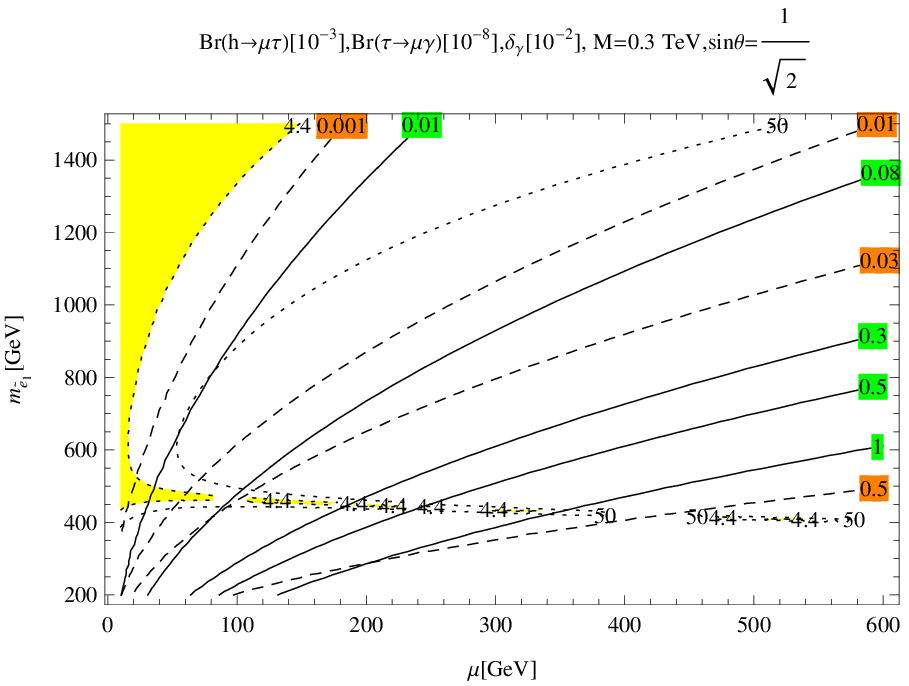} & \includegraphics[width=8cm]{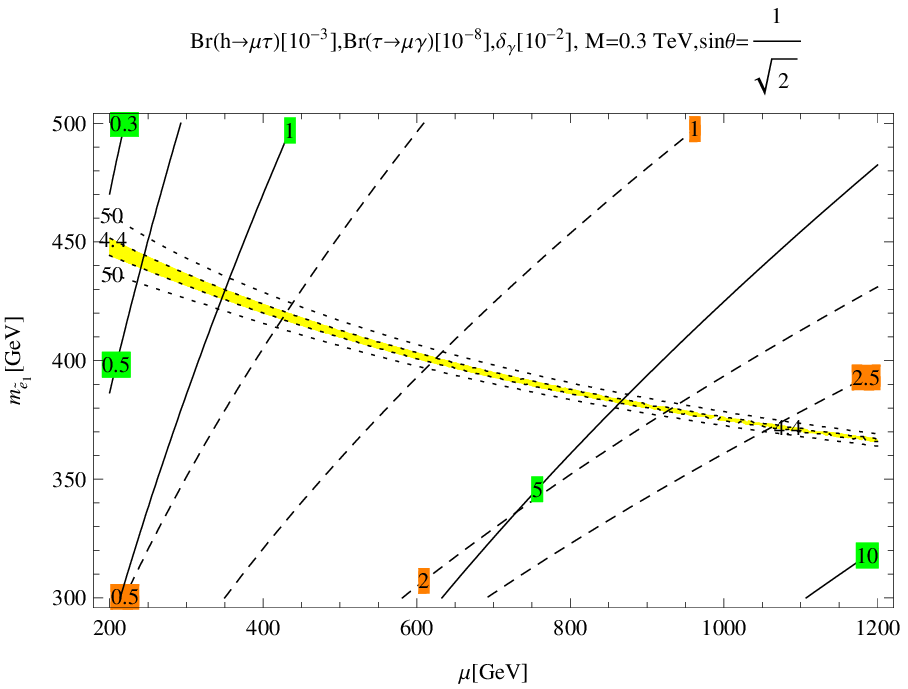} \\
\includegraphics[width=8cm]{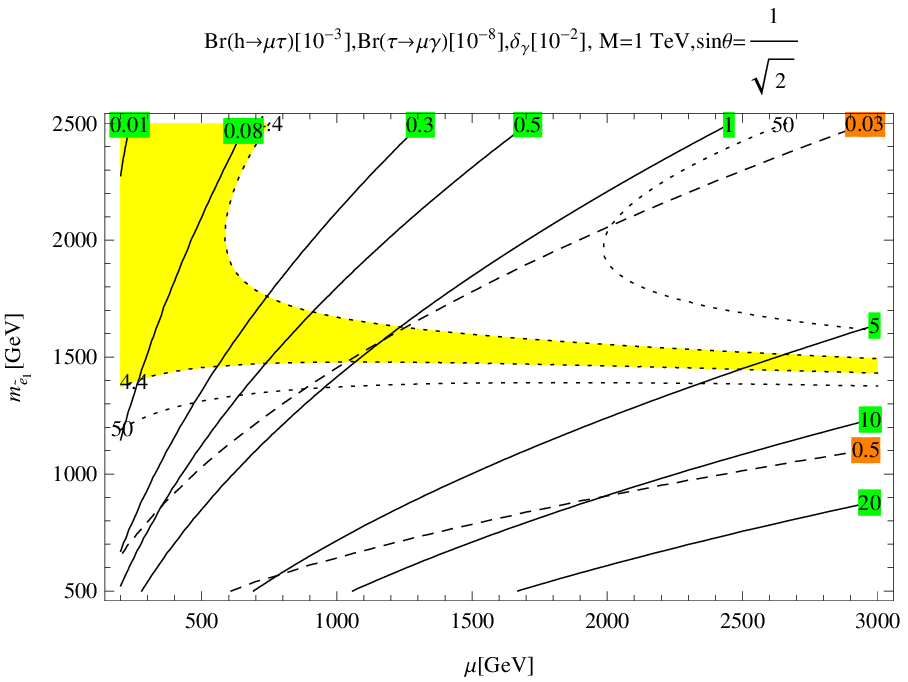} & \includegraphics[width=8cm]{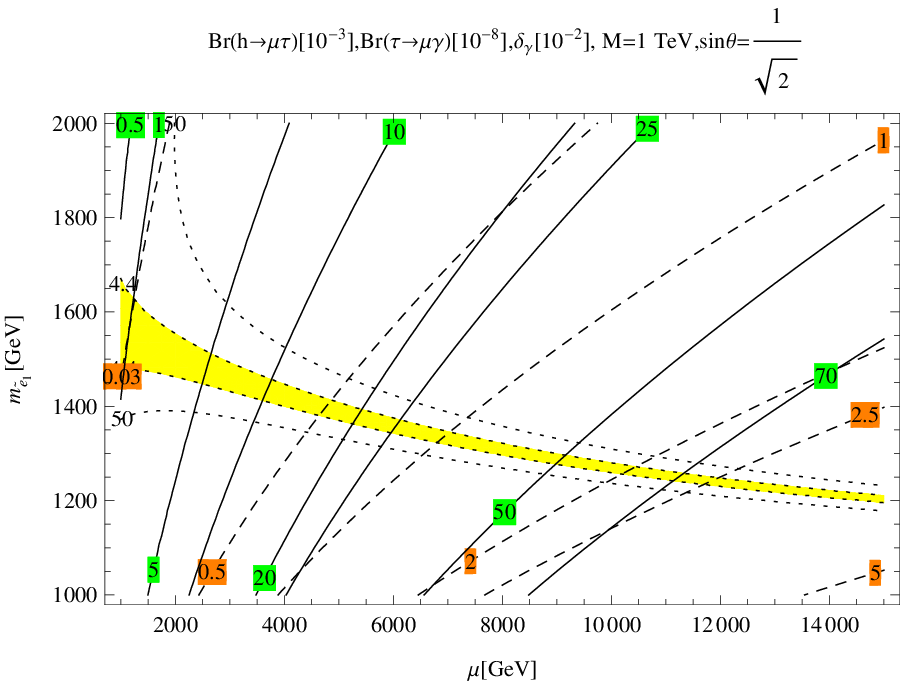} \\
  \end{tabular}
   \caption{Contour plots as functions of  $\mu$ and $m_{\tilde{e}_1}$ in the maximal mixing limit. Conventions are the same as those given in Fig. \ref{conmum1d}}\label{conmum1m}
\end{figure}

 Illustrations are shown in Figs. \ref{conmum1d} and  \ref{conmum1m}, corresponding to the two decoupling and maximal mixing limits.
 There are common properties shown in the two figures. In each figure, the allowed region  from the constraint Br$(\tau\rightarrow\mu\gamma)$ consists of two distinguishable parts:  (i)  large $m_{\tilde{e}_1}$  and  small $\mu$, and (ii) $m_{\tilde{e}_1}$ is around the value of $M$ while $\mu$ is arbitrarily large. The first part, which lies in the upper-left region of the left panel, corresponds to very small $\mu$ or $\sqrt{2}v\mu\sin2\theta$. Therefore it gives Br$(h\rightarrow\mu\tau)$  smaller than $10^{-5}$ with small $M=300$ GeV, and $10^{-3}$ with large $M=1$ TeV. While the second part gives much larger BR of LFVHD, it is still constrained by $h\gamma\gamma$ coupling deviation.  For $M=300$ GeV, the largest Br$(h\rightarrow\mu\tau)$ can reach order $10^{-3}$, very close to the recent experimental value.  For $M=1$ TeV, the Br can reach values of  $10^{-2}$, which is an order larger than the case of $M=300$ GeV.  The constraint from $h\gamma\gamma$ coupling gives a less strict constraint on $\mu$ than that from the LFVHD. So the information of  free parameters depends on the experimental bounds of the LFVHD with large $M$ and $m_{\tilde{e}_1}\simeq M$.

\begin{figure}[h]
 \centering
\begin{tabular}{cc}
\includegraphics[width=8cm]{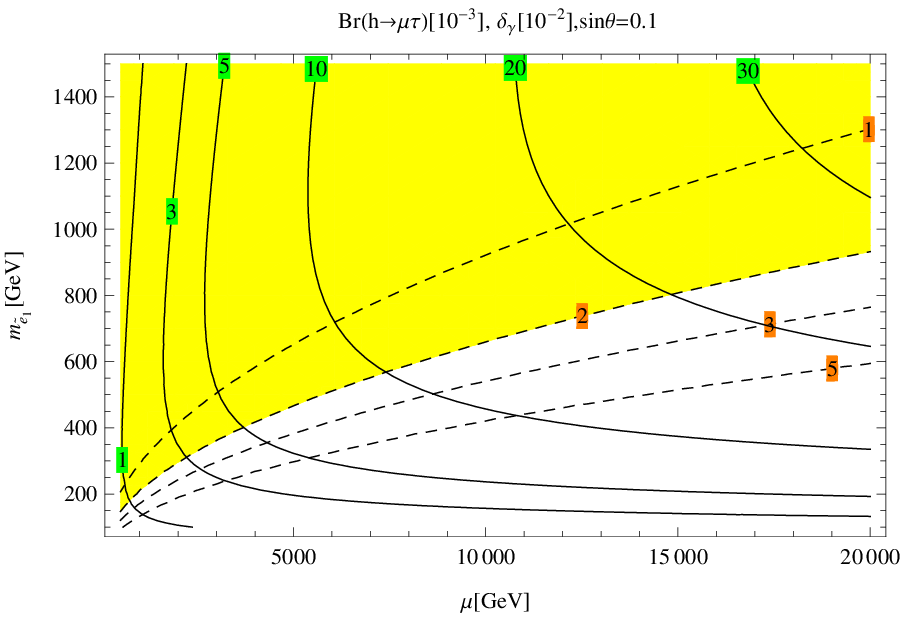} & \includegraphics[width=8cm]{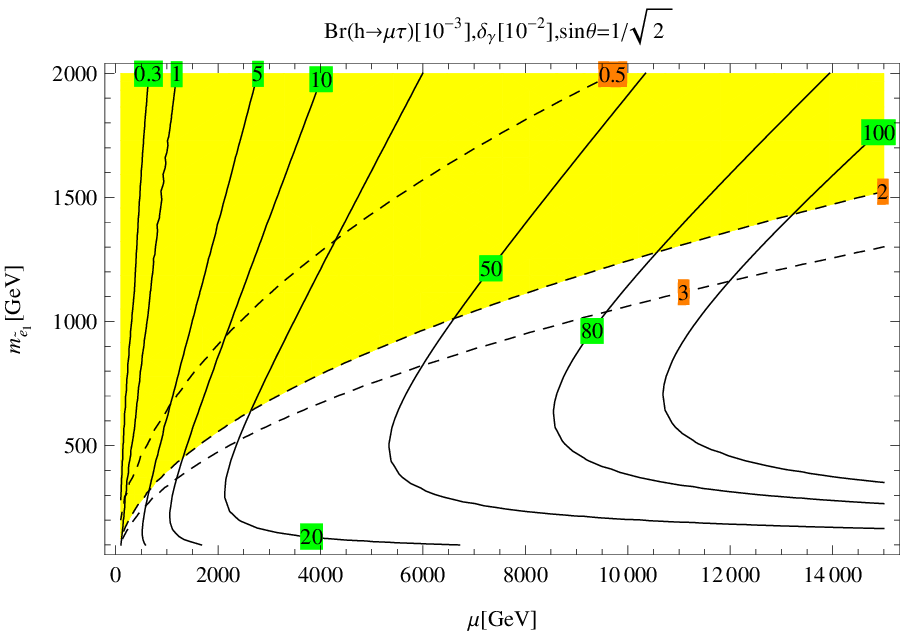} \\
  \end{tabular}
   \caption{Contour plots as functions of  $\mu$ and $m_{\tilde{e}_1}$ in the limit of $M=m_{\tilde{e}_1}<m_{\tilde{e}_2}$. The Yellow regions satisfy $\delta_{\gamma}<0.2$. }\label{conmax}
\end{figure}

From above discussion, one can conclude that if $M\leq 1$ TeV the most interesting region giving large BR$(h\rightarrow\mu\tau)$ corresponds to the degeneration  of $M$ and the lighter slepton, i.e.,  $M=m_{\tilde{e}_1}\ll m_{\tilde{e}_2}$ or $x_1=1\ll x_2$. Now we will focus on this special case.  As mentioned above, because  $\lim_{x_1\rightarrow 1} F_2(x_1)=0$ and $\lim_{x_2\rightarrow \infty} F_2(x_2)=0$,  resulting  in very suppressed Br$(\tau\rightarrow\mu\gamma)$,   the constraint now comes from the Higgs coupling $c_{\gamma}$.   Illustrations are drawn in Fig. \ref{conmax}.
 Both of the coupling and decoupling limits  can explain the experimental LFVHD value of $5\times 10^{-3}$ when  $M=m_{\tilde{e}_1}\geq 400$ GeV.  And the constraint from $h\gamma\gamma$ coupling deviation gives lower bound on these masses. The parameter $\mu$ can be determined rather strictly from the information of LFVHD values.  With the recent constraint of Br$(h\rightarrow\mu\tau)\leq 10^{-2}$, $\mu$ should be smaller than a few TeV if the  Majorana mass $N$ is below 1 TeV.

Because the masses of  the DM candidate $N$ and $ m_{\tilde{e}_1}$, especially $M\simeq  m_{\tilde{e}_1}$, are consistent with values discussed in \cite{lfvhloop1},  the electroweak scale of $M$ and $m_{\tilde{e}_1}$ can give the correct relic density of DM. And this conclusion does not depend on $\mu$.  But in order to satisfy both the condition of  large Br$(h\rightarrow\mu\tau)$ and $c_{\gamma}$ constraint, we indicate that the value of $\mu$ should not be larger than 2 TeV, which is smaller than the values used in \cite{lfvhloop1} for investigating direct searches of DM, $\mu\geq 5$ TeV. It can be seen that the DM-nucleon scattering rate decreases with decreasing values of $\mu$  \cite{lfvhloop1}.  In particular,  the DM-nucleon scattering  is generated by radiative corrections via mediations of the virtual photon and Higgs boson.  The contribution from photon mediation is significant only for  lighter sleptons $\tilde{e}^\pm_1$, but  is insensitive to $\mu$.  The  scattering rate will be much smaller than the current LUX sensitivity if only photon mediation is considered and $M$ is in range  $\mathcal{O}(100 \;\mathrm{GeV})$ \cite{lfvhloop1,LUX}. The contribution from the Higgs mediation can be presented by the  effective coupling $\lambda_{hN}(0)$ being proportional to $\mu$ \cite{lfvhloop1}.  Because the  scattering rate can reach close to the current LUX sensitivity  for  $\mu\geq 5$ TeV, the smaller value of $\mu$ results in a smaller scattering rate. Hence, it is harder to detect DM from DM-nucleon scattering with small $\mu$ indicated by our investigation.

The  region of  parameter space we discussed above, with degeneration of masses of $M$ and $ m_{\tilde{e}_1}$,  is an especially interesting   explanation for gamma ray peak being internal bremsstrahlung in DM annihilation through a charged t-channel mediator $\tilde{e}^{\pm}_i$ \cite{LUX}.  This parameter  region is also the most promoting region for finding signals of Majorana DM from planned XENON1T \cite{XENON1T} and LUXZEPLIN \cite{LUX-ZEPLIN} experiments \cite{LUX}.

\subsection{Productions  of new particles at colliders}
As we  have shown,  at least the masses of the $\tilde{e}^{\pm}_{1}$ and  lepton-flavored Majorana DM $N$  can be smaller  than 1 TeV. Therefore, they may be detected at current running energies of the LHC or near future $e^+e^-$ colliders such as the International Linear Collider (ILC) \cite{ILC1,ILC2} and the Compact Linear Collider (CLIC) \cite{CLIC1, CLIC2}. Interestingly, the sleptons defined in the model under consideration couple to SM particles in a very similar way to the sleptons in  supersymmetric (SUSY) extensions of the SM.  In addition, all new particles in both kinds of models are odd under $Z_2$ symmetries.  Hence the lightest neutral particle $N$ is a DM candidate, and plays the same role as the lightest  neutralino in the SUSY models with $R$-parity conservation.  In general, the new particle spectrum in the considering model can be seen as a simplified version of  the superpartner spectrum, which  has been hunted by LHC \cite{PDG2014}, especially  searches for  slepton productions \cite{ATLAS1,ATLAS1p,ATLAS2,ATLAS3}. For SUSY models, current channels of experimental searches are   $pp\rightarrow \tilde{\ell} \tilde{\ell}\rightarrow ( \ell \widetilde{\chi}^0_1) (\ell \widetilde{\chi}^0_1)$,   where $\ell^\pm$ denotes an SM lepton state:  $e,\mu,\tau$. In notation of the model under consideration, these channels correspond to processes  $pp\rightarrow \tilde{e}^+_i \tilde{e}^-_j\rightarrow ( \ell^+\overline{N}) (\ell^-N)$.  The slepton productions at the LHC happen via virtual  gauge bosons, i.e., $pp\rightarrow \gamma^*, Z^*\rightarrow \widetilde{\ell}^0\widetilde{\ell}^{0*},\widetilde{\ell}^+\widetilde{\ell}^{-}$;  or $pp\rightarrow W^{\pm*}\rightarrow \widetilde{\ell}^\pm\widetilde{\ell}^{0}$, because the gauge bosons are always lighter than the new particles.  Based on  couplings of new particles at final states, we see that the signals of detection of new particles in both kinds of models, namely  SUSY  and the model studied in our work,  are of the same order.  Couplings of SUSY particles are given in detail  in many textbooks, e.g.,   \cite{SUSYcoup1}. The relevant couplings of sleptons and $N$ predicted by the model under consideration are collected in Table \ref{newcoup}. Note that $\phi^0_\ell$ is the neutral component of the slepton doublet, $\phi_\ell=(\phi^+_\ell,\; \phi^0_\ell)^T$.
\begin{table}[ht]
  \centering
  \begin{tabular}{|c|c|c|c|}
    \hline
    Coupling & Vertex & Coupling & Vertex \\
      \hline
    $h\tilde{e}^{+}_1\tilde{e}^{-}_1$ & $-is_{2\theta}\frac{\mu}{\sqrt{2}}$  &   $h\tilde{e}^{+}_2\tilde{e}^{-}_2$ &$is_{2\theta}\frac{\mu}{\sqrt{2}}$  \\
    \hline
     $h\tilde{e}^{+}_1\tilde{e}^{-}_2$&$-i c_{2\theta}\frac{\mu}{\sqrt{2}}$ & $h\tilde{e}^{-}_1\tilde{e}^{+}_2$&$-i c_{2\theta}\frac{\mu}{\sqrt{2}}$\\
    \hline
     $\tilde{e}^{-}_1\overline{e_a} N$ & $i\left(s_{\theta}y_{Ra}P_L- c_{\theta}y_{La}P_R\right)$& $\tilde{e}^{-}_2\overline{e_a} N$ & $-i\left( c_{\theta}y_{Ra}P_L+ s_{\theta}y_{La}P_R\right)$\\
    \hline
     $Z_\mu\left( \tilde{e}^-_1\partial^{\mu}\tilde{e}^+_1- \partial^{\mu}\tilde{e}^-_1\tilde{e}^+_1\right)$ & $ \frac{-g(c^2_{\theta}-2s^2_W)}{2 c_W}$& $Z_\mu\left( \tilde{e}^-_2\partial^{\mu}\tilde{e}^+_2- \partial^{\mu}\tilde{e}^-_2\tilde{e}^+_2\right)$&  $ \frac{-g(s^2_{\theta}-2s^2_W)}{2 c_W}$\\
    \hline
     $Z_\mu\left( \tilde{e}^-_i\partial^{\mu}\tilde{e}^+_j- \partial^{\mu}\tilde{e}^-_i\tilde{e}^+_j\right)$ & $ \frac{-g}{4 c_W}s_{2\theta}$&  $A_\mu\left( \tilde{e}^-_i\partial^{\mu}\tilde{e}^+_i- \partial^{\mu}\tilde{e}^-_i\tilde{e}^+_i\right)$& $-gs_W$\\
    \hline
     $W^{\mp}_\mu\left( \phi^{0*}_\ell\partial^{\mu}\tilde{e}^{\pm}_1- \partial^{\mu}\phi^{0*}_\ell\tilde{e}^{\pm}_1\right)$   & $\mp\frac{ g}{\sqrt{2}}c_{\theta}$&   $W^{\mp}_\mu\left( \phi^{0*}_\ell\partial^{\mu}\tilde{e}^{\pm}_2- \partial^{\mu}\phi^{0*}_{\ell}\tilde{e}^{\pm}_2\right)$& $\mp \frac{ g}{\sqrt{2}}s_{\theta}$\\
    \hline
  \end{tabular}
  \caption{Couplings of new particles in the model introduced in \cite{lfvhloop1} . Here $i,j=1,2$ and $i\neq j$.}\label{newcoup}
\end{table}

Precise  properties of couplings  are  as follows.  Couplings of SM gauge bosons with sleptons in both SUSY and the model under consideration are of the same order as the gauge coupling $g$.  The coefficients of slepton-lepton-neutralino couplings $\widetilde{\ell} \ell \chi^0_1$ in SUSY models are also of  order of the gauge couplings, the same with the order of  the Yukawa coefficients $Y_{L,Ra}$ chosen  for   $\tilde{e}^{-}_i\overline{e_a} N$ vertices.   Regarding  h-slepton-slepton couplings,  the vertex coefficients are proportional to $\mu$ in case of the nondecoupling limit, but $\mu$ is constrained from above because of the constraint of the $h\gamma\gamma$ coupling.  Anyway, the model predicts a promising signal of the slepton production from gluon fusion channel $gg\rightarrow h_0^1\rightarrow \tilde{e}^+\tilde{e}^-$ relating  to a top quark loop.

The above discussion shows that searches for SUSY sleptons can be applied for sleptons in the model under consideration. Current lower bounds of  sleptons are a few hundred GeV, which do not exclude  the light sleptons and $N$ in the  region of parameter space we discussed above. And they may be detected   at the LHC \cite{LHCPredict,LHCPredict1,ILCPredict6}. For example, with the condition of very small differences between masses of stau and the lightest neutralino (not larger than 1 GeV), Ref. \cite{LHCPredict} suggested that the expected number of staus may be several hundred at 8 and  14TeV LHC run with light masses larger than 450 GeV.

The sleptons and $N$ can also be searched for in future $e^+e^-$ colliders with collision energies up to 3 TeV, such as the ILC and CLIC. Predictions for signals of  sleptons and DM  were indicated in  SUSY models   \cite{ILCPredict1,ILCPredict2,ILCPredict3,ILCPredict4} and  models with lepton-flavored DM \cite{ILCPredict5,ILCPredict6}. Similarly to the LHC, slepton productions will be  searched through $s$ channels of $e^+e^-\rightarrow Z^*,\gamma^*\rightarrow \tilde{e}^+_i  \tilde{\ell}^-_j, \phi^0_{\ell}\phi^{0*}_{\ell}$.  In contrast to the LHC, where quarks and gluons do not couple to $\tilde{e}_i$ and $N$,  there are  additional  $t(u)$ channels through the exchange of $N$, leading to enhancements of slepton production at the ILC. In addition, $e^+e^-$ colliders give a direct channel of DM search  $e^+e^-\rightarrow \overline{N}N\gamma$, corresponding to a  signal of a mono-photon plus missing energy. Another indirect search  is the one-loop contribution, where  sleptons and $N$ run in loops, to lepton pair production, $e^+e^-\rightarrow l^+_il^-_i$,  and multiflavor lepton final state $e^+e^-\rightarrow l^+_il^-_j$ ($i\neq j$) \cite{ILCPredict6}. Recent bounds of new particle masses obtained from the LEP  are a few hundred GeV \cite{ILCPredict6,LEP1,LEP2}, which is consistent with results obtained from the LHC.

In summary, the lower constraints of masses of sleptons and $N$ under recent experimental results are a few hundred GeV. The parameter region of the lepton-flavor DM we discussed in this work is still valid, and is very predictive for  many future projects of (in)direct  searches for these new particles.

\section{\label{Con}Conclusion}
The one-loop contributions to LFV decays of neutral Higgs bosons are now very interesting in many models beyond the SM, where many new particles may inherit masses that are not far from the electroweak scale. In some models, even the top quarks can play the role of LFV mediations in the loop. These one-loop contributions can be conveniently written in terms of the one-loop-three point $C$-functions, before taking any approximations for more precise forms used for numerical investigations. We have shown that  numerical results obtained from the analytical forms of  the $C$-functions introduced in Ref. \cite{hue}  are  in great agreement with those evaluated by LoopTools. This conclusion is true for all ranges  of mass values in the loops, even with loops containing active neutrino masses smaller than a few eV. We  have compared this with the two other analytic approximations given in \cite{lfvhloop1} and \cite{apo1}.  We have found that the latter two expressions are still safe with all masses in the loops large than 1 TeV for the case of studying LFVHD of the SM-like Higgs boson. But they fail with masses in the loops below a few hundred GeV. Furthermore, they can not be applied for LFVHD of new heavy neutral Higgs bosons appearing in many models beyond the SM. However, the results in \cite{hue} still work very well.

  Based on the above conclusions, the analytic formulas of $C$-functions given in \cite{hue} have been used to reinvestigate the LFVHD mentioned in Ref. \cite{lfvhloop1}, focused on the regions of small masses of Majorana dark matter $M$ and slepton masses $m_{\tilde{e}_{1,2}}$. We stress that these regions could not be accurate with the approximation used in previous works. We found many interesting results that are not metioned in \cite{lfvhloop1}. In particular,  large Br$(h\rightarrow\mu\tau)$ depends strongly on $M$, namely it enhances with increasing values of $M$.  Even when  constraints of both Br$(\tau\rightarrow\mu\gamma)$ and $h\gamma\gamma$ coupling deviation are included, the LFVHD can be arbitrary large with very large $M$ if the following condition is satisfied:  $M=m_{\tilde{e}_1}\ll m_{\tilde{e}_2}$. In the case of $M$ below 300 GeV, the large Br of LFVHD near the recent experimental report  occurs only in the region having  $M=m_{\tilde{e}_1}$. The BR of LFVHD is constrained by the  $h\gamma\gamma$ coupling deviation, where the largest value is of order $10^{-3}$. With $M$  around 1 TeV, the LFVHD constraint from experiment leads to the consequence that $\mu$  should be smaller than few TeV. The parameter region discussed in this work  can be tested by the LHC and the ILC in coming years.

\section*{Acknowledgments}
This work is funded by Hanoi Pedagogical University 2 under grant number C.2016-18-03.
\appendix
\section{\label{aPV}Analytic expressions of PV-functions}
 Here we list the analytic expressions for calculating one-loop contributions to  LFVHD in the 't Hooft Feynman gauge. They are from \cite{hue}. We would like to stress that these PV-functions were derived from the general form given in \cite{Hooft}, using only the conditions of very small masses of tauon and muon. They are consistent with \cite{bardin}. A more precise and detailed explanation is given in \cite{Ninhthes}. The denominators of the propagators  are denoted by  $D_0=k^2-M_0^2+i\delta$, $D_1=(k-p_1)^2-M_{1}^2+i\delta$, and $D_2=(k+p_2)^2-M_2^2+i\delta$, where $\delta$ is  infinitesimally a positive real quantity. The scalar integrals are defined as
 \bea
 B^{(1)}_0 &\equiv&\frac{\left(2\pi\mu\right)^{4-D}}{i\pi^2}\int \frac{d^D k}{D_0D_1},
 \hs  B^{(2)}_0\equiv \frac{\left(2\pi\mu\right)^{4-D}}{i\pi^2}\int \frac{d^D k}{D_0D_2},  \crn
  B^{(12)}_0 &\equiv& \frac{\left(2\pi\mu\right)^{4-D}}{i\pi^2}\int \frac{d^D k}{D_1D_2}, \hs
 C_0\equiv  C_{0}(M_0,M_1,M_2) =\frac{1}{i\pi^2}\int \frac{d^4 k}{D_0D_1D_2},
 \label{scalrInte}\eea
 where $i=1,2$.
   In addition, $D=4-2\epsilon \leq 4$ is the dimension of the integral;  $M_0,~M_1,~M_2$ are masses of virtual particles in the loop. The momenta satisfy conditions: $p^2_1=m^2_{1},~p^2_2=m^2_{2}$ and $(p_1+p_2)^2=m^2_{h}$. In this work, with $m_1$ and $m_2$ are the respective masses of muon and tauon, and $m_h$ is the SM-like Higgs mass. The tensor integrals are
 \bea
C^{\mu} &=&C^{\mu}(M_0,M_1,M_2)=\frac{1}{i\pi^2}\int \frac{d^4 k\times k^{\mu}}{D_0D_1D_2}\equiv  C_1 p_1^{\mu}+C_2 p_2^{\mu},
 \label{oneloopin1}\eea
where  $B^{(i)}_{0}$ and $C_{0,1,2}$   are PV-functions.  It is well known that $C_{0,1,2}$  are finite,  while  $B^{(i)}_{0}$ and $B^{(12)}_{0}$  are divergent. We define $\Delta_{\epsilon}\equiv \frac{1}{\epsilon}+\ln4\pi-\gamma_E+\ln\mu^2$, where $\gamma_E$ is the  Euler constant.  The divergent parts of the $B$-functions can be determined as $\mathrm{Div}[B^{(i)}_0]= \mathrm{Div}[B^{(12)}_0]= \Delta_{\epsilon}$, then the  finite parts   depend  on the scale of the $\mu$ parameter with the same coefficient of the divergent parts. In order to be consistent with LoopTools, we choose $\mu=1$ GeV.  The analytic formulas of the above PV-functions are:
 \be B^{(i)}_{0}= \mathrm{Div}[B^{(i)}_{0}]+ b^{(i)}_{0,1}, \hs   B^{(12)}_{0}= \mathrm{Div}[B^{(12)}_{0,1,2}]+ b^{(12)}_{0}. \label{B01i}\ee
In the limit $p_1^2,p_2^2\simeq0$ we have
 \bea  b^{(i)}_0 &=& 1-\ln( M_i^2)+\frac{M_0^2}{M_0^2-M_i^2}\ln\frac{M_i^2}{M_0^2},\hs \mathrm{and}\crn
 b_0^{(12)}&=&-\ln (M_1^2) +2 + \sum_{k=1}^2 x_k\ln\left(1-\frac{1}{x_k}\right),\label{b0i}\eea
where $x_k,~(k=1,2)$ are solutions of the equation
 \be  x^2-\left(\frac{m_h^2-M_1^2+M_2^2}{m_h^2}\right)x+\frac{M_2^2-i\delta}{m_h^2}=0.\label{squeq}\ee
 The $C_0$-function was given in \cite{hue} consistent with that discussed in \cite{bardin}, namely
 \be  C_0=\frac{1}{m_h^2}\left[R_0(x_0,x_1)+ R_0(x_0,x_2)-R_0(x_0,x_3)\right] , \label{C0fomula1}\ee
 where
 \be R_0(x_0,x_i) \equiv Li_2(\frac{x_0}{x_0-x_i})- Li_2(\frac{x_0-1}{x_0-x_i}), \label{r0function}\ee
 $Li_2(z)$ is the di-logarithm function,  $x_{1,2}$ are solutions  of  Eq.  (\ref{squeq}),  and $x_{0,3}$ are given as
  \be x_0=\frac{M_2^2-M_0^2}{m_h^2},\hs x_3=\frac{-M_0^2+i\delta}{M_1^2-M_0^2}. \label{x03}\ee
In the limit $p_1^2,p_2^2\rightarrow 0$ the $C_{1,2}$-functions are
{\footnotesize\bea
 C_1 &=& \frac{1}{m_h^2}   \left[b^{(1)}_0 -b_0^{(12)}+(M_2^2-M_0^2)C_0\right],\hs
  C_2 =  -\frac{1}{m_h^2}   \left[b^{(2)}_0 -b_0^{(12)}+(M_1^2-M_0^2)C_0\right]. \label{PV_func}\eea
  }
  If $M_1=M_2$, it can be seen that $b^{(1)}_0=b^{(1)}_0$ and $C_1=-C_2$. The mentioned PV-function is enough to discuss the LFVHD of the models mentioned in this work

 \section{\label{C0C0p}Proving $C'_0(M_1,M_0,M_2)=C_0(M_0,M_1,M_2)$.}
 The  parameterization  of  $C'_0$ is chosen as 
 \be \frac{1}{D'_1D'_2D'_3}=\Gamma(3)\int_0^1dx\int_0^{1-x}\frac{dy}{\left[x D'_1+(1-x-y)D'_2+yD'_3\right]^3},\nn\ee
where $D'_{123}=x D'_1+(1-x-y)D'_2+yD'_3=x\left( q^2-m_0^2\right)+ (1-x-y)\left[ (q+p'_1)^2-m_1^2\right]+ y\left[ (q+p'_1+p'_2)^2-m_1^2\right]$. From the equalities $ (p'_1+p'_2)^2=m_h^2$ and $2p'_1.p'_2=(p'_1+p'_2)^2-p'_1-p'_2=m_h^2-p'_1-p'_2$, we get
$$ C'_0=-\int^1_0dx\int^{1-x}_0 \frac{dy}{(1-x-y)m_1^2+xm_0^2+ym_2^2-xym_h^2}.$$
Comparing with $C_0$ shown in \cite{hue}, namely
$$ C_0= -\int^1_0dx\int^{1-x}_0 \frac{dy}{(1-x-y)M_0^2+xM_1^2+yM_2^2-xym_h^2},$$
we have the same conclusion as shown in (\ref{C0func1}).

\section{ \label{apo1}Analytic approximation from \cite{apo1} }
Here we list the needed approximation $\frac{m_h^2}{4 m_V^2}, \frac{m_h^2}{4 m_N^2}\ll 1$.
The general $C_0$ is defined as in (\ref{scalrInte}). After using the Feynman parameterization we get an expression of  $C_0$ that is the same as mentioned above, and the $C_{1,2}$-functions are
{\bea  C_1(M_0,M_1,M_2)&=&- C_2(M_0,M_1,M_2)\crn
& =&-\int_0^1dx\int_0^{1-x}  \frac{xdy}{(1-x-y)M_0^2+x M_1^2+y M_2^2- xy m_h^2}.\label{C0xy} \eea
  }
The approximation in some special cases are
{\footnotesize
\bea  C'''_0(m_N,m_V,m_V)&=& -\frac{1}{m_V^2}\left(\frac{1}{1-\lambda_N}+ \frac{\lambda_N\ln\lambda_N}{(1-\lambda_N)^2} \right. \crn
&+&\left. \frac{m_h^2}{4 m_V^2}\times \frac{1-6\lambda_N+3\lambda_N^2+2\lambda_N^3-6\lambda_N^2\ln\lambda_N}{2(1-\lambda_N)^4} \right)+  \mathcal{O}\left(\left[\frac{m_h^2}{4 m_V^2}\right]^2\right),\crn
C'''_0(m_V,m_N,m_N)&=&-\frac{1}{m_V^2(1-\lambda_N)^2}\left(\frac{}{}-1+\lambda_N-\ln\lambda_N \right. \crn
&+&\left. \frac{m_h^2}{4 m_V^2}\times \frac{2 +3\lambda_N -6\lambda_N^2 +\lambda_N^3 +6\lambda_N\ln\lambda_N}{3\lambda_N (1-\lambda_N)^2} \right)+  \mathcal{O}\left(\left[\frac{m_h^2}{4 m_V^2}\right]^2\right), \crn
C'''_0(M_0,M_0,M_2)&=& -\frac{1}{M^2_0}\left(\frac{1-\lambda_N +\lambda_N\ln\lambda_N }{(\lambda_N-1)^2}-\frac{m_h^2}{4M_0^2}\times \frac{1+4 \lambda_N-5\lambda_N^2+2\lambda_N(2+\lambda_N)\ln \lambda_N}{(\lambda_N-1)^4} \right),  \nn\eea
\bea
 C'''_1(m_V,m_N,m_N)&=&- C'''_2(m_V,m_N,m_N)=-\frac{1}{4m_V^2(1-\lambda_N)^3}\left(3 -4\lambda_N +\lambda_N^2 +2\ln\lambda_N \right. \crn
&+&\left. \frac{m_h^2}{ m_V^2}\times \frac{-3 -10\lambda_N +18\lambda_N^2 -6\lambda_N^3 +\lambda_N^4 -12\lambda_N\ln\lambda_N}{9\lambda_N(1-\lambda_N)^2}\right) +\mathcal{O}\left(\left[\frac{m_h^2}{4 m_V^2}\right]^2\right), \crn
C'''_1(m_N,m_V,m_V)&=&-C'''_2(m_N,m_V,m_V)=-\frac{1}{4m_V^2(1-\lambda_N)^3}\left(-1 +4\lambda_N -3\lambda_N^2 +2\lambda_N^2\ln\lambda_N \right. \crn
&+&\left. \frac{m_h^2}{ m_V^2}\times \frac{-1+6\lambda_N -18\lambda_N^2 +10\lambda_N^3 +3\lambda_N^4 -12\lambda_N^3\ln\lambda_N}{9 (1-\lambda_N)^2}\right)  +\mathcal{O}\left(\left[\frac{m_h^2}{4 m_V^2}\right]^2\right). \crn   \label{apoapp1}\eea
}


\end{document}